\documentclass[10pt]{article}
\usepackage{multirow}
\usepackage[nottoc]{tocbibind}
\usepackage{geometry}
\usepackage[affil-it, auth-sc]{authblk}
\usepackage[utf8]{inputenc}
\usepackage{graphicx}
\usepackage{amsthm}
\usepackage{amssymb}
\usepackage{amsmath}
\usepackage{color, soul}
\usepackage{verbatim}
\usepackage{lineno}
\usepackage{todonotes}
\usepackage{subcaption}
\usepackage{ulem}
\usepackage{float}
\usepackage[pdftex,colorlinks,bookmarks,linkcolor=darkred, citecolor=darkgreen]{hyperref}
\usepackage[ampersand]{easylist}
\usepackage{babel}
\usepackage{bm}
\usepackage{feynmf}
\usepackage[sort&compress,numbers]{natbib}
\usepackage{doi}
\usepackage{eso-pic}
\presetkeys{todonotes}{inline, color=blue!30, size=\small}{}
\definecolor{darkred}{rgb}{0.9, 0.0, 0.0}
\definecolor{darkgreen}{rgb}{0.0, 0.5, 0.0}
\allowdisplaybreaks

\def\slash#1{#1\!\!\!/}
\newcommand{\nl}{\nonumber \\ }
\newcommand{\order}{{\cal O}}
\newcommand{\x}{x}
\newcommand{\y}{y}
\newcommand{\numl}{n_\ell}

\begin{document}

\AddToShipoutPictureFG*{\AtPageUpperLeft{\put(-60,-75){\makebox[\paperwidth][r]{FERMILAB-PUB-21-378-T,~LA-UR-21-30632,~USTC-ICTS/PCFT-22-13}}}}

\title{\Large\bf Theory of QED radiative corrections to neutrino scattering at accelerator energies}

\author[1,2,3]{Oleksandr~Tomalak}
\author[1,4,5]{Qing~Chen}
\author[1,2]{Richard~J.~Hill}
\author[6]{Kevin~S.~McFarland}
\author[6]{Clarence~Wret}
\affil[1]{University of Kentucky, Department of Physics and Astronomy, Lexington, KY 40506 USA \vspace{1.2mm}}
\affil[2]{Fermilab, Theoretical Physics Department, Batavia, IL 60510, USA \vspace{1.2mm}}
\affil[3]{Theoretical Division, Los Alamos National Laboratory, Los Alamos, NM 87545, USA \vspace{1.2mm}}
\affil[4]{Interdisciplinary Center for Theoretical Study, University of Science and Technology of China, Hefei, Anhui 230026, China \vspace{1.2mm}}
\affil[5]{Peng Huanwu Center for Fundamental Theory, Hefei, Anhui 230026, China \vspace{1.2mm}}
\affil[6]{University of Rochester, Department of Physics and Astronomy, Rochester, NY 14627, USA \vspace{1.2mm}}

\date{\today}

\maketitle
\begin{abstract}
Control over quantum electrodynamics (QED) radiative corrections is critical for precise determination of neutrino oscillation probabilities from observed (anti)neutrino detection rates. It is particularly important to understand any difference between such corrections for different flavors of (anti)neutrinos in charged-current interactions. We provide theoretical foundations for calculating these corrections. Using effective field theory, the corrections are shown to factorize into soft, collinear, and hard functions. The soft and collinear functions contain large logarithms in perturbation theory but are computable from QED. The hard function parametrizes hadronic structure but is free from large logarithms. Using a simple model for the hard function, we investigate the numerical impact of QED corrections in charged-current (anti)neutrino-nucleon elastic cross sections and cross-section ratios at GeV energies. We consider the implications of mass singularity theorems that govern the lepton-mass dependence of cross sections for sufficiently inclusive observables and demonstrate the cancellation of leading hadronic and nuclear corrections in phenomenologically relevant observables.
\end{abstract}

\newpage
\tableofcontents
\newpage

\section{Introduction}

The program of studying neutrino oscillations to answer fundamental questions about neutrinos requires precise interaction cross sections. In particular, the ratio of cross sections for electron versus muon (anti)neutrinos is critical to phenomenology and difficult to constrain by direct experimental measurements. This ratio is impacted by electromagnetic radiative corrections~\cite{DeRujula:1979grv,Day:2012gb}, and it is thus imperative to compute these corrections and to constrain their uncertainties. In this paper, we construct the theoretical framework to determine radiative corrections to neutrino cross sections and cross-section ratios.\footnote{For a concise summary of the theoretical formalism and implications for $\nu_e/\nu_\mu$ ratios see Ref.~\cite{Tomalak:2021qrg}.} We consider the experimental conditions under which quantum electrodynamics (QED) radiative corrections must be computed and discuss the practical implications for neutrino oscillation analyses.

We formulate the problem of QED radiative corrections to (anti)neutrino scattering processes in the framework of effective field theory. This allows a clean separation between corrections that are rigorously computable within QED and corrections that depend on hadronic physics. When the (anti)neutrino energy $E_\nu$ is much larger than the charged lepton mass $m_\ell$ (i.e., $E_\nu \gg m_\ell$) the former, soft and collinear, corrections contain flavor-dependent large logarithm enhancements, $\ln(E_\nu/m_\ell)$; these corrections depend on detailed experimental conditions but can be computed perturbatively. The latter, hard, corrections are subject to hadronic uncertainty but are independent of the charged lepton mass and cancel in ratios of cross sections for different lepton flavors (i.e., electron and muon) involving the same hadronic kinematics. Using effective field theory methods, we resum large logarithms in perturbation theory and estimate the uncertainty from neglected higher-order corrections. We investigate the insensitivity of certain cross-section ratios to hadronic uncertainty at the nucleon level. While the detailed incorporation of nucleon-level corrections into nuclear cross sections is beyond the scope of this work, we also discuss how nuclear effects impact these flavor ratios.

The paper is organized as follows. Section~\ref{sec:overview} provides an overview of the experimental setups and analysis strategies for current and near-future experiments with GeV energy (anti)neutrino beams, which are the primary focus of our work. Section~\ref{sec:static} studies the limit of a nonrelativistic nucleon, $m_\ell \ll E_\nu \ll M$, where $M$ denotes the nucleon mass. This formal limit is used to introduce a class of observables that incorporates real photon radiation and to motivate the more general theoretical formalism that follows. In Sec.~\ref{sec:factorization}, we discuss factorization for hard scattering exclusive processes. We establish the relevant factorization theorem and evaluate the separate contributions of soft, collinear, and hard photons. Section~\ref{sec:inclusive} considers the inclusion of noncollinear energetic photons in experimental observables. Section~\ref{sec:pheno} illustrates the phenomenological consequences of our analysis; we study observables representing electron- and muon-induced electromagnetic showers at neutrino detectors, present cross sections and cross-section ratios, and compare them with existing experimental data. Section~\ref{sec:summ} contains a summary and outlook.

\section{Experimental signatures of events with photons \label{sec:overview}}

Our focus is on accelerator neutrino experiments with $E_\nu \sim 1 {\rm~GeV}$. In the typical setup, a beam consisting primarily of muon flavor neutrinos or antineutrinos is created, and one of the goals in oscillation experiments is to determine the rate at which electron flavor (anti)neutrinos are observed in the beam after propagating some distance. In a simplified description, near detectors are used to constrain (anti)neutrino flux and muon (anti)neutrino cross sections, while theory is needed to determine the effect of differences between electron and muon (anti)neutrino cross sections.

We consider two basic classes of observables: exclusive and inclusive. Exclusive observables are subdivided into two general classes: first, the charged-current elastic process accompanied by only soft photons of energy smaller than some value $\Delta E$; second, the same as the first, but including also energetic collinear photons, within a cone of angular size $\Delta \theta$. In both cases, distributions are computed with respect to the invariant momentum transfer $Q^2$ between the initial (anti)neutrino and the final-state lepton jet. For small $\Delta E$ and small $\Delta \theta$, $Q^2$ may be identified with the hadronic momentum transfer for elastic (anti)neutrino-nucleon scattering. For the inclusive case, we consider the class of observables that also includes energetic noncollinear photons in the cross section.

The parameter $\Delta E$ is a soft-photon cutoff parameter: photons with energy smaller than $\Delta E$ are assumed to be unseen by the detector. The appropriate choice of $\Delta E$ depends on the specific process and observable under consideration and on properties of the detector. For example, photons that interact repeatedly by Compton scattering in a diffuse way throughout the detector, and photons that destructively interact by $e^+e^-$ pair production and therefore appear spatially and directionally local within the detector, will leave very different signatures. Photons in the former class are effectively ``invisible" as separate objects because they do not leave a localized signature, although they might contribute to a calorimetric measurement of nonlepton energy in some detectors. In polystyrene scintillator or water, Compton scattering dominates the photon cross section from energies between $20$~keV and $30$ or $25$~MeV, respectively, whereas in liquid argon detectors, Compton scattering dominates photon interactions only for energies between $150$~keV and $12$~MeV~\cite{https://doi.org/10.18434/t48g6x}. Therefore, $\Delta E$ will be between ten and tens of MeV for experiments using these detectors.

The appropriate choice of the parameter $\Delta\theta$ depends on the flavor of the (anti)neutrino and the detector technology. In nearly all detector technologies deployed in oscillation experiments at accelerator energies, electrons and positrons ($e^\mp$) from $\nu_e$ and $\bar{\nu}_e$ charged-current interactions are observed through electromagnetic showers, i.e., a cascade of bremsstrahlung photons from electrons and positrons and $e^+e^-$ pair production from these photons. Radiated photons that are roughly collinear with the initial $e^-$ or $e^+$ will generally not be distinguished from this electromagnetic shower and must be included as part of the observable. For a given charged lepton direction, we consider this observable to consist of all events with a specified total lepton jet energy, after summing the charged lepton energy and the energy of photons within a cone of angular size $\Delta \theta$. To estimate the appropriate $\Delta\theta$, we consider the Moli\`{e}re radius of the electromagnetic shower initiated by the primary $e^\pm$ and the length of the mean shower maximum and find the angle which would place the photon within the Moli\`{e}re radius at shower maximum~\cite{Zyla:2020zbs}. This angle decreases logarithmically with the primary $e^\pm$ energy. Consider for example a typical $500$~MeV $e^\pm$ energy; for polystyrene scintillator, water, and liquid argon, the angle is $9^\circ$, $10^\circ$, and $16^\circ$, respectively. For all three materials, the angle is a factor $\sim 2$ lower at $3$~GeV $e^\pm$ energies. Accordingly, a range of $\Delta\theta$ is interesting to consider: from $5^\circ$ for the NO$\nu$A experiment, to $10^\circ$ for T2K, Hyper-K, and DUNE, to $20^\circ$ for the short-baseline neutrino program at Fermilab~\cite{SBN:2015bmn}. For low-density tracking detectors in a magnetic field, such as in the existing T2K and proposed DUNE near detectors, photons would generally not be clustered together with the primary $e^\pm$.

In a tracking target detector, such as segmented scintillator, a gaseous tracking detector in a magnetic field, or in a liquid argon time projection chamber, the energy of final-state muons from $\nu_\mu$ or $\bar{\nu}_\mu$ interactions is typically determined using the range of the muon in the detector or its curvature in the magnetic field. Accordingly, collinear photons would usually not contribute to the reconstructed muon energy. On the other hand, in a water Cherenkov detector, photons would contribute to the observed energy associated with a muon if their angle from the muon direction were consistent with multiple scattering over the range of the muon. We determine the relevant cone size by computing the Gaussian core of multiple scattering~\cite{Zyla:2020zbs} for a muon. A similar criterion would hold for identification of a photon in the detector; the photon typically would not be reconstructed as distinct from the muon if it is emitted along the path of the muon. The resulting angles are similar in the relevant materials (polystyrene scintillator, water and liquid argon), with water 10\% larger than polystyrene and liquid argon another 10\% larger. The angles depend weakly on the kinetic energy of the muon, with angles of $\approx 2^\circ$ for the few hundred MeV kinetic energies relevant for T2K in scintillator, dropping to $\approx 1.5^\circ$ at few GeV energies. Accordingly, a cone of size $\Delta \theta = 2^\circ$ around the muon is a reasonable separation between photons in Cherenkov detectors\footnote{This analysis omits an important detail for large water Cherenkov detectors such as Super-Kamiokande, where large photosensors and scattering of Cherenkov light play significant roles in the angular separation between adjacent features~\cite{Abe:2013gga}. Such effects in a real Cherenkov detector could result in merging of angular features on scales significantly larger than the $2^\circ$ inherent resolution calculated above, but a detailed detector simulation would be required to assess this effect.} that would be identified as collinear with muons and those which could be seen as separate particles in the reconstruction.

\section{Static limit in (anti)neutrino-nucleon scattering \label{sec:static}}

\begin{figure}[tb]
\centering
\includegraphics[width=0.2\textwidth]{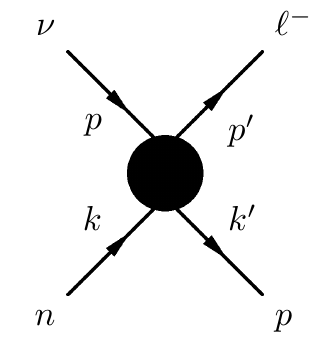}
\includegraphics[width=0.2\textwidth]{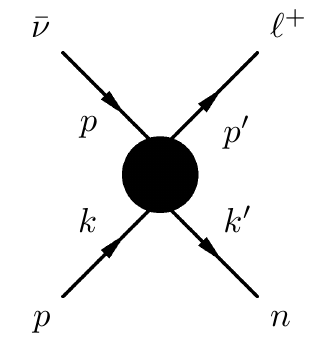}
\caption{Kinematics of charged-current elastic scattering for neutrino (left) and antineutrino (right). \label{fig:CCQE_kin}}
\end{figure}

The tree-level neutrino scattering process is displayed in Fig.~\ref{fig:CCQE_kin}:
\begin{align}
 \nu_\ell(p) + n(k) \to \ell^-(p^\prime) + p(k^\prime) \,. \label{eq:nuprocess}
\end{align}
We consider also the related antineutrino process:
\begin{align}
 \bar{\nu}_\ell(p) + p(k) \to \ell^+(p^\prime) + n(k^\prime) \,. \label{eq:antinuprocess}
\end{align}

It is instructive to first study charged-current elastic scattering in the formal static limit, $m_\ell \ll E_\nu \ll M$. This is an important exactly calculable limit for the more general case involving nontrivial hadronic structure, analogous to the well-known McKinley-Feshbach correction for electron-proton scattering~\cite{McKinley:1948zz,Dalitz:1951ah}. Beyond this formal utility, however, the static limit captures the leading logarithmically enhanced contributions to radiative corrections in perturbation theory and thus correctly describes the leading logarithm approximation for flavor ratios of interest to neutrino oscillation experiments. In the following subsections, we compute this benchmark cross section.

\subsection{Lagrangian and leading-order cross section \label{sec:static_pheno}}

The static limit $M\to \infty$, as well as $E_\nu/M$ power corrections, may be computed systematically in nonrelativistic (NR) effective field theory. The appropriate effective theory consists of NRQED for relativistic charged leptons and nonrelativistic nucleons~\cite{Caswell:1985ui,Pachucki:1996zza,Pineda:1997bj,Jentschura:2005xu,Hill:2012rh}, supplemented by the four-fermion interaction between leptons and heavy nucleons:
\begin{align}
{\cal L}_{\rm eff} &= - \sqrt{2} \mathrm{G}_{\rm F} V_{ud} \, \bar{\ell} \gamma^\mu \mathrm{P}_\mathrm{L} \nu_\ell \, \bar{h}_v^{(p)} \gamma_\mu \big[ c_V + c_A \gamma_5 \big] h_v^{(n)} + {\rm h.c.}\,, \label{eq:static_L}
\end{align}
with $\mathrm{P}_\mathrm{L} = \left( 1 - \gamma_5 \right)/2$. Here $h_v^{(p)}$ and $h_v^{(n)}$ denote heavy particle fields for the proton and neutron, respectively~\cite{Caswell:1985ui,Neubert:1993mb,Manohar:2000dt, Hill:2016gdf}; subscript $v$ denotes the laboratory frame reference vector $v^\mu=(1,0,0,0)$. We have expressed the operator coefficients in terms of the scale-independent Fermi constant $\mathrm{G}_{\rm F}$ and the Cabibbo-Kobayashi-Maskawa(CKM) matrix element $V_{ud}$. At tree level, the effective operator coefficients are related to familiar nucleon structure parameters: $c_V \to g_V \approx 1$ and $c_A \to g_A \approx -1.27$. Beyond tree level, we define operators and coefficients in the $\overline{\rm MS}$ scheme at renormalization scale $\mu$. The anomalous dimension of both effective operators is readily calculated:
\begin{align}
 \frac{\mathrm{d} \ln c_i}{\mathrm{d}\ln \mu^2} = -\frac{3}{8} \frac{\alpha}{\pi} + \order \left( \alpha^2 \right) .
\end{align}
The tree-level differential cross section for the processes~(\ref{eq:nuprocess}) and~(\ref{eq:antinuprocess}) is given by~\cite{Fukugita:2004cq,Raha:2011aa}
\begin{align}
 \frac{\mathrm{d}\sigma_{\rm LO}}{\mathrm{d} Q^2} &= \frac{\mathrm{G}_{\rm F}^2 |V_{ud}|^2}{2\pi} \bigg[ c_V^2 + c_A^2 - (c_V^2 - c_A^2) \frac{Q^2+m^2_\ell}{4E_\nu^2} \bigg] \,, \label{eq:static_diff}
\end{align}
where $Q^2=-(p^\prime-p)^2$ is the momentum transfer between initial and final lepton states. The total cross section at leading order is
\begin{align}
 \sigma_{\rm LO} &= \frac{\mathrm{G}_{\rm F}^2 |V_{ud}|^2}{\pi} \left( c_V^2 + 3 c_A^2 \right) E_\nu \sqrt{E_\nu^2-m_\ell^2}\,. \label{eq:static_tot}
\end{align}
Note that Eqs.~(\ref{eq:static_diff}) and (\ref{eq:static_tot}) are valid for arbitrary lepton mass; below we will focus on the limit $m_\ell/E_\nu \to 0$. The cross sections for neutrino and antineutrino scattering are identical in the static limit.

\subsection{Charged-current elastic process with one soft photon}

\begin{figure}[t]
 \centering
 \includegraphics[height=0.32\textwidth]{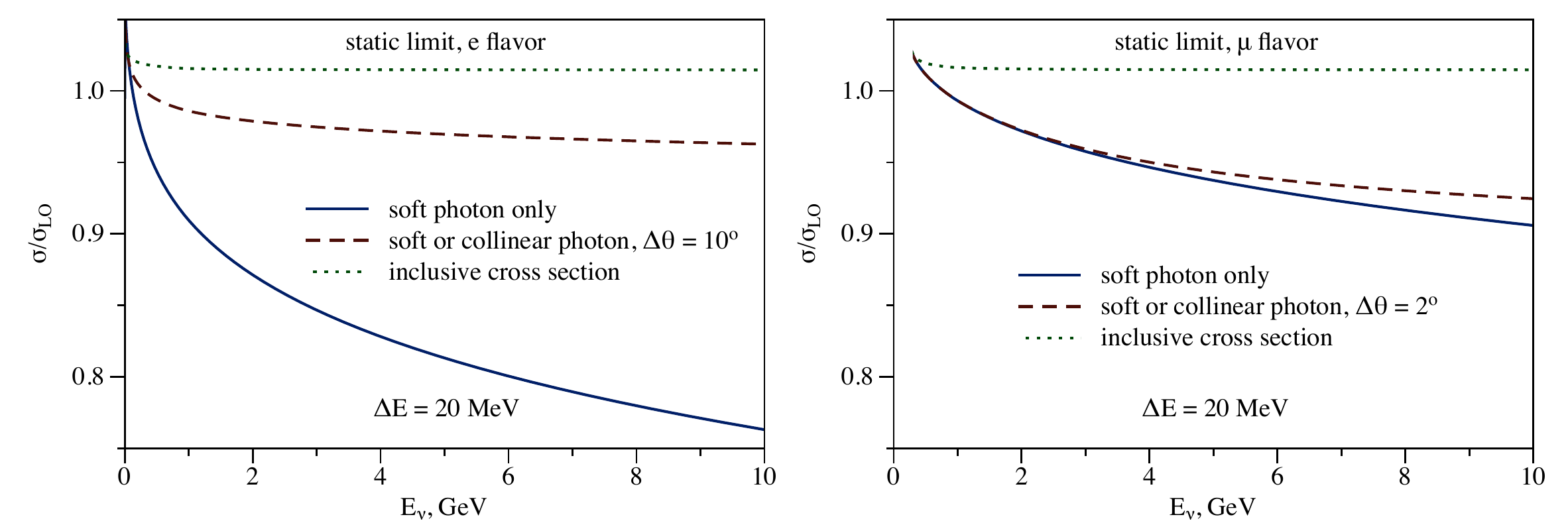}
 \caption{Ratio of one-loop cross section to the tree-level result in the static limit, for scattering of electron (anti)neutrinos (left plot) and muon (anti)neutrinos (right plot). The observable including only the radiation of one soft photon of energy below $\Delta E = 20~\mathrm{MeV}$ is shown by the blue solid curve. The correction including also radiation of one collinear hard photon is represented by the red dashed line, with cone sizes $\Delta \theta=10^\circ$ and $2^\circ$ appropriate for electrons and muons, respectively. Inclusive cross sections are shown by the green dotted line. All curves are shown with the choice of the scale corresponding to the physical nucleon mass $\mu = M$. \label{fig:static_limit}}
\end{figure}

In the static limit, radiative corrections to the tree-level result in Eq.~(\ref{eq:static_diff}) can be computed using the Feynman rules of the effective Lagrangian of Eq.~(\ref{eq:static_L}). Consider first the $\order ( \alpha )$ correction to the observable that does not distinguish between the elastic process and the process with radiation of one photon with energy below $\Delta E$:
\begin{equation}
 \mathrm{d}\sigma^{\rm soft} = \left[ 1 + \frac{\alpha}{\pi} \delta^{\rm soft} \right] \mathrm{d}\sigma_\mathrm{LO} \,. \label{eq:static_soft}
\end{equation}
The correction $\delta^{\rm soft}$ is the sum of virtual corrections $\delta_v$ and real soft-photon corrections $\delta_s$. Note that the superscript ``soft" here denotes the observable including radiation of only soft photons and does not imply restriction to the soft region of virtual diagrams. In the limit of small charged lepton mass $m_\ell/E_\nu \ll 1$, we have
\begin{align}
 \delta_v & = 2 \left( 1 - \ln \frac{2 E_\nu}{m_\ell} \right) \ln \frac{m_\ell}{\lambda_\gamma} - \ln^2 \frac{2 E_\nu}{m_\ell} + \ln \frac{2 E_\nu}{m_\ell} + \frac{3}{4} \ln \frac{\mu^2}{m^2_\ell} + \frac{5}{6} \pi^2\,, \\
 \delta_s & = 2 \left( 1 - \ln \frac{2 E_\nu}{m_\ell} \right)\ln \frac{\lambda_\gamma}{2 \Delta E} - \ln^2 \frac{2 E_\nu}{m_\ell} + \ln \frac{2 E_\nu}{m_\ell} + 1 - \frac{1}{6} \pi^2 \,, \label{fig:soft_photon_static}
\end{align}
with the photon mass regulator $\lambda_\gamma$. The total correction is thus
\begin{align}
\delta^{\rm soft} = \delta_s + \delta_v = 2 \left( 1 - \ln \frac{2 E_\nu}{m_\ell} \right) \ln \frac{E_\nu}{\Delta E} + \frac{3}{4} \ln \frac{\mu^2}{m^2_\ell} + 1 + \frac{2}{3}\pi^2 \,, \label{eqn:delta_soft_static}
\end{align}
where, as required, the result does not depend on the photon mass regulator $\lambda_\gamma$~\cite{Bloch:1937pw,Yennie:1961ad,Kinoshita:1962ur,Lee:1964is}. This correction is shown as a function of (anti)neutrino energy by the blue solid line in Fig.~\ref{fig:static_limit}.\footnote{It is interesting to note the cancellation of terms in Eq.~(\ref{eqn:delta_soft_static}) involving double logarithms in the lepton mass, $\sim \ln^2(m_\ell)$. This is a consequence of the Kinoshita–Lee–Nauenberg (KLN) theorem~\cite{Bloch:1937pw,Yennie:1961ad,Kinoshita:1962ur,Lee:1964is} and finiteness at $m_\ell\to 0$ of the cross section including real collinear radiation.} The cross-section corrections are larger in the case of the electron flavor, which is subject to larger kinematic logarithms.

\subsection{Charged-current elastic process with one soft or collinear photon \label{sec:static_soft_or_collinear}}

Let us similarly consider the charged-current elastic observable that includes energetic collinear radiation. For definiteness, the cone observable is defined by fixing the charged lepton direction and allowing collinear radiation within $\Delta \theta$ of this fixed direction. An explicit computation yields the relative correction $\delta^\mathrm{jet}$, including soft and virtual parts as above, to the leading-order cross section:
\begin{align}
 \mathrm{d} \sigma^\mathrm{jet}& = \left( 1+ \frac{\alpha}{\pi} \delta^\mathrm{jet}\right) \mathrm{d} \sigma^\mathrm{LO} , \\
\delta^\mathrm{jet} &= 2 \left( 1 - \ln \frac{2 E_\nu}{m_\ell} \right) \ln \frac{E_\nu}{\Delta E} + \frac{3}{4} \ln \frac{\mu^2}{m^2_\ell} + \frac{13}{4} + \frac{2\pi^2}{3} + \left( \ln \left( 1 + \eta^2 \right) - \frac{\eta^2}{1+\eta^2} \right) \ln \frac{E_\nu}{\Delta E} -\left(\mathrm{tan}^{-1} \eta \right)^2 \, \nl
&+ \frac{1}{2} \mathrm{Li}_2 \left( -\eta^2 \right) + \frac{1}{4} \ln^2 \left( 1 + \eta^2 \right) + \frac{1}{4} \left(\frac{2}{1+\eta^2} - \frac{1}{\eta^2} -3 \right) \ln \left( 1 + \eta^2 \right) - \left( 1 + \frac{1}{1+\eta^2}\right) \frac{\mathrm{tan}^{-1} \eta }{\eta} \,, \label{eqn:delta_jet_static}
\end{align}
with the dimensionless parameter $\eta = \Delta \theta E_\nu/m_\ell$. At $\Delta\theta \to 0$, the cross section reduces smoothly to the soft-photon-only result, $\delta^{\rm jet} \to \delta^{\rm soft}$. In the limit of very small charged lepton mass $m_\ell/E_\nu \ll \Delta\theta$, the cross section becomes
\begin{equation}
\delta^\mathrm{jet} \underset{m_\ell \ll \Delta \theta E_\nu}{\longrightarrow} \left( 1+ \ln \frac{\left(\Delta \theta\right)^2}{4}\right) \ln \frac{E_\nu}{\Delta E} - \frac{3}{4} \ln \frac{\left( \Delta \theta\right)^2}{4} +\frac{3}{4} \ln \frac{\mu^2}{4 E^2_\nu} +\frac{\pi^2}{3} + \frac{13}{4}. \label{eqn:delta_jet_static_limit_m}
\end{equation}
Note that, in contrast to Eq.~(\ref{eqn:delta_soft_static}), the collinear singularity $\sim \ln m_\ell$ in Eq.~(\ref{eqn:delta_jet_static_limit_m}) is regulated by the jet size $\Delta \theta$. Flavor dependence of radiative corrections enters via the parameter $\eta$ appearing in Eq.~(\ref{eqn:delta_jet_static}) but cancels in the massless limit $m_\ell\to 0$, in agreement with the KLN theorem~\cite{Bloch:1937pw,Yennie:1961ad,Kinoshita:1962ur,Lee:1964is}. The correction to the jet observable in Eq.~(\ref{eqn:delta_jet_static}) is shown as a function of (anti)neutrino energy by the red dashed line in Fig.~\ref{fig:static_limit}.

\subsection{Inclusive cross section in the static limit}

Finally, let us describe the contribution of noncollinear energetic photons. Integrating over the total phase space for photons above $\Delta E$ in energy, we obtain the real hard-photon contribution:
\begin{align}
 \sigma_h
 = \frac{\alpha}{\pi} 
 \delta_h 
 \sigma_\mathrm{LO} \,, \qquad
 \delta_h
 = - 2 \left( 1- \ln \frac{2 E_\nu}{m_\ell} \right)\ln \frac{E_\nu}{\Delta E} - \frac{3}{2} \ln \frac{2 E_\nu}{m_\ell} + \frac{13}{4} - \frac{\pi^2}{3} .
\end{align}
The resulting inclusive cross section $\sigma^{\rm tot}$ is factorizable:
\begin{equation}
 \sigma^{\rm tot} = \left( 1 + \frac{\alpha}{\pi} \delta^{\rm tot} \right) \sigma_\mathrm{LO}.
\end{equation}
In the limit of small charged lepton mass $m_\ell/E_\nu \ll 1$, the correction $\delta^{\rm tot}$ does not depend on lepton flavor and is given by
\begin{align}
 \delta^{\rm tot} & = \delta_v + \delta_s + \delta_h = \frac{3}{4} \ln \frac{\mu^2}{4 E^2_\nu} + \frac{17}{4} + \frac{\pi^2}{3}. \label{eqn:deltaSimple}
\end{align}
This simple calculation provides another example (cf.~\cite{Tomalak:2019ibg}) of the cancellation of Sudakov double logarithms~\cite{Sudakov:1954sw} considering the inclusive cross section with both soft and hard radiation. We present the ratio of the inclusive cross section to the tree-level result in Fig.~\ref{fig:static_limit} by the green dotted line and observe a cancellation of virtual and real contributions entering the cross section with opposite signs. Lepton-mass corrections enter with $m^2_\ell$ suppression in agreement with the KLN theorem~\cite{Bloch:1937pw,Yennie:1961ad,Kinoshita:1962ur,Lee:1964is}.\footnote{The correction to the total cross section of Eq.~(\ref{eq:static_tot}) with a finite charged lepton mass can be written in terms of the charged lepton velocity $\beta$ as
\begin{equation}
 \delta = \left( 4 + \frac{2}{\beta} \ln \frac{1+\beta}{1-\beta} \right) \ln \frac{1+\beta}{2\beta} +\frac{4}{\beta} \mathrm{Li}_2 \frac{1-\beta}{1+\beta} + \frac{ 7 -16 \beta + 3 \beta^2}{8 \beta} \ln \frac{1+\beta}{1-\beta} + \frac{3}{4} \ln \frac{\mu^2}{m^2_\ell} + \frac{17}{4} + \frac{\pi^2}{3\beta}. \label{eqn:deltaFiniteLeptonMass}
\end{equation}}

In the static limit, there is no dependence on hadronic structure (beyond the numerical values of $c_V$ and $c_A$), and hence the entire calculation may be performed perturbatively. For the general case, where $E_\nu$ is not small compared to hadronic scales, it is imperative to disentangle the hard scale, containing nonperturbative hadronic physics, from the perturbative soft and collinear scales.

\section{Factorization \label{sec:factorization}}

The process under consideration is charged-current (anti)neutrino-nucleon elastic scattering: $\nu_\ell n \to \ell^- p (\gamma)$, $\bar{\nu}_\ell p \to \ell^+ n (\gamma)$, where ``$(\gamma)$" denotes possible photon emission. Our default observable includes only soft or collinear radiation of real photons, with the precise definition for soft and collinear given in Sec.~\ref{sec:overview}. Throughout the analysis, we consider fixed initial- and final-state nucleon momenta. In a final step for phenomenological applications, we may integrate over the final-state nucleon kinematics.

For both virtual and real radiative corrections to the tree-level process, we may decompose the total momentum integration into regions. It is convenient to work in the light-cone basis, decomposing an arbitrary four-vector $L^\mu$ as
\begin{align}
 L^\mu = n\cdot L \frac{\bar{n}^\mu}{2} + \bar{n}\cdot L \frac{n^\mu}{2} + L_\perp^\mu \equiv L_+^\mu + L_-^\mu + L_\perp^\mu \leftrightarrow (n\cdot L, \, \bar{n}\cdot L,\, L_\perp) \,,
\end{align}
where $n^\mu$ and $\bar{n}^\mu$ are conventional four-vectors satisfying $n^2=\bar{n}^2=0$ and $n\cdot \bar{n} = 2$. For a charged lepton moving in the $z$ direction, we choose $n^\mu = (1,0,0,1)$ and $\bar{n}^\mu=(1,0,0,-1)$, with the four-velocity of the initial-state nucleon in its rest frame $v^\mu=(n^\mu + \bar{n}^\mu)/2=(1,0,0,0)$. The relevant momentum regions are
\begin{align}
 {\rm soft:} \quad & L^\mu \sim \Lambda (\lambda, \lambda, \lambda) \,, \nl
 {\rm collinear:} \quad & L^\mu \sim \Lambda (\lambda, 1, \sqrt{\lambda} ) \,, \nl
 {\rm hard:} \quad & L^\mu \sim \Lambda ( 1,1,1) \,,
\end{align}
where the hard scale is $\Lambda \sim M \sim E_\nu \sim Q$. The dimensionless expansion parameter $\lambda\sim \Delta E/\Lambda$ is determined by experimental conditions that dictate the fraction of energy allowed in soft radiation. The lepton mass satisfies $m_\ell \lesssim \sqrt{\lambda}\Lambda$, and the jet angular resolution satisfies $\Delta \theta \lesssim \sqrt{\lambda}$.

In our default factorization analysis represented by Eq.~(\ref{eq:factorization_formula}) below, we employ the formal power counting $m_\ell/\Lambda_{\rm hard} \ll 1$. For the muon, an alternative counting $\Delta E \ll m_\mu \sim \Lambda_{\rm hard}$ would group the muon mass with other ``hard" scales, resulting in a simpler soft-hard factorization formula without collinear function. Since ``large logarithms" are not extremely large for the muon, it may seem preferable to adopt this simpler description. However, muon power corrections are well described by our power-counting analysis and are numerically small compared to other uncertainties. Moreover, important phenomenological features, such as the insensitivity of flavor ratios to hadronic and nuclear uncertainties, can be readily seen by adopting a unified description of electron and muon observables. A more detailed effective theory analysis of soft-hard versus soft-collinear-hard factorization is left to future work.

\subsection{Factorization in the static limit \label{sec:static_factorization}}

\begin{figure}[t]
 \begin{center}
 \begin{align}
 \parbox{40mm}{\includegraphics[width=0.2\textwidth]{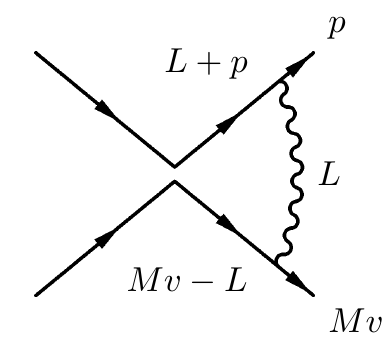}} &= \parbox{40mm}{\includegraphics[width=0.2\textwidth]{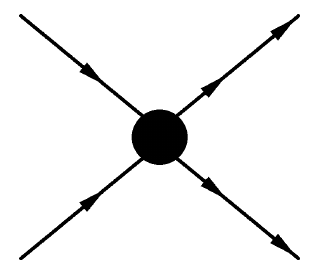}} + \parbox{40mm}{\includegraphics[width=0.2\textwidth]{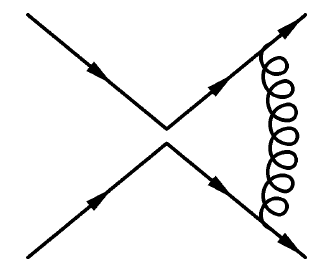}} + \parbox{40mm}{\includegraphics[width=0.2\textwidth]{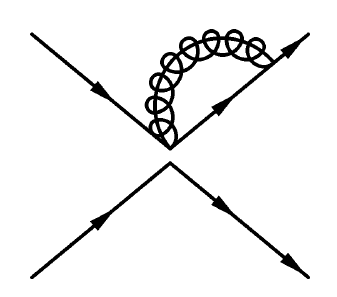}} \nonumber
 \end{align}
 \caption{Decomposition of full theory diagram into hard, soft and collinear momentum regions. \label{fig:static_diagrams}}
\end{center}
\end{figure}
In preparation for the more general case, let us examine the contributions from different momentum regions to the cross section in the static limit. As an example, consider the one-loop virtual correction depicted in Fig.~\ref{fig:static_diagrams} (in Feynman gauge, with photon mass regulator $\lambda_\gamma$):
\begin{align}
 {\cal M} &= i e^2 \int \frac{d^dL}{ (2\pi)^d} \frac{1}{ L^2 + 2L\cdot p} \frac{1}{ -v\cdot L} \frac{1}{ L^2-\lambda_\gamma^2} \, \slash{v} \big( \slash{L} + \slash{p} + m_\ell \big) \Gamma_\ell \otimes \Gamma_h \,,
\end{align}
where the tree-level amplitude from Eq.~(\ref{eq:static_L}) is conveniently expressed as the product of leptonic $\Gamma_\ell$ and hadronic $\Gamma_h$ Dirac structures: ${\cal M}^{\rm tree} = \Gamma_\ell \otimes \Gamma_h$. The hard contribution to the scattering amplitude, ${\cal M}_\mathrm{H}$, is given in dimensional regularization by setting $m_\ell=0$:
\begin{align}
 {\cal M}_\mathrm{H} &= i e^2 \int \frac{d^dL}{ (2\pi)^d} \frac{ 1}{ L^2 + 2L\cdot p} \frac{1}{ -v\cdot L} \frac{1}{ L^2} \, \slash{v} \big( \slash{L} + \slash{p} \big) \Gamma_\ell \otimes \Gamma_h \bigg|_{p^2=0} \,. 
\end{align}
The soft contribution ${\cal M}_\mathrm{S}$ is
\begin{align}
 {\cal M}_\mathrm{S} &= i e^2 \int \frac{d^dL}{ (2\pi)^d} \frac{ 1}{ 2L\cdot p} \frac{1}{ -v\cdot L} \frac{1}{ L^2-\lambda^2} \, \slash{v} \big( \slash{p} + m_\ell \big) \Gamma_\ell \otimes \Gamma_h \bigg|_{p^2=m_\ell^2} \,. 
\end{align}
Finally, the remaining collinear contribution ${\cal M}_\mathrm{J}$ is
\begin{align}
 {\cal M}_\mathrm{J} &= i e^2 \int \frac{d^dL}{ (2\pi)^d} \frac{ 1}{ L^2 + 2L\cdot p} \frac{1}{ -v\cdot L_-} \frac{1}{ L^2} \, \slash{v} \big( \slash{L} + \slash{p} + m_\ell \big) \Gamma_\ell \otimes \Gamma_h \bigg|_{p^2=m_\ell^2} \,. 
\end{align}
It is straightforward to evaluate these integrals explicitly and to include the soft and collinear contributions to wave-function renormalization. The radiative corrections to the cross section of Eq.~(\ref{eq:static_soft}) are found to be
\begin{align}
 \delta^{\rm soft} &= \delta_\mathrm{S} + \delta_\mathrm{J} + \delta_\mathrm{H} \,,
\end{align}
where (after $\overline{\rm MS}$ renormalization~\footnote{The bare and renormalized couplings are related as $\frac{e_{\rm bare}^2}{4\pi} (4\pi)^\epsilon e^{-\gamma_E \epsilon}= \alpha_{\rm bare} = \mu^{2\epsilon}\alpha(\mu)[1+\order(\alpha)]$.} and including real soft-photon radiation) the hard ($\delta_\mathrm{H}$), soft $(\delta_\mathrm{S}$) and collinear ($\delta_\mathrm{J})$ contributions are, respectively,
\begin{align}
 \delta_\mathrm{H} &= -\frac14 \ln^2\frac{4E^2_\nu}{ \mu^2} + \frac12 \ln\frac{4E^2_\nu}{ \mu^2} - 1 + \frac{19\pi^2}{ 24} \,, \nonumber \\
 \delta_\mathrm{S} &= \left(1 - \ln \frac{2E_\nu }{ m_\ell} \right) \left( \ln \frac{\mu^2 }{ \left(\Delta E\right)^2} + \ln \frac{2E_\nu }{ m_\ell} \right) + 1 - \frac{\pi^2 }{ 6} \,, \nonumber \\ 
 \delta_\mathrm{J} &= \frac14 \ln^2 \frac{\mu^2 }{ m^2_\ell} + \frac14 \ln \frac{\mu^2}{ m^2_\ell} + 1 + \frac{\pi^2}{ 24} \,. \label{eq:deltaSJH}
\end{align}
For the observable that includes energetic photon emission within angle $\Delta\theta$ of the charged lepton direction, radiative corrections in Eq.~(\ref{eqn:delta_jet_static}) are given by
\begin{align}
 \delta^{\rm jet} &= \delta_\mathrm{S} + \delta_\mathrm{J} + \delta_\mathrm{J}^\prime + \delta_\mathrm{H} \,,
\end{align}
where the additional contribution from the radiation of the real photon is obtained from collinear photon Feynman diagrams and is given explicitly from Eqs.~(\ref{eqn:delta_jet_static}) and (\ref{eqn:delta_soft_static}) by
\begin{align}
 \delta_\mathrm{J}^\prime = \delta^{\rm jet} - \delta^{\rm soft} &= \frac94 + \left( \ln \left( 1 + \eta^2 \right) - \frac{\eta^2}{1+\eta^2} \right) \ln \frac{E_\nu}{\Delta E} -\left(\mathrm{tan}^{-1} \eta \right)^2 - \left( 1 + \frac{1}{1+\eta^2}\right) \frac{\mathrm{tan}^{-1} \eta }{\eta} \, \nl
 &+ \frac{1}{2} \mathrm{Li}_2 \left( -\eta^2 \right) + \frac{1}{4} \ln^2 \left( 1 + \eta^2 \right) + \frac{1}{4} \left(\frac{2}{1+\eta^2} - \frac{1}{\eta^2} -3 \right) \ln \left( 1 + \eta^2 \right) \,. \label{eqn_delta_jetprime_static}
\end{align}
We see that $\delta_\mathrm{H}$ contains no large logarithms when $\mu$ is of order of the hard scale, $\mu \sim E_\nu = \order(\Lambda)$. Similarly, $\delta_\mathrm{J}$ contains no large logarithms when $\mu$ is of order of the collinear scale, $\mu \sim m_\ell = \order(\sqrt{\lambda}\Lambda)$. The correction $\delta_\mathrm{J}^\prime$ contains a logarithm $\ln(E_\nu/\Delta E)$, as required for cancellation of $\ln{m_\ell}$ singularities~\cite{Kinoshita:1962ur,Lee:1964is} in the limit $m_\ell\to 0$. The expected eikonal logarithms remain in $\delta_\mathrm{S}$ when $\mu$ is of order of the soft scale, $\mu\sim \Delta E = \order(\lambda \Lambda)$. We have thus achieved a scale separation, isolating the contributions from each of the relevant hard, collinear and soft momentum regions. Below, we state the general factorization theorem beyond the static limit. The jet and soft functions are straightforward generalizations of the static limit quantities computed above and reduce to $\delta_\mathrm{S}$, $\delta_\mathrm{J}$, and $\delta_\mathrm{J}^\prime$ in the appropriate limits. The hard function becomes a nonperturbative quantity that we describe using phenomenological form factors. In the formal static limit $E_\nu/M \to 0$ (and $m_\ell/E_\nu \sim \Delta\theta \ll 1)$, the hard function would reduce to $\delta_\mathrm{H}$ in Eq.~(\ref{eq:deltaSJH}), including the appropriate combination of Wilson coefficients $c_V$ and $c_A$ from Eq.~(\ref{eq:static_L}).

\subsection{Factorization beyond the static limit}

The above analysis of momentum regions, formalized in soft-collinear effective theory (SCET), does not rely on the static limit and can be straightforwardly generalized. The key difference for $E_\nu \sim M$ is that the hard function becomes sensitive to hadronic structure and cannot be computed in perturbation theory. Including also the general kinematic dependence in the soft and collinear factors, we have the following factorization theorem for the cross section:\footnote{Our model of Sec.~\ref{sec:model} provides an explicit demonstration of the factorization theorem at one-loop order. It reproduces correctly the soft and collinear functions, leaving a hard function independent of IR scales.}
\begin{multline}
 \frac{d\sigma}{ dQ^2} \propto H\left( \frac{E_\nu}{ M}, \frac{Q^2}{ M^2}, \frac{\mu}{M} \right) \bigg[ J\left( \frac{\mu}{m_\ell} \right) R\left(\frac{\mu}{m_\ell}, v_\ell\cdot v_p\right) S\left( \frac{\mu}{\Delta E}, v_{\ell} \cdot v_p, v\cdot v_{\ell}, v\cdot v_p \right) \\
 + \int_{0}^{1-\frac{\Delta E }{ E_{\ell}^{\rm tree}}} dx \, j\left( \frac{\mu}{m_\ell}, x, v\cdot v_{\ell} \Delta \theta \right) R\left(\frac{\mu}{m_\ell}, x v_\ell\cdot v_p\right) S\left( \frac{\mu}{\Delta E}, x\, v_{\ell} \cdot v_p, x\, v\cdot v_{\ell}, v\cdot v_p \right) \bigg] \,, \label{eq:factorization_formula}
\end{multline}
valid up to power corrections in $\lambda$. Here $v_\ell$ and $v_p$ denote the charged lepton and proton four-velocities, respectively, entering the tree-level process, $Q^2=-(k^\prime-k)^2 = 2M^2( v_n \cdot v_p - 1)$ is the momentum transfer between initial and final nucleon states, $E_\ell^{\rm tree} = E_\nu + M-v\cdot k^\prime = m_\ell v\cdot v_\ell$ is the lepton energy for the tree-level process, and $x = E_\ell/E_\ell^{\rm tree}$ denotes the fraction of the total jet energy carried by the charged lepton (the total jet energy is defined as the energy carried by the charged lepton plus collinear photons). The quantities $\Delta E$ and $\Delta \theta$ denote soft energy and angular acceptance parameters depending on experimental conditions, and $\mu$ defines the renormalization scale. For completeness, we have included the remainder function $R$ in the factorization formula~\cite{Hill:2016gdf}; this function relates the running electromagnetic coupling in the QED theory with and without the dynamical charged lepton $\ell$ and contributes numerically small corrections starting at two-loop order. All energy components are taken in the laboratory frame. To obtain various observable predictions, the expression in Eq.~(\ref{eq:factorization_formula}), for appropriate choices of $\Delta E$ and $\Delta \theta$, can be integrated with respect to $Q^2$. The second term in square brackets, involving $j$, is omitted for observables with soft-photon radiation only.\footnote{For simplicity, we define the jet to contain a single charged lepton plus any number of collinear or soft photons, ignoring contributions of additional $e^+e^-$ pairs generated by ``internal" conversion of a virtual photon; such effects are of two-loop order. The region of $x$ integration, $x=0..1-\Delta E/E_\ell^{\rm tree}$ results from $x=m_\ell/E_\ell^{\rm tree} .. 1$, upon omitting the region near $x=1$ that is contained in the soft-photon term and noticing that the region at $x=0$ is power suppressed. The soft function multiplying $j$ corresponds to the charged lepton of energy $E_\ell = x E_\ell^{\rm tree}$; the replacement $v\cdot v_\ell \to x v\cdot v_\ell$ is by definition exact, and the replacement $v_\ell \cdot v_p \to x v_\ell \cdot v_p$ is valid up to power corrections. The integration region near $x=1$, corresponding to soft photons, is described exactly up to power corrections in our explicit expressions for $j$, which use only the approximation $m_\ell/E_\ell \ll 1$.}

The factorization formula of Eq.~(\ref{eq:factorization_formula}) is valid up to power-suppressed contributions in the power-counting parameter $\lambda$: $\lambda \sim \Delta E/\Lambda_\mathrm{hard} \sim m_\ell^2/\Lambda_\mathrm{hard}^2 \sim \left( \Delta \theta \right)^2$, with the hard scale $\Lambda_\mathrm{hard} \sim M \sim E_\nu \sim Q$ in general. In this paper, we include lepton-mass corrections in tree-level cross-section expressions. Consequently, all neglected power-suppressed contributions are additionally suppressed by electromagnetic coupling constant $\alpha$, although logarithmic enhancements are still possible. The power-counting parameter turns out to be at or below the percent size at accelerator neutrino experiments for all three determining quantities, as one can estimate from discussions of experimental details in Sec.~\ref{sec:overview}. For example, for photon energy resolution $\Delta E= 20~\mathrm{MeV}$ and cone size $\Delta \theta = 10^\circ$, we have $\Delta E/\Lambda_{\rm hard} \sim 0.02$, $\left( \Delta \theta \right)^2 \sim 0.03$, $m_e^2/\Lambda_\mathrm{hard}^2 \sim 3\times 10^{-7}$, and $m_\mu^2/\Lambda_\mathrm{hard}^2 \sim 0.01$, assuming $\Lambda_\mathrm{hard} \sim 1~\mathrm{GeV}$.

In the following sections, we compute the separate soft ($S$), collinear ($J$), and hard ($H$) functions appearing in the factorization formula. We then collect results and compute several illustrative observables.

\subsection{Soft function \label{sec:soft}}

The soft function appearing in Eq.~\eqref{eq:factorization_formula} accounts for physics below the energy scale of particle masses. It is independent of hadronic structure and depends only on particle velocities, electric charges, the energy cutoff variable $\Delta E$ and the renormalization scale $\mu$. Including virtual soft corrections and real soft-photon radiation with energy below $\Delta E$ yields the process-independent soft function through one-loop order~\cite{Hill:2016gdf}:
\begin{align}
 S\left( \frac{\mu}{\Delta E},\, v_\ell \cdot v_p,\, v\cdot v_\ell,\, v\cdot v_p \right) = 1 + \frac{\alpha}{\pi} \bigg[ 2 \bigg(1 - v_\ell\cdot v_p f(v_\ell\cdot v_p)\bigg) \ln \frac{\mu}{2 \Delta E} + G \left( v_\ell \cdot v_p, v\cdot v_\ell, v\cdot v_p \right) \bigg], \label{eq:soft_function}
\end{align}
where $v^\mu$ defines the laboratory frame in which $\Delta E$ is measured [by default $v^\mu=(1,0,0,0)$], $v_\ell^\mu = p^{\prime\mu}/m_\ell$ and $v_p^\mu$ are the charged lepton and proton velocity vectors, respectively, and the functions $f$ and $G$ are given, respectively, by~\cite{tHooft:1978jhc,Hill:2016gdf}
\begin{align}
f(w) &= \frac{\ln w_+}{\sqrt{w^2-1}} \,, \nl
G(w,\x,\y) &= \frac{\x }{ \sqrt{ \x^2-1} } \ln{\x_+} + \frac{\y }{\sqrt{ \y^2 -1}} \ln{y_+} + \frac{w}{ \sqrt{w^2-1}} \bigg[ \ln^2 \x_+ -\ln^2 \y_+ \nl
 &\quad + {\rm Li}_2\left( 1 - \frac{ \x_+ }{ \sqrt{w^2-1}} ( w_+ \x - \y ) \right) + {\rm Li}_2\left( 1 - \frac{ \x_- }{ \sqrt{w^2-1}} ( w_+ \x - \y ) \right) \nl
 &\quad - {\rm Li}_2\left( 1 - \frac{ \y_+ }{ \sqrt{w^2-1}} ( \x - w_- \y ) \right) - {\rm Li}_2\left( 1 - \frac{ \y_- }{ \sqrt{w^2-1}} ( \x - w_- \y ) \right) \bigg] \,,
\end{align}
with $a_\pm = a \pm \sqrt{a^2-1}$. The soft function of Eq.~(\ref{eq:soft_function}) applies to both the neutrino process in Eq.~(\ref{eq:nuprocess}), where $v_p$ denotes the final-state proton velocity, and to the antineutrino process in Eq.~(\ref{eq:antinuprocess}), where $v_p$ denotes the initial-state proton velocity. The soft function is infrared (IR) finite and is universal for the different operator structures contributing to the hard function in Eq.~(\ref{eq:factorization_formula}). We remark that for $v\cdot v_p\to 1$, i.e., when the proton is at rest in the laboratory frame, the soft function reduces to~\cite{Tomalak:2019ibg}
\begin{align}
 S\left( \frac{\mu}{\Delta E},\, w,\, w,\, 1 \right) = 1 &+ \frac{\alpha}{\pi} \bigg[ 1 +2 \left(1 - \frac{1}{2\beta} \ln \frac{1+\beta}{1-\beta} \right) \ln \frac{\mu}{2 \Delta E} + \frac{1}{2 \beta} \ln \frac{1+\beta}{1-\beta}\left( 1 + \ln \frac{ 1+ \beta }{4 w \beta^2} \right) \, \nl
 &+\frac{1}{ \beta}\left( \mathrm{Li}_2 \frac{1-\beta}{1+\beta} - \frac{\pi^2}{6} \right) \bigg], \label{eq:softantinu}
\end{align}
where $\beta=\sqrt{1-1/w^2}$ is the final lepton velocity in the proton rest frame. In particular, the limit (\ref{eq:softantinu}) applies for the case of antineutrino-proton scattering with $\Delta E$ measured in the proton rest frame.

Large logarithms appear in the limit of small charged lepton mass. For example, in the proton rest frame,
\begin{align}
 S\left( \frac{\mu}{\Delta E},\, \frac{E_\ell}{ m_\ell},\, \frac{E_\ell}{ m_\ell},\, 1 \right) \underset{ E_\ell \gg m_\ell}{\longrightarrow} 1 + \frac{\alpha}{\pi} \left[ -\ln^2 \frac{2 E_\ell}{m_\ell} + \ln \frac{2 E_\ell}{m_\ell} + 2 \left( 1 - \ln \frac{2 E_\ell}{m_\ell} \right) \ln \frac{\mu}{2 \Delta E}+ 1 - \frac{\pi^2}{6} \right]\,, \label{eq:soft_function_massless}
\end{align}
with the recoil charged lepton energy in the laboratory frame $E_\ell$. For GeV (anti)neutrino energies, the double-logarithmic corrections for the electron case, $\ell= e$, are large and higher-order contributions must be included for percent-level accuracy. The result in Eq.~(\ref{eq:soft_function_massless}) coincides with the one-loop soft correction $\delta_\mathrm{S}$ in the static nucleon limit with an ultrarelativistic lepton, Eq.~(\ref{eq:deltaSJH}).

The renormalization group evolution of the soft function is described through two-loop order by~\cite{Korchemsky:1987wg,Kilian:1992tj,Hill:2016gdf}
\begin{equation}
 \frac{\mathrm{d} S\left( \mu \right)}{\mathrm{d} \ln \mu^2} = \gamma_\mathrm{S} \left( \mu \right) S\left( \mu \right), \qquad \gamma_\mathrm{S} \left( \mu \right) = \left( \gamma_0 +\gamma_1 \frac{\alpha\left( \mu \right)}{\pi} \right) \left( 1- \frac{1}{2\beta} \ln \frac{1 + \beta}{1-\beta}\right)\frac{\alpha\left( \mu \right)}{\pi} \,, \label{eq:running_soft}
\end{equation}
with $\gamma_0 = 1$ and $\gamma_1 = - 5 \numl /9$, including $\numl$ virtual dynamical charged leptons. Exponentiating the one-loop result, we evaluate the complete resummed soft function through $\order(\alpha)$ in the counting $\alpha \ln^2(2E_\ell/m_\ell) = \order(1)$ [i.e., $\ln(2E_\ell/m_\ell) = \order(\alpha^{-1/2})$]. Some expressions for the next order in $\alpha$ can be found in Refs.~\cite{Burgers:1985qg,Kniehl:1989kz,Mastrolia:2003yz,Arbuzov:2015vba,Hill:2016gdf}.

\subsection{Collinear function \label{sec:collinear}}

The collinear contributions represented by $J$ and $j(x)$ in Eq.~(\ref{eq:factorization_formula}) describe physics at energy scales of order of the charged lepton mass. Like the soft function, these contributions are independent of hadronic structure. As discussed in Sec.~\ref{sec:overview}, energetic photons radiated within a cone of size $\Delta \theta$ around the final lepton direction can be included in the definition of the observable depending on the lepton flavor and detector details.

Consider first the case when no energetic collinear radiation is emitted. We have derived the collinear function both by momentum region decomposition of the virtual correction, and by exploiting Feynman rules in the collinear sector of SCET. The one-loop result is given by
\begin{align}
 J\left( \frac{\mu}{m_\ell} \right) = 1 + \frac{\alpha}{4 \pi} \left( \ln^2 \frac{\mu^2}{m_\ell^2} + \ln \frac{\mu^2}{m_\ell^2} + 4 + \frac{\pi^2}{6}\right) \,. \label{eq:jet_function_at_fixed_x}
\end{align}
Since the jet function is independent of hadronic structure, the result of Eq.~(\ref{eq:jet_function_at_fixed_x}) is identical to the one-loop correction $\delta_\mathrm{J}$ in the static nucleon limit with ultrarelativistic lepton, Eq.~(\ref{eq:deltaSJH}). Some two-loop results for collinear functions in QED were derived in Refs.~\cite{Hoang:1995ex,Bernreuther:2004ih,Becher:2007cu,Hill:2016gdf}.

Jet observables include energetic collinear photons. Through one-loop order, the contribution to the cross section is given by ($\eta \equiv \Delta\theta E_\ell^{\rm tree}/m_\ell$)
\begin{align}
 j\left(\frac{\mu}{m_\ell}, x, \eta\right) = \frac{\alpha}{\pi} \left[ \frac{1}{2}\frac{1+x^2}{1-x}\ln(1+x^2\eta^2) -\frac{x}{ 1-x} \frac{x^2\eta^2 }{ 1 + {x^2\eta^2}} \right]\,. \label{eq:SCETjet}
\end{align}
The expression in Eq.~(\ref{eq:SCETjet}) is obtained using SCET Feynman rules. After integrating over $x$, the contribution is identical to $\delta_{J}^\prime$ of Eq.~(\ref{eqn_delta_jetprime_static}) in the static limit (when $E_\ell^{\rm tree} = E_\nu$):
\begin{align}
\int_0^{1-\Delta E/E_\ell^{\rm tree}} \mathrm{d} x\, j \left(\frac{\mu}{m_\ell}, x, \eta \right) &= \frac{\alpha}{\pi} \left[ \frac{1}{2} \mathrm{Li}_2 \left( -\eta^2 \right) + \frac{1}{4} \ln^2 \left( 1 + \eta^2 \right)+ \frac{1}{4} \left(\frac{2}{1+\eta^2} - \frac{1}{\eta^2} -3 \right) \ln \left( 1 + \eta^2 \right) + \frac{9}{4} \right. \, \nl
&\left. + \left( \ln \left( 1 + \eta^2 \right) - \frac{\eta^2}{1+\eta^2} \right) \ln \frac{E_\ell^{\rm tree}}{\Delta E} -\left(\mathrm{tan}^{-1} \eta \right)^2 - \left( 1 + \frac{1}{1+\eta^2}\right) \frac{\mathrm{tan}^{-1} \eta }{\eta}\right] \,. \label{eqn:delta_jet_analytic}
\end{align}
We remark that, in the limit of small charged lepton mass,
\begin{align}
j\left(\frac{\mu}{m_\ell}, x, \eta \right) \underset{m_\ell \ll \Delta\theta E_\ell^{\rm tree}}{\longrightarrow} \frac{\alpha}{\pi} \left( \frac{1+x^2}{1-x} \ln\frac{ x E_\ell^{\rm tree}\Delta\theta}{ m_\ell} - \frac{x}{1-x} \right) \,, \label{eq:jet_as_function_of_x}
\end{align}
where the coefficient of the logarithm is identified with the well-known splitting function~\cite{Gribov:1972ri,Gribov:1972rt,Lipatov:1974qm,Dokshitzer:1977sg,Altarelli:1977zs,Dokshitzer:1978hw,Peskin:1995ev}.

\begin{figure}[th]
\centering
\includegraphics[height=0.3\textwidth]{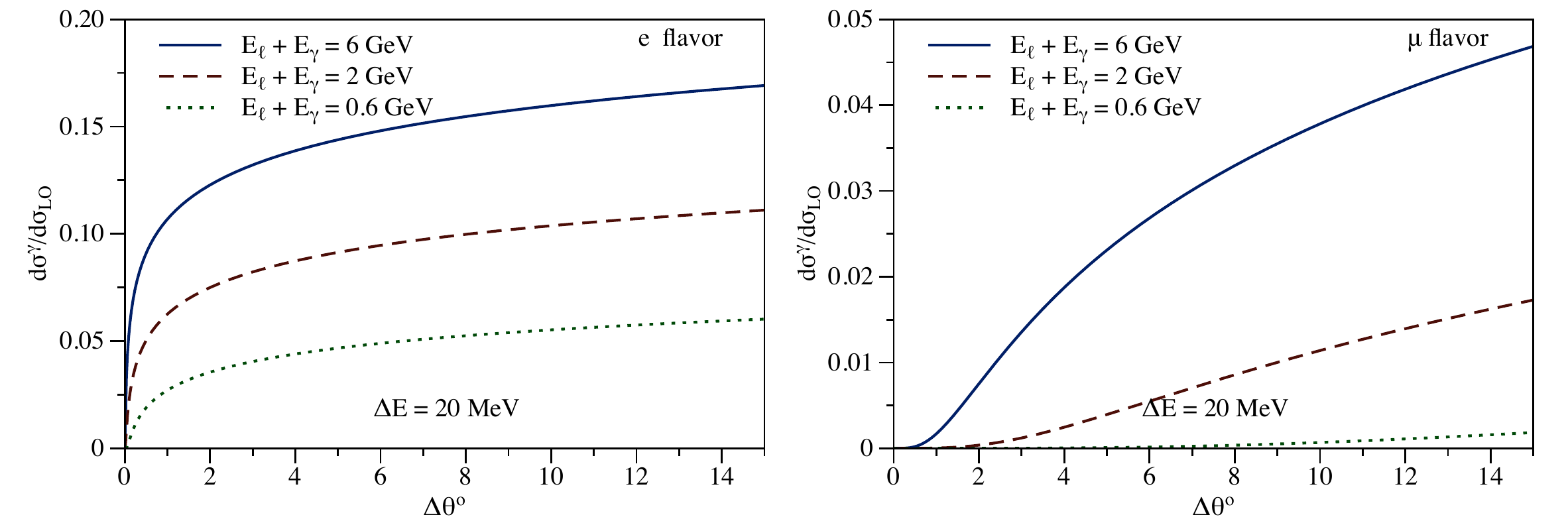}
 \caption{Differential cross section $\mathrm{d} \sigma^\gamma$ for one radiated photon of energy above $\Delta E = 20~\mathrm{MeV}$ within the angle $\Delta \theta$ to the lepton direction, divided by the tree-level charged-current elastic cross section $\mathrm{d} \sigma_\mathrm{LO}$. The ratio $\mathrm{d} \sigma^\gamma/\mathrm{d} \sigma_\mathrm{LO}$ is computed using Eq.~(\ref{eqn:delta_jet_analytic}) for the scattering of electron (anti)neutrinos (left plot) and muon (anti)neutrinos (right plot). The bottom, middle, and top curves correspond to electromagnetic jet energy 600~MeV, 2~GeV, and 6~GeV, respectively. \label{fig:static_limit_angle}}
 \centering
 \includegraphics[height=0.3\textwidth]{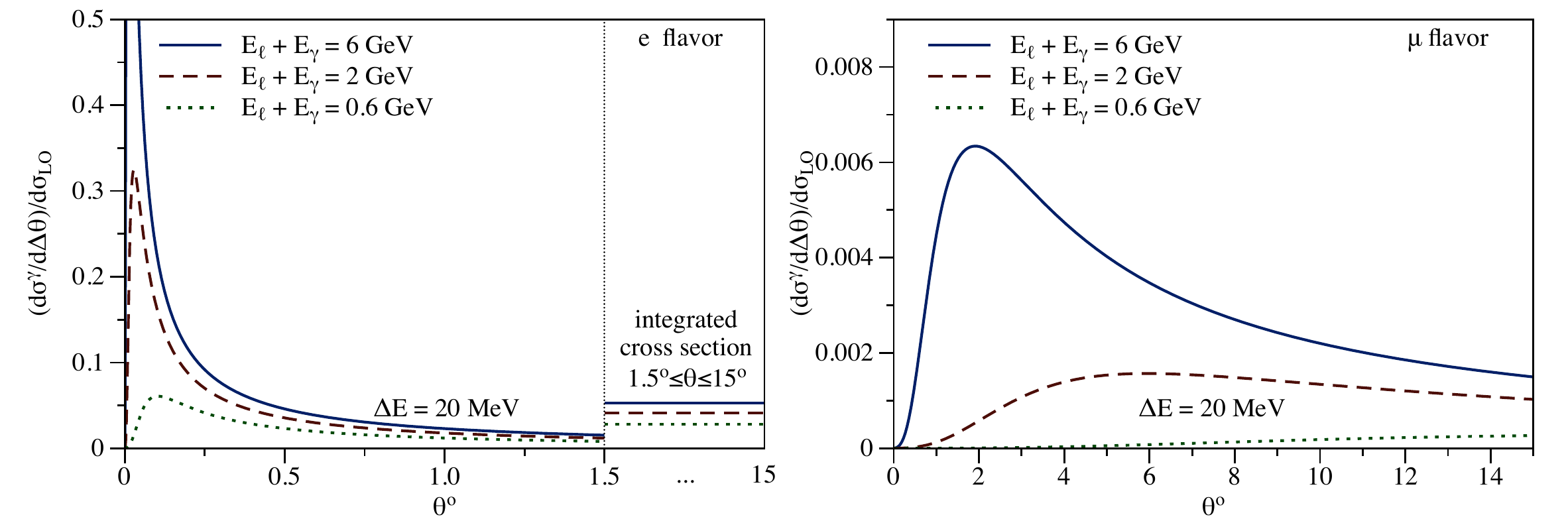}
 \caption{Derivative of $\mathrm{d} \sigma^\gamma$ in Fig.~\ref{fig:static_limit_angle} with respect to $\Delta\theta$, divided by the tree-level charged-current elastic cross section $\mathrm{d} \sigma_\mathrm{LO}$. \label{fig:static_limit_angle2}}
\end{figure}

Muon and electron jets for accelerator neutrino energies are quite different objects. Photon radiation is highly collimated in the direction of the outgoing lepton for charged-current processes with electron flavor, but only a relatively small contribution is contained within a $1-2^\circ$ cone for the muon flavor. We illustrate this dependence of radiated events on the jet angle by plotting Eq.~(\ref{eqn:delta_jet_analytic}), and the derivative of Eq.~(\ref{eqn:delta_jet_analytic}) with respect to $\Delta\theta$, in Figs.~\ref{fig:static_limit_angle} and~\ref{fig:static_limit_angle2}, respectively.

Two limiting cases are of interest. The limit of very small charged lepton mass gives (recall $\eta = \Delta \theta E_\ell^{\rm tree}/m_\ell$)
\begin{equation}
\int_0^{1-\Delta E/E_\ell^{\rm tree}} \mathrm{d} x\, j\left(\frac{\mu}{m_\ell}, x, \eta \right) \underset{\eta \gg 1}{\longrightarrow} \frac{\alpha}{\pi} \left[ \left( 2 \ln \eta -1 \right) \ln \frac{E_\ell^{\rm tree}}{\Delta E} - \frac{3}{2} \ln \eta -\frac{\pi^2}{3} + \frac{9}{4} + \frac{\pi}{2 \eta} + \order \left( \frac{1}{\eta^2} \right) \right]. \label{eqn:delta_jet_limit_m}
\end{equation}
Adding the virtual soft and collinear corrections, the jet angular size regulates the collinear singularity when $m_\ell \ll \Delta \theta \left( E_\ell + E_\gamma \right)$, as observed for the static limit case in Eq.~(\ref{eqn:delta_jet_static_limit_m}). For example, $\eta \gtrsim 10$ is realized for $E_\ell^{\rm tree} \gtrsim 30~\mathrm{MeV}$ for electrons and $E_\ell^{\rm tree} \gtrsim 6~\mathrm{GeV}$ for muons when the cone angle is $\Delta \theta = 10^\circ$. In the opposite limit of very narrow jet, the collinear radiation is suppressed by phase space and vanishes as
\begin{equation}
\int_0^{1-\Delta E/E_\ell^{\rm tree}} \mathrm{d} x\, j\left(\frac{\mu}{m_\ell}, x, \eta \right) \underset{\eta \ll 1}{\longrightarrow} \frac{\alpha}{\pi} \left[\frac{\eta^2}{24} + \left( \ln \frac{E_\ell^{\rm tree}}{\Delta E} - \frac{23}{10} \right) \frac{\eta^4}{2} + \order \left(\eta^6 \right) \right]. \label{eqn:jetSmallConeLimit}
\end{equation}
\noindent

The two-loop renormalization group evolution of the collinear function is given by~\cite{Becher:2009qa,Becher:2009kw,Hill:2016gdf}
\begin{align}
 \frac{\mathrm{d} J\left( \mu \right)}{\mathrm{d} \ln \mu^2} &= \gamma_\mathrm{J} \left( \mu \right) J\left( \mu \right)\,, \nl
 \gamma_\mathrm{J} \left( \mu \right) &= \left( 2 \ln \frac{\mu^2}{m_\ell^2} + 1 \right) \left(\gamma_0+ \gamma_1 \frac{\alpha\left(\mu\right)}{\pi} \right)\frac{\alpha\left( \mu \right)}{4\pi} + \gamma_2 \left(\frac{\alpha\left( \mu \right)}{4\pi} \right)^2, \label{eq:running_collinear}
\end{align}
where $\gamma_2 = \frac{3}{2}+\frac{50 \numl}{27} - 2 \left( 1 + \frac{\numl}{3}\right) \pi^2 + 24 \zeta_3 $ with the number of dynamical charged leptons in the theory $\numl$.

In the factorization theorem of Eq.~(\ref{eq:factorization_formula}), the soft and jet functions appear as
\begin{align}
 J R(1) S(1) + \int \mathrm{d} x \, j(x) R(x) S(x) \,,\label{eq:Jn}
\end{align}
where the $x$ integration limits depend on the observable. In place of Eq.~(\ref{eq:Jn}), we use the simplified form $J R(1) S(1) \exp\big[ \int \mathrm{d} y\, j_1(y) \big]$ in the phenomenological analysis, where $j_1(x)$ denotes the one-loop contribution. This exponentiation of the one-photon collinear correction accounts for the potentially enhanced leading phase-space double logarithms $ \left(\frac{2 \alpha}{\pi} \ln \frac{\Delta E}{E_\ell} \ln \frac{ \Delta \theta E_\ell}{m_\ell} \right)^{n_c}$ arising from the radiation of two or more ($n_c$) collinear photons.

\subsection{Hard function \label{sec:hadronic}}

The hard function $H$ appearing in Eq.~(\ref{eq:factorization_formula}) represents a matching coefficient when the full theory including hadronic physics\footnote{For definiteness, we consider the full theory as $n_f=3$ flavor quantum chromodynamics (QCD) in the presence of four-fermion electroweak operators~\cite{Hill:2019xqk}, i.e., the Standard Model after integrating out electroweak vector bosons, $W^\pm$, $Z^0$, Higgs field, $h$, and heavy quarks $t$, $b$, and $c$.} is matched onto soft-collinear effective theory (where hadronic physics is integrated out). The matching is performed by equating the amplitudes displayed in Fig.~\ref{fig:CCQE_kin}, computed in both full theory and effective theory, accounting for tree-level and one-loop corrections. We begin in Sec.~\ref{sec:invariant} by parametrizing the hard function by invariant amplitudes. In Sec.~\ref{sec:model}, we then introduce a default model for the nonperturbative matching condition. Finally, we consider the renormalization group evolution for the hard matching coefficient within the effective theory in Sec.~\ref{sec:rge_evolution}.

\subsubsection{Invariant amplitudes \label{sec:invariant}}

At leading power, charged lepton masses may be ignored in the hard matching condition. The matrix element of the charged-current elastic process with massless charged lepton can then be expressed as%
\footnote{We use the shorthand notation $\bar{\ell}^- (\dots ) \nu_{\ell} = \bar{u}^{(\ell)}(p^\prime) (\dots ) u^{(\nu)}(p)$ and $\overline{\bar{\nu}}_\ell (\dots) \ell^+= \bar{v}^{(\nu)}(p^\prime) (\dots ) v^{(\ell)}(p)$ for the usual Dirac spinors with momentum assignment in Fig.~\ref{fig:CCQE_kin}.}
\begin{align}
T_{\nu_\ell n \to \ell^- p} = \sqrt{2}\mathrm{G}_\mathrm{F} V_{ud} \, \bar{\ell}^- \gamma^\mu \mathrm{P}_\mathrm{L} \nu_{\ell}\, \bar{p} \left(f_1 \gamma_\mu + f_2 \frac{i \sigma_{\mu \rho} q^\rho}{2M} + f_A \gamma_\mu \gamma_5 - f^{3}_A \frac{{K}_\mu}{M} \gamma_5 \right) n\,, \nl
T_{\bar{\nu}_\ell p \to \ell^+ n} = \sqrt{2}\mathrm{G}_\mathrm{F} V^*_{ud} \, \overline{\bar{\nu}}_\ell \gamma^\mu \mathrm{P}_\mathrm{L} \ell^+ \,
\bar{n} \left( \bar{f}_1 \gamma_\mu + \bar{f}_2 \frac{i \sigma_{\mu \rho} q^\rho}{2M} + \bar{f}_A \gamma_\mu \gamma_5 + \bar{f}^{3}_A \frac{{K}_\mu}{M} \gamma_5 \right) p \,. \label{eq:CCQE_amplitude}
\end{align}
Here $ q = p-p^\prime = k^\prime - k$, $K = (k+k^\prime)/2$, $\mathrm{G}_\mathrm{F}$ is the Fermi constant, $V_{ud}$ is the CKM matrix element and $M = (M_p + M_n)/2$ is the average nucleon mass. The four independent invariant amplitudes are functions of two kinematic variables: $Q^2=-t=-q^2$ and $\nu = \left( s - u \right)/4 = M E_\nu - \left(Q^2+m_\ell^2 \right)/4$, where $s$, $t$, and $u$ are the usual Mandelstam invariants. For the antineutrino case, the invariant amplitudes are given by $\bar{f}_i \left( \nu+i0, Q^2 \right) = f_i \left( -\nu-i0, Q^2 \right)^*$, where $f_i$ stands for one of $f_1$, $f_2$, $f_A$, and $f_A^3$. The quantities $f_i$ are ultraviolet (UV) finite and IR divergent when virtual QED corrections are included. These IR divergences cancel in the matching between full theory and effective theory.

Using the representation of Eq.~(\ref{eq:CCQE_amplitude}), the charged-current elastic cross section (without radiation) in the laboratory frame is expressed in terms of invariant amplitudes as~\cite{LlewellynSmith:1971uhs}\footnote{We neglect the relative difference in nucleon masses, $(M_n-M_p)/(M_n+M_p)$, and electroweak power corrections suppressed by the $W$-boson mass, $Q^2/M_W^2$; these effects contribute at the permille level.}
\begin{equation}
\frac{d\sigma}{dQ^2} (E_\nu, Q^2) = \frac{\mathrm{G}_\mathrm{F}^2 |V_{ud}|^2}{2\pi} \frac{M^2}{E_\nu^2} \left[ \left( \tau + r^2 \right)A(\nu,~Q^2) - \frac{\nu}{M^2} B(\nu,~Q^2) + \frac{\nu^2}{M^4} \frac{C(\nu,~Q^2)}{1+ \tau} \right] \,, \label{eq:xsection_CCQE}
\end{equation}
where $\tau = Q^2/(4M^2)$ and $r = m_\ell/(2M)$. The quantities $A$, $B$, and $C$ are given, respectively, by\footnote{The sign in front of $f_A^3$ in $B$ differs from Ref.~\cite{LlewellynSmith:1971uhs} for antineutrino scattering, as noted in Ref.~\cite{Kuzmin:2007kr}.}
\begin{align}
A &= \tau | g_M |^2 - | g_E |^2 + (1+ \tau) | f_A |^2- \tau (1+ \tau) | f_A^3 |^2 - r^2 \left( | g_M |^2 + | f_A + 2 F_P |^2 - 4 \left( 1 + \tau \right) F_P^2 \right) \,, \nl
B &= \mathrm{Re} [ \pm 4 \tau f^*_A g_M - 2 r^2 \left( f_A - 2 \tau F_P \right)^* f_A^3 ]\,, \nl
C &= \tau |g_M|^2 + |g_E|^2 + (1+ \tau) |f_A|^2 + \tau (1+ \tau) |f_A^3|^2\,.
\end{align}
For numerical results in this paper, we include all mass corrections within the hadronic model of Sec.~\ref{sec:model}.\footnote{In terms of Eq.~(\ref{eq:factorization_formula}), the matching calculation includes power-suppressed terms in the hard function. This procedure does not achieve a complete scale separation beyond the leading power (both hard and collinear momentum regions contribute power-suppressed terms), but the complete lepton-mass dependence in the hadronic model is retained through $\order(\alpha)$.} At tree level, these corrections are accounted for by including the pseudoscalar form factor in the expressions for $A$, $B$ and $C$ and retaining the lepton-mass dependence in kinematic prefactors. We will use a standard ansatz (partially conserved axial current and the assumption of pion pole dominance) for the pseudoscalar form factor: $F_P(Q^2) = 2M^2 F_A(Q^2)/\left(m_\pi^2 + Q^2\right)$. Electric and magnetic amplitudes $g_E$ and $g_M$ are defined from $f_1$ and $f_2$ as, respectively,
\begin{align}
 g_E = f_1 - \tau f_2, \qquad g_M = f_1 + f_2.
\end{align}
For antineutrino-proton scattering, the sign in the first term of $B$ is negative.

When one-loop radiative corrections are included, the full-theory cross section in Eq.~(\ref{eq:xsection_CCQE}) receives contributions originating from soft, collinear, and hard regions of virtual photon momentum. The soft and collinear region contributions are reproduced by the effective theory and cancel in the matching; in the computation of observables, these contributions are replaced by the complete soft and collinear functions computed within the effective theory, incorporating real photon radiation and perturbative resummation. The hard region contribution determines the hard function appearing in the factorization formula of Eq.~(\ref{eq:factorization_formula}).

\subsubsection{Hadronic model \label{sec:model}}

The matching condition depends on the nonperturbative quantities $f_1$, $f_2$, $f_A$, and $f_A^3$ appearing in Eq.~(\ref{eq:CCQE_amplitude}). At the leading order in $\alpha$, these invariant amplitudes are determined by quark current operators taken between nucleon states. Assuming isospin symmetry, they can be expressed in terms of isovector electromagnetic form factors ${F}_{V1,V2} = {F}_{1,2}^p - {F}_{1,2}^n$, extracted from electron scattering data with constraints from atomic spectroscopy~\cite{dudelzakphd,janssens66,bartel66,albrecht66,frerejacque66,albrecht67,goitein67,litt70,goitein70,berger71,price71,ganichot72,bartel73,kirk73,borkowski74,murphy74,borkowski75,stein75,simon80,simon81,bosted90,rock92,sill93,walker94,andivahis94,dutta03,christy04,qattan05,bernauer14,milbrath99,pospischil01,gayou01,gayou02,strauch02,punjabi05,maclachlan06,jones06,crawford07,ron11,zhan11,puckett10,paolone10,puckett12,puckett17,meyerhoff94,eden94,passchier99,herberg99,rohe99,golak01,schiavilla01,zhu01,bermuth03,madey03,warren04,glazier05,geis08,riordan10,schlimme13,rock82,lung93,gao94,anklin98,kubon02,anderson07,lachniet09,Xiong:2019umf,Mohr:2015ccw,Schneider:2017lff,Ye:2017gyb,pohl10,antognini13,Kopecky:1995zz,Kopecky:1997rw,Lee:2015jqa}, and the axial form factor ${F}_\mathrm{A}$ as~\cite{LlewellynSmith:1971uhs}
\begin{equation}
 f_1 (\nu, Q^2) \to F_{V1}(Q^2) \,, \qquad f_2 (\nu, Q^2) \to F_{V2}(Q^2) \,, \qquad f_A (\nu, Q^2) \to F_A(Q^2) \,, \qquad f^3_A (\nu, Q^2) \to 0.
\end{equation}

\begin{figure}[t]
\centering
\includegraphics[width=0.2\textwidth]{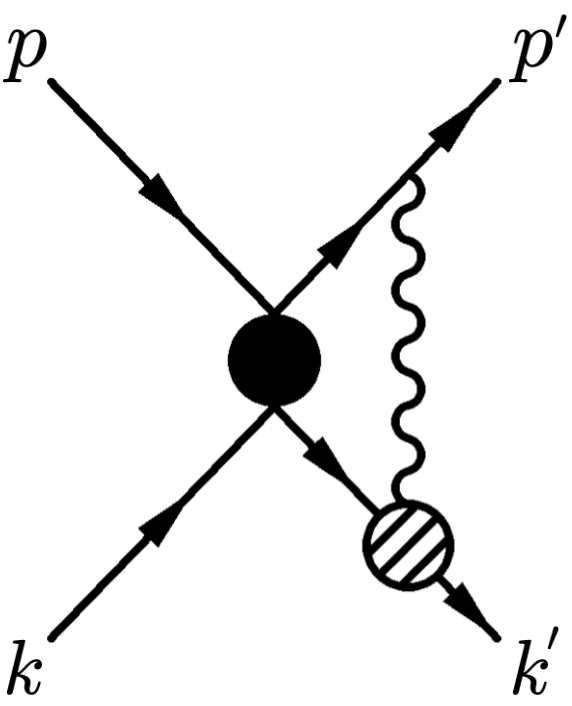}
\qquad \qquad
\includegraphics[width=0.2\textwidth]{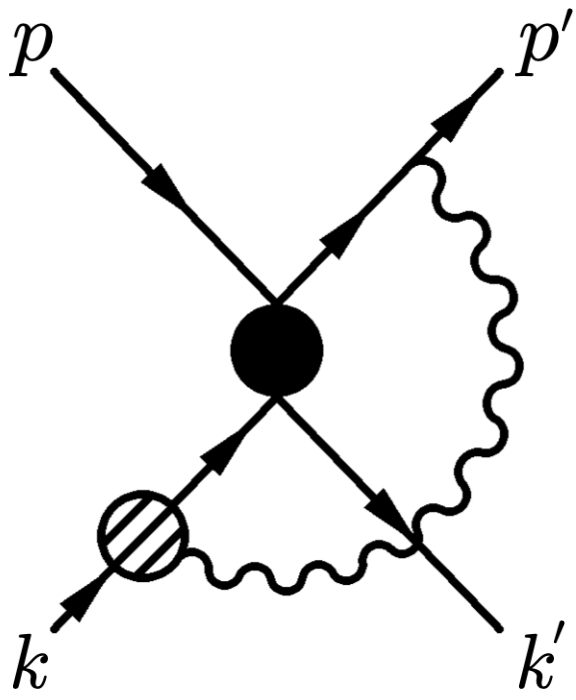}
\caption{Hadronic model for hard matching at one-loop level. The photon is exchanged between the charged lepton and nucleon lines. \label{fig:SIFF}}
\end{figure}

Radiative corrections modify each of the four invariant amplitudes and introduce nonperturbative information beyond the form factors $F_{1,2,A}(Q^2)$. In principle, sufficiently precise (anti)neutrino-nucleon scattering measurements could be used to extract this information~\cite{Mann:1973pr,Barish:1977qk,Miller:1981fa,Baker:1981su,Kitagaki:1983px,Belikov:1983kg,Bodek:2007ym,Kuzmin:2007kr,Bhattacharya:2011ah,Meyer:2016oeg,Alvarez-Ruso:2018rdx,Megias:2019qdv}. Alternatively, future lattice QCD calculations could perform a first-principles evaluation starting from the quark-level Lagrangian~\cite{Green:2017keo,Alexandrou:2017hac,Capitani:2017qpc,Gupta:2017dwj,Bali:2018qus,Jang:2019vkm,Alexandrou:2020okk,Djukanovic:2021cgp}. Since neither of these options is currently available, we use a simple hadronic model to represent the $\order(\alpha)$ corrections to the invariant amplitudes in the matching condition for purposes of illustration. The model is based on a form-factor insertion ansatz whereby point-particle Feynman-t'Hooft diagrams are dressed with on-shell form factors at the hadronic vertices. This procedure is illustrated in Fig.~\ref{fig:SIFF}. This hadronic model generalizes a commonly used estimate of two-photon exchange corrections in elastic lepton-proton scattering~\cite{Maximon:2000hm,Blunden:2003sp,Graczyk:2013fha,Tomalak:2014dja}.

The resulting vertex correction to the tree-level charged-current elastic process on nucleons is given by [suppressing the prefactor $\sqrt{2}\mathrm{G}_\mathrm{F} V_{ud}$ common with the tree-level expression in Eq.~(\ref{eq:CCQE_amplitude})]
\begin{align}
 T^v_{\nu_\ell n \to \ell^- p} &= e^2 \int \hspace{-0.1cm} \frac{\mathrm{d}^d L}{ \left( 2 \pi \right)^d} \bar{\ell} \gamma_\mu \frac{ - \slash{p}' - \slash{L} }{\left(L+p'\right)^2-m_\ell^2} \gamma^\sigma \mathrm{P}_\mathrm{L} \nu_{\ell} \mathrm{\Pi}^{\mu \nu} \left( L \right) \bar{p}\left( \Gamma^p_\nu \frac{\slash{k}' - \slash{L} + M}{\left( L - k'\right)^2 - M^2} \Gamma_\sigma + \Gamma_\sigma \frac{\slash{k} + \slash{L} + M}{\left( L + k\right)^2 - M^2} \Gamma^n_\nu \right) n\,, \nl
 T^v_{\bar{\nu}_\ell p \to \ell^+ n} &= e^2 \int \hspace{-0.1cm} \frac{\mathrm{d}^d L}{ \left( 2 \pi \right)^d} \bar{\bar{\nu}}_\ell \gamma^\sigma \mathrm{P}_\mathrm{L} \frac{ \slash{p}'+ \slash{L} }{\left(L+p'\right)^2-m_\ell^2} \gamma_\mu \bar{\ell} \mathrm{\Pi}^{\mu \nu} \left( L \right) \bar{n}\left( \Gamma^n_\nu \frac{\slash{k}' - \slash{L} + M}{\left( L - k'\right)^2 - M^2} \Gamma_\sigma + \Gamma_\sigma \frac{\slash{k} + \slash{L} + M}{\left( L + k\right)^2 - M^2} \Gamma^p_\nu \right) p\,. \label{eq:contribution_of_boxes}
\end{align}
Here
\begin{align}
 \mathrm{\Pi}^{\mu \nu} \left( L \right) &= \frac{i}{ L^2 - \lambda_\gamma^2}\left( - g^{\mu \nu} + \left( 1 - \xi_\gamma \right) \frac{L^\mu L^\nu}{L^2 - a \xi_\gamma \lambda_\gamma^2}\right), \label{eq:photon_propagator}
\end{align}
is the momentum-space photon propagator with the photon mass regulator $\lambda_\gamma$, the gauge-fixing parameter $\xi_\gamma$, and an arbitrary constant $a$, and
\begin{equation}
 \Gamma^N_\nu =\gamma_\nu {F}_1^N \left(-L^2\right) + \frac{i \sigma_{\nu \rho} L^\rho}{2M} {F}_2^N \left(-L^2\right), \label{eq:vector_vertex}
\end{equation}
is the electromagnetic vertex, where $N$ denotes proton ($p$) or neutron ($n$), while ${F}_1^N$ and ${F}_2^N$ are Dirac and Pauli electromagnetic form factors, respectively. The charged-current weak vertex is
\begin{equation}
 \Gamma_\sigma = \gamma_\sigma {F}_{V1}\left(Q^2\right) + \frac{i \sigma_{\sigma \rho} q^\rho}{2M} {F}_{V2}\left(Q^2\right) + \gamma_\sigma \gamma_5 {F}_\mathrm{A}\left(Q^2\right) + \frac{q_\sigma}{M} \gamma_5 \mathrm{F}_{P}\left(Q^2\right) \,. \label{eq:axial_vertex}
\end{equation}
We remark that a naive implementation of the form-factor insertion ansatz would involve form factors evaluated at momentum transfer $(q \pm L)^2$ rather than $q^2$, as in Eq.~(\ref{eq:axial_vertex}). Such an ansatz would violate electromagnetic gauge invariance, giving rise to spurious collinear singularities.\footnote{For example, current conservation requires a cancellation between amplitudes for photon emission from the charged lepton line and from the proton line. The former contribution would involve $F_i(-q^2)$, but the latter would involve $F_i(-(q\pm L)^2)$, spoiling the cancellation.} We thus adopt Eq.~(\ref{eq:axial_vertex}) as our default model. For this model calculation of virtual diagrams with Eq.~(\ref{eq:contribution_of_boxes}), exploiting the on-shell vertex of Eq.~(\ref{eq:vector_vertex}), we express the Dirac and Pauli form factors as [recall $M = (M_p + M_n)/{2}$ is the average nucleon mass and $\tau = {Q^2}/({4M^2})$]
\begin{align}
{F}^{p,n}_\mathrm{1} \left( Q^2 \right)&= \frac{G^{p,n}_\mathrm{E} + \tau G^{p,n}_\mathrm{M}}{1+ \tau}, \quad {F}^{p,n}_\mathrm{2} \left( Q^2 \right) = \frac{G^{p,n}_\mathrm{M} - G^{p,n}_\mathrm{E}}{1+ \tau},
\end{align}
and employ a simple dipole form for the proton electric, proton magnetic, and neutron magnetic form factors, and an ansatz for the neutron electric form factor constrained by its charge radius~\cite{Kopecky:1995zz,Kopecky:1997rw} at low $Q^2$ and by $\sim Q^{-4}$ perturbative QCD behavior at high $Q^2$~\cite{Galster:1971kv,Lepage:1980fj,Kelly:2004hm}:
\begin{align}
    G^{p}_\mathrm{E}\left( Q^2 \right) = \frac{G^{p}_\mathrm{M} \left( Q^2 \right)}{\mu_p} = \frac{G^{n}_\mathrm{M} \left( Q^2 \right)}{\mu_n} = \frac{1}{\left( 1 + \frac{Q^2}{\Lambda^2} \right)^2} \,,
    \quad
    G^{n}_\mathrm{E} \left( Q^2 \right)&= \frac{-<r^2_\mathrm{E}> Q^2}{6\left( 1+ b \tau \right)}
    G^{p}_\mathrm{E} \left( Q^2 \right) \,, 
    \label{eq:dipole4}
\end{align}
with $<r^2_\mathrm{E}> = - 0.1161~\mathrm{fm}^2$, $\mu_p = 2.7928$, $\mu_n = -1.9130$, $\Lambda^2 = 0.71\,{\rm GeV}^2$, and $b=4.6$. We have performed the evaluation of the UV-finite amplitudes from Eqs.~(\ref{eq:contribution_of_boxes}) in $d=4$ and cross-checked imaginary parts of all terms in Eqs.~(\ref{eq:contribution_of_boxes}) by independent calculation exploiting unitarity.\footnote{In contrast to elastic charged lepton-proton scattering, the high-energy behavior of imaginary parts in the charged-current elastic process does not allow us to write down unsubtracted dispersion relations for any of the invariant amplitudes. Performing crossing and charge conjugation, the amplitudes of (anti)neutrino-nucleon elastic scattering $\nu_\ell n \to \ell^{-} p$ ($\overline{\nu}_\ell p \to \ell^{+} n$) are not related to amplitudes of a (anti)neutrino scattering reaction. Thus, we cannot use crossing relations for amplitudes in the same channel to suppress the high-energy behavior and obtain convergence of dispersive integrals.} This section represents a model for the real part of the nucleon contribution and a discussion of uncertainties owing to parametric inputs and neglected inelastic contributions.

The amplitudes in the hadronic model can be represented as
\begin{align}
 f_1 (\nu, Q^2) &= \sqrt{Z_\ell Z_h^{(p)}} \left( F_{V1}(Q^2) + f^v_1 (\nu, Q^2) \right)\,, \nl
 f_2 (\nu, Q^2) &= \sqrt{Z_\ell Z_h^{(p)}} \left(F_{V2}(Q^2) + f^v_2 (\nu, Q^2) \right)\,, \nl
 f_A (\nu, Q^2) &= \sqrt{Z_\ell Z_h^{(p)}} \left( F_A(Q^2) + f^v_A (\nu, Q^2)\right)\,, \nl
 f^3_A (\nu, Q^2) &= \sqrt{Z_\ell Z_h^{(p)}} \left(f^3_A \right)^v (\nu, Q^2) \,, \label{eq:hadronic_model}
\end{align}
where the charged lepton field renormalization factor $Z_\ell$ is given by the standard QED expression:
\begin{equation}
 Z_\ell = 1 - \frac{\alpha}{4\pi} \left( \ln \frac{\mu^2}{m_\ell^2} + 2\ln \frac{\lambda_\gamma^2}{m_\ell^2} + 4 \right) + \frac{\alpha}{4\pi}\left( 1 - \xi_\gamma \right)\left( \ln \frac{\mu^2}{\lambda_\gamma^2} + 1 + \frac{a \xi_\gamma \ln (a \xi_\gamma)}{1-a \xi_\gamma} \right) \,.
\end{equation}
For the external proton, we include the on-shell wave-function renormalization constant of the heavy spin-1/2 fermion (cf. Ref.~\cite{Hill:2016gdf} for the analogous discussion of Born amplitudes in electron-proton scattering)
\begin{equation}
 Z_h^{(p)} = 1 + \frac{\alpha}{2 \pi} \ln \frac{\mu^2}{\lambda_\gamma^2} + \frac{\alpha}{4 \pi} \left( 1 - \xi_\gamma \right) \left( \ln \frac{\mu^2}{\lambda_\gamma^2} + 1 + \frac{a \xi_\gamma \ln (a \xi_\gamma) }{1 - a \xi_\gamma } \right)\,.
\end{equation}

Amplitudes in Eqs.~(\ref{eq:hadronic_model}) do not depend on the arbitrary regularization parameter $a$ and have correct IR behavior. The expressions in Eqs.~(\ref{eq:hadronic_model}) provide a conventional, and model-independent, definition of charged-current ``Born" form factors $F_{1,2,A}(Q^2)$, once the functions $f_{1,2,A}^v$ are specified. For definiteness, we take $\mu=M$. In the absence of $CP$ violation (which we neglect), we have defined the overall phase of the amplitudes in Eq.~(\ref{eq:CCQE_amplitude}) such that the Born form factors are real. We remark that the separation of the gauge-independent hard matching contribution into ``Born" and ``non-Born" terms is necessarily QED gauge dependent; for definiteness, we specify Feynman-'t Hooft gauge, i.e., $\xi_\gamma = 1$.\footnote{Alternatively, one can define gauge-independent ``Born" form factors by moving local gauge-dependent virtual contributions to field renormalization factors of the external proton and neutron. Such a prescription differs from our default calculation by small structure-dependent contributions to the nucleon self-energies, which are independent of the lepton mass.} Our hadronic model identifies $f_{1,2,A}^v$ with the contributions from Eqs.~(\ref{eq:contribution_of_boxes}). For $F_{V1,V2}(Q^2)$, we use the isospin rotation of Born form factors%
\footnote{We have neglected isospin-violating effects in the relation of charged-current form factors to electromagnetic form factors and in the omission of second-class current contributions to the tree-level process~\cite{Weinberg:1958ut,Shiomi:1996np,Govaerts:2000ps,Minamisono:2001cd}. Isospin-breaking effects are expected from constituent quark model estimates to be of permille level~\cite{Behrends:1960nf,Dmitrasinovic:1995jt,Miller:1997ya,Lewis:2007qxa}. According to a chiral perturbation theory-based calculation~\cite{Kubis:2006cy}, they can reach the percent level for the magnetic form factor; while this is potentially significant compared to the precision of electron-proton scattering data, the tree-level uncertainty of the charged-current process is dominated by axial form factor. Radiative corrections are neglected in existing extractions of $F_A(Q^2)$.} extracted from experimental $e-p$ and $e-n$ scattering data, measurements of the neutron scattering length, and $\mu H$ spectroscopy data~\cite{Borah:2020gte}, i.e., $F_{V1,2}(Q^2) = F^p_{1,2}(Q^2) - F^n_{1,2}(Q^2)$, where $p$ stands for the proton and $n$ stands for the neutron. For $F_A(Q^2)$, we use the Born form factor extracted from experimental $\nu_\mu-n$ scattering data~\cite{Meyer:2016oeg}.

\begin{figure}[t]
 \centering
 \includegraphics[height=0.5\textwidth]{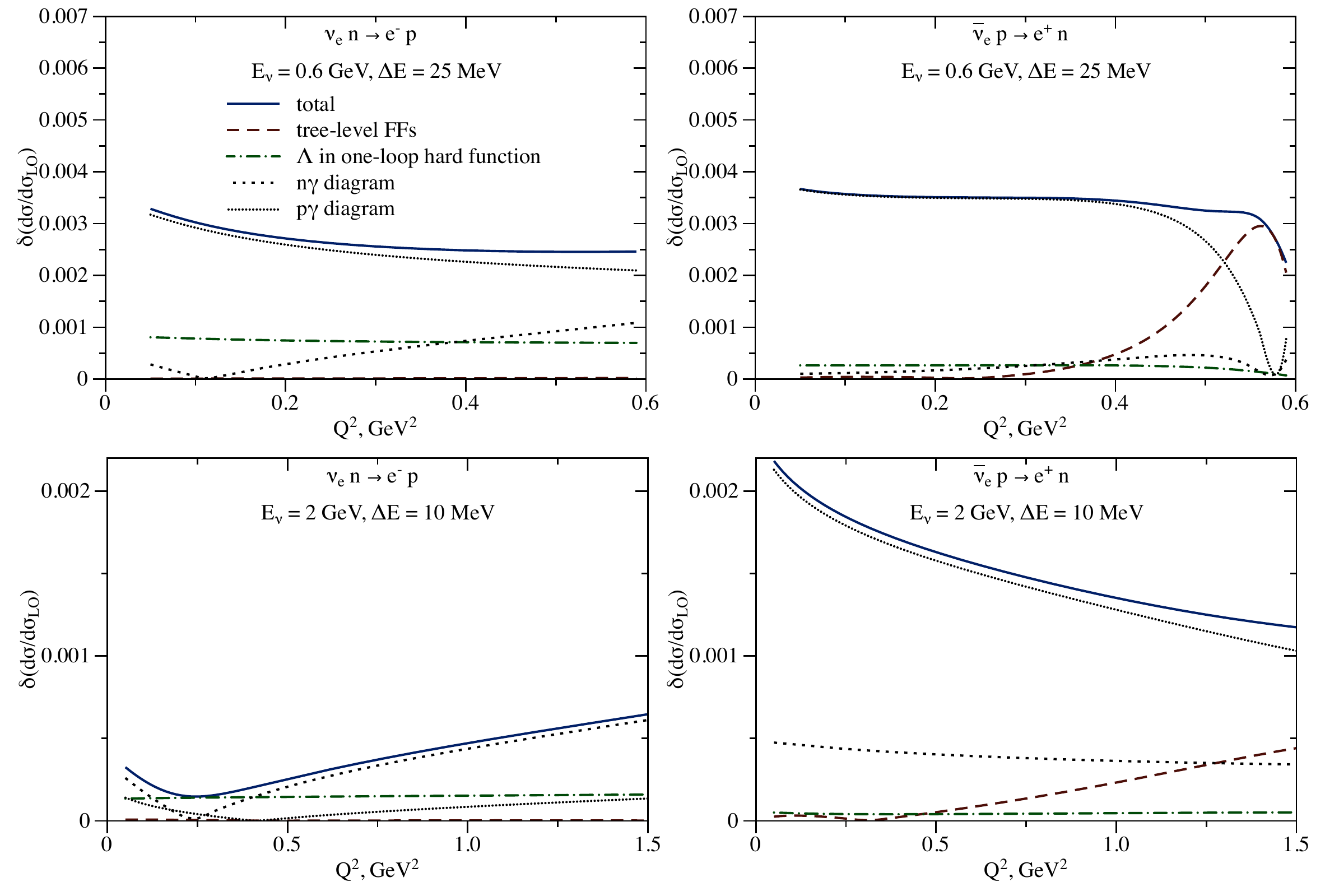}
 \caption{Uncertainties from the hard function on the ratio of the differential cross section, with only one soft photon radiated, to the tree-level result in $\nu_e n \to e^- p $ (left plots) and $\bar{\nu}_e p \to e^+ n $ (right plots). For initial (anti)neutrino energies $E_\nu = 600~\mathrm{MeV}$ (upper plots) and $E_\nu = 2~\mathrm{GeV}$ (lower plots), the soft-photon energy cutoff is $\Delta E = 25~\mathrm{MeV}$ and $\Delta E = 10~\mathrm{MeV}$, respectively. The uncertainty propagated from the tree-level form factors is shown by the red dashed line. The difference obtained by varying nucleon electromagnetic form-factor parameter $\Lambda^2$ as $\Lambda^2 \to 1.1\Lambda^2$ in the loop diagram is shown by the green dash-dotted line. As discussed in the text, the total contribution of $G_E^n$ and $G_M^n$ at the neutron electromagnetic vertex and the variation of the proton electromagnetic vertex obtained by adding $G_E^p \to G_E^p +G_E^n$ and $G_M^p \to G_M^p +G_M^n$ represent an estimate for neglected contributions in the hadronic model. These variations are shown by the black dotted line and by the black fine-dotted line, respectively. The sum of all variations in quadrature is shown by the blue solid line. \label{fig:hm_electron_error2}}
\end{figure}

\begin{figure}[t]
 \centering
 \includegraphics[height=0.5\textwidth]{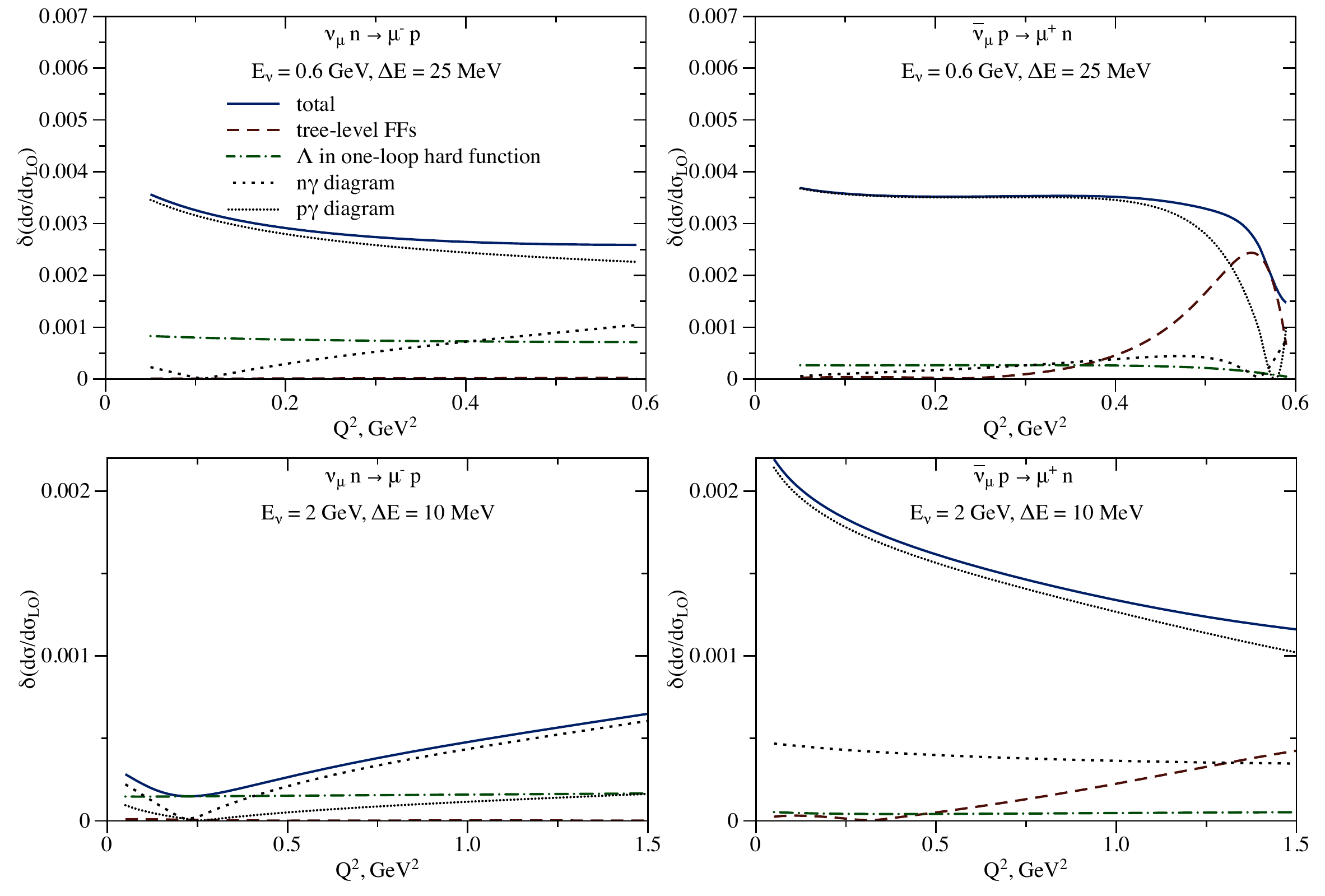}
 \caption{The same as Fig.~\ref{fig:hm_electron_error2} but for the muon flavor. \label{fig:hm_muon_error2}}
\end{figure}

For the phenomenological analysis, we consider the following assignment of uncertainty to the hard function matching condition. Consider first the Born form factors $F_{1,2,A}(Q^2)$. For tree-level amplitudes (but not for the calculation of loop diagrams), these form factors and corresponding uncertainties are taken as default fits of data with $Q^2 < 1~\mathrm{GeV}^2$ from the analysis of scattering and spectroscopy data~\cite{Meyer:2016oeg,Borah:2020gte}, taking the isospin-decomposed fit for the vector form factors; all form factors are presented in $z$-expansion form while uncertainties are described by the covariance matrix of $z$-expansion parameters. In ratios to the leading-order cross section, this dominant source of uncertainty is largely reduced. Second, for the amplitudes $f^v_{1,2,A}(\nu,Q^2)$, we consider separate variations in the hadronic model to account for parametric inputs and for neglected contributions. Note that the hadronic model involves form factors $F_{1,2,A}(Q^2)$ appearing inside virtual loop diagrams; here we use a simpler dipole ansatz for electromagnetic form factors as specified in Eqs.~(\ref{eq:dipole4}). For the parametric inputs, we vary the nucleon electromagnetic form factors appearing in loop diagrams by shifting the pole parameter $\Lambda^2$ in Eqs.~(\ref{eq:dipole4}) as $\Lambda^2 \to (1\pm 0.1) \Lambda^2$; this range covers experimentally allowed values for electromagnetic form factors~\cite{Borah:2020gte}. Model dependence due to the insertion of on-shell hadronic vertices and the neglect of inelastic intermediate states is represented by a simple ansatz that adds the neutron electric and magnetic form factors to each of the neutron and proton electromagnetic vertices.\footnote{This ansatz is motivated by the observation that two-photon exchange corrections in elastic electron-proton scattering from the proton Pauli form factor were shown to have a similar size to inelastic contributions~\cite{Tomalak:2015aoa}.} Based on this ansatz, relative contributions to the uncertainty of the cross-section ratio to the tree-level result from the hard function are displayed for the electron flavor in Fig.~\ref{fig:hm_electron_error2} and for the muon flavor in Fig.~\ref{fig:hm_muon_error2}. Each separate source of uncertainty is at permille level or even below.

Our definition of Born form factors in Eqs.~(\ref{eq:hadronic_model}) corresponds to a conventional $\overline{\rm MS}$ scheme at scale $\mu=M$~\cite{Hill:2016gdf} and $\xi_\gamma = 1$. As discussed in Ref.~\cite{Hill:2016gdf}, a variety of other schemes have been used to extract Born form factors in the literature~\cite{Maximon:2000hm,Lee:2015jqa} but differ at a level not resolved by data. To account for this discrepancy, we add in quadrature a perturbative uncertainty estimated by varying the scale of matching, $\mathrm{min}\left( Q^2, M^2, E_\nu^2 \right)/2 \lesssim \mu^2 \lesssim 2 \mathrm{max}\left( Q^2, M^2, E_\nu^2 \right)$, and assuming that the experimental extraction $F_i (Q^2)$ represents the Born amplitude at $\mu=M$.

\subsubsection{Matching and renormalization group evolution \label{sec:rge_evolution}}

The hard function $H$ in Eq.~(\ref{eq:factorization_formula}) is determined by matching the factorization formula to the cross section of Eq.~(\ref{eq:xsection_CCQE}) evaluated in the hadronic model. Contributions from soft and collinear momentum regions in the hadronic model are the same, by construction, as the corresponding one-loop expressions in the effective theory derived in Secs.~\ref{sec:soft} and~\ref{sec:collinear}. By choosing the hard matching scale $\mu_\mathrm{H} \sim \Lambda_\mathrm{hard} \sim M \sim Q \sim E_\nu$, we ensure that $H$ is free from large logarithms. In order to resum large logarithms in the final observable cross section, we evolve the hard function to the low scale $\mu_\mathrm{L}$, writing
 \begin{equation}
 \mathrm{d} \sigma \sim H\left(\mu_\mathrm{H} \right) \left[\frac{H\left( \mu_\mathrm{L}\right)}{H\left(\mu_\mathrm{H}\right)}\right] J \left( \mu_\mathrm{L}\right) S \left( \mu_\mathrm{L}\right).
 \end{equation}
Performing the running at two-loop level with the anomalous dimension $\gamma_\mathrm{H} = - \gamma_\mathrm{J} - \gamma_\mathrm{S}$, with $\gamma_\mathrm{S}$ and $\gamma_\mathrm{J}$ from Eqs.~(\ref{eq:running_soft}) and~(\ref{eq:running_collinear}), respectively, we resum leading and subleading large logarithms and obtain $H\left( \mu_\mathrm{L}\right)$ at renormalization scale $\mu_\mathrm{L}\ll \mu_\mathrm{H}$. The low scale could be further resolved into separate soft and collinear scales, $\mu_\mathrm{S} \sim \lambda \Lambda_{\rm hard}$ and $\mu_\mathrm{J} \sim \lambda^\frac12 \Lambda_{\rm hard}$, writing $J(\mu_\mathrm{L})= J(\mu_\mathrm{J}) [J(\mu_\mathrm{L})/J(\mu_\mathrm{J})]$ and $S(\mu_\mathrm{L}) = S(\mu_\mathrm{S}) [S(\mu_\mathrm{L})/S(\mu_\mathrm{S})]$. Since the impact of such further resummation is numerically small, we use a common scale $\mu_\mathrm{S} = \mu_\mathrm{J} = \mu_\mathrm{L}$ for simplicity. Residual perturbative uncertainty is estimated by varying $\mathrm{min} \left(m_\ell^2, \left(\Delta E \right)^2 \right)/2 \lesssim \mu_\mathrm{L}^2 \lesssim 2\, \mathrm{max} \left(m_\ell^2, \left(\Delta E \right)^2 \right)$ and $\mathrm{min} \left( Q^2, M^2, E_\nu^2 \right)/2 \lesssim \mu_\mathrm{H}^2 \lesssim 2 \mathrm{max}\left( Q^2, M^2, E_\nu^2 \right)$.

\begin{figure}[t]
 \centering
 \includegraphics[height=0.5\textwidth]{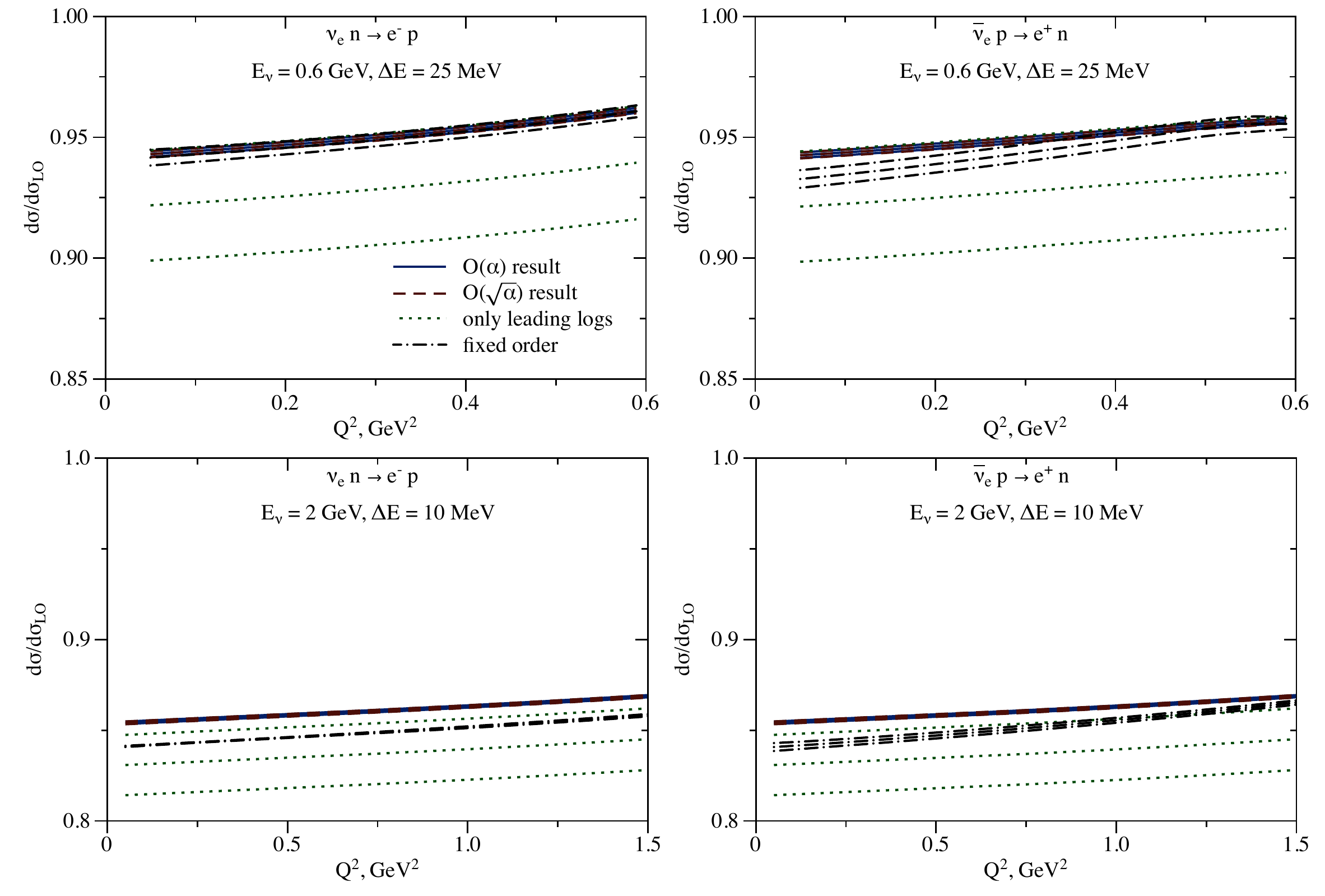}
 \caption{Ratio of the differential cross section, with only soft-photon radiation, to the tree-level result in $\nu_e n \to e^- p $ (left plots) and $\bar{\nu}_e p \to e^+ n $ (right plots). For incident (anti)neutrino energies $E_\nu = 600~\mathrm{MeV}$ (upper plots) and $E_\nu = 2~\mathrm{GeV}$ (lower plots), the soft-photon energy cutoff is $\Delta E = 25~\mathrm{MeV}$ and $\Delta E = 10~\mathrm{MeV}$, respectively. The resummed result including all corrections of $\order \left( \alpha \right)$ is shown by the blue solid lines. It is compared to the resummation result including only the leading logarithms, shown by the green dotted lines, and to the result including subleading logarithms of order $\order \left( \sqrt{\alpha} \right)$, shown by the red dashed lines. To illustrate the net effect of resummation, we also present the fixed-order result using the same hadronic model by the black dash-dotted lines. Hadronic uncertainty, described in Sec.~\ref{sec:model}, is displayed for the fixed-order calculation. Only perturbative uncertainty, estimated by scale variation as described in the text, is presented for the first three lines. \label{fig:pert_electron_error}}
\end{figure}

To illustrate convergence of the perturbative expansion, we consider the theory with one dynamical charged lepton $\numl = 1$, i.e., electron loops only.\footnote{For numerical results in following sections, we specify $\numl = 0$ below the electron mass scale, $\numl=1$ between electron and muon mass scales, and $\numl=2$ above.} The electromagnetic coupling constant at the scale $\mu$ is related to its on-shell value $\alpha_0\approx 1/137.036$ as~\cite{Hill:2016gdf}
\begin{equation}
\alpha = \alpha_0 \left[ 1 + \frac{1}{3} \ln \frac{\mu^2}{m_e^2} \left( \frac{\alpha_0}{\pi} \right) + \left( \frac{1}{9} \ln^2 \frac{\mu^2}{m_e^2} + \frac{1}{4} \ln \frac{\mu^2}{m_e^2} + \frac{15}{16} \right) \left( \frac{\alpha_0}{\pi} \right)^2\right].
\end{equation}
The running of the hard function from the hard scale $\mu_\mathrm{H}$ to the soft scale $\mu_\mathrm{L}$ can be solved analytically and can be expressed in terms of the charged lepton energy in the proton rest frame, $\varepsilon_\ell = m_\ell \left( v_\ell \cdot v_p \right)$, as
\begin{align}
\ln \frac{H \left( \mu_\mathrm{L} \right)}{H \left( \mu_\mathrm{H} \right)} &= \gamma_0 \ln \frac{\mu_\mathrm{H}^2}{\mu_\mathrm{L}^2} \left( \frac{1}{4} \ln \frac{\mu_\mathrm{H}^2}{\mu_\mathrm{L}^2} + \frac{1}{2} \ln \frac{\mu_\mathrm{L}^2}{4 \varepsilon_\ell^2} + \frac{5}{4} \right) \left(\frac{\alpha_0}{\pi} \right) + \frac{\gamma_0}{18} \ln \frac{\mu_\mathrm{H}^2}{\mu_\mathrm{L}^2} \left( \ln^2 \frac{\mu_\mathrm{H}^2}{\mu_\mathrm{L}^2} +3 \ln \frac{\mu_\mathrm{H}^2}{\mu_\mathrm{L}^2} + 6 \ln \frac{\mu_\mathrm{L}^2}{m_e^2} \right) \left(\frac{\alpha_0}{\pi} \right)^2 \nonumber \\
&+ \frac{\gamma_0}{6}\ln \frac{\mu_\mathrm{H}^2}{\mu_\mathrm{L}^2} \left( \ln \frac{\mu_\mathrm{H}^2}{4 \varepsilon_\ell^2} \ln \frac{\mu_\mathrm{L}^2}{m_e^2} + \frac{1}{2}\ln \frac{\mu_\mathrm{H}^2}{\mu_\mathrm{L}^2} \ln \frac{m_e^2}{4 \varepsilon^2_\ell} + \frac{1}{4} \ln \frac{\mu_\mathrm{H}^2}{m_e^2} +\frac{1}{4} \ln \frac{\mu_\mathrm{L}^2}{m_e^2} - \frac{3\gamma_1}{2\gamma_0} \ln \frac{\mu_\mathrm{H}^2}{\mu_\mathrm{L}^2} - \frac{3\gamma_1}{\gamma_0} \ln \frac{\mu_\mathrm{L}^2}{4 \varepsilon_\ell^2} \right) \left(\frac{\alpha_0}{\pi} \right)^2 \nonumber \\
&+ \frac{\gamma_0}{18} \ln \frac{\mu_\mathrm{H}^2}{\mu_\mathrm{L}^2} \left( \frac{1}{4} \ln^3 \frac{\mu_\mathrm{H}^2}{\mu_\mathrm{L}^2} + \frac{1}{2} \ln \frac{\mu_\mathrm{H}^2}{\mu_\mathrm{L}^2} \ln^2 \frac{\mu_\mathrm{L}^2}{m_e^2}+ \frac{1}{3} \ln^2 \frac{\mu_\mathrm{H}^2}{\mu_\mathrm{L}^2} \ln \frac{\mu_\mathrm{L}^2}{4\varepsilon^2_\ell} + \frac{2}{3} \ln^2 \frac{\mu_\mathrm{H}^2}{\mu_\mathrm{L}^2} \ln \frac{\mu_\mathrm{L}^2}{m_e^2} \right)\left(\frac{\alpha_0}{\pi} \right)^3 \nonumber \\
&+ \frac{\gamma_0}{18} \ln \frac{\mu_\mathrm{H}^2}{\mu_\mathrm{L}^2} \ln \frac{\mu_\mathrm{L}^2}{m_e^2} \left( \ln \frac{\mu_\mathrm{L}^2}{m_e^2} \ln \frac{\mu_\mathrm{L}^2}{4 \varepsilon^2_\ell} + \ln \frac{\mu_\mathrm{H}^2}{\mu_\mathrm{L}^2} \ln \frac{\mu_\mathrm{L}^2}{4\varepsilon^2_\ell} \right)\left(\frac{\alpha_0}{\pi} \right)^3 + \order \left(\alpha_0^\frac32\right) \,, \label{eq:running_hard_function}
\end{align}
where we keep terms contributing through order $\alpha^1$, when $\alpha \ln^2(\mu_\mathrm{H}/\mu_\mathrm{L}) = \order(1)$. We illustrate the convergence of the perturbation theory by evaluating the cross section for the electron flavor charged-current elastic process, including only soft radiation, in Fig.~\ref{fig:pert_electron_error}. The successive curves represent resummations correct through ${\cal O}(1)$, ${\cal O}(\sqrt{\alpha})$, and ${\cal O}(\alpha)$, respectively, with perturbative uncertainties of order ${\cal O}(\sqrt{\alpha})$, ${\cal O}(\alpha)$, and ${\cal O}(\alpha\sqrt{\alpha})$; resummation corrections beyond ${\cal O}(\sqrt{\alpha})$ are numerically small. We also compare our result to the fixed-order calculation of the same process.

\subsection{Factorization for inelastic processes}

The factorization formalism extends naturally to other exclusive processes at the nucleon level, for example radiation of noncollinear hard photons ($\nu_\ell N \to \ell N \gamma$) or pion emission ($\nu_\ell N \to \ell N \pi$). In each case, soft and jet functions are perturbatively calculable, depending only on particle masses and velocities and kinematic variables. Hadronic physics is encoded in a basis of hard functions. Semi-inclusive processes that sum over hadronic channels ($\nu_\ell N \to \ell X$) can be similarly analyzed. Such processes are left to future work.

\section{Inclusive cross sections \label{sec:inclusive}}

In this section, we describe the inclusion of noncollinear hard photons to scattering cross sections and discuss model-independent properties of inclusive cross sections on nucleons and nuclei.

\subsection{Hadronic model for hard real photon emission \label{sec:inclusive_model}}

Let us consider the process $\nu_\ell n \to \ell^- p (\gamma)$ [or the analogous antineutrino process $\bar{\nu}_\ell p \to \ell^+ n (\gamma)$] including arbitrary photon kinematics. To describe the radiation of noncollinear hard photons, we exploit the same hadronic model as in Sec.~\ref{sec:model}. Recall that our prescription is equivalent to employing free-particle propagators for intermediate states, with the electroweak vertex evaluated using external-leg kinematics, as in Eq.~(\ref{eq:axial_vertex}). This ansatz ensures electromagnetic gauge invariance and avoids spurious collinear singularities of a naive form-factor insertion model. In the phenomenological analysis, we demonstrate how such spurious singularities would impact cross-section predictions for near-collinear kinematics (Fig.~\ref{fig:real_radiation_misidentified}). For generic, noncollinear, kinematics, the difference between our default gauge invariant and a local non-gauge-invariant model, where the electroweak vertex is evaluated in kinematics of a local field theory, can be interpreted as a simple measure of hadronic model dependence. This difference is illustrated in Figs.~\ref{fig:real_radiation_hard_noncollinear_photons_electron} and \ref{fig:real_radiation_hard_noncollinear_photons_muon}.

\subsection{Expansion in small lepton mass \label{sec:mass_expansion}}

We have observed previously that sufficiently inclusive observables have a finite limit at vanishing lepton mass $m_\ell\to 0$. In the static limit, this behavior is seen explicitly in the jet observable at fixed $\Delta \theta$ [cf. Eq.~(\ref{eqn:delta_jet_static_limit_m})] and in the inclusive cross section [cf. Eq.~(\ref{eqn:deltaSimple})]. In general, the $m_\ell\to 0$ limit must be finite for any observable including real photon radiation that is kinematically degenerate with the charged lepton in the absence of radiation~\cite{Bloch:1937pw,Yennie:1961ad,Kinoshita:1962ur,Lee:1964is}.

The finiteness of the $m_\ell\to 0$ limit has important implications for flavor ratios of charged-current (anti)neutrino cross sections. Consider, e.g., the total inclusive cross section for $\nu_\ell n \to \ell^- p (\gamma)$ as a function of the lepton mass:
\begin{align}
 \sigma(m_\ell) = A + B_0 \frac{m_\ell^2}{\Lambda^2} + B_1 \frac{m_\ell^2}{\Lambda^2} \ln\frac{m_\ell^2}{\Lambda^2} + \dots \,,
 \label{eq:sigmaKLN}
\end{align}
where $\Lambda$ denotes a conventional hard scale and the ellipsis denotes terms of order $m_\ell^4$ and terms containing $m_\ell^2\ln^2(m_\ell)$ that arise from two-loop QED corrections; such terms are negligibly small in the numerical analysis. Here $A$, $B_0$, and $B_1$ are independent of $m_\ell$ but depend on the neutrino energy and hadronic parameters. The ratio of muon to electron cross sections is then
\begin{align}
 \frac{\sigma(m_\mu) }{ \sigma(m_e)} & = 1 + {\cal B}_0 \frac{m_\mu^2}{\Lambda^2} + {\cal B}_1 \frac{m_\mu^2}{\Lambda^2} \ln\frac{m_\mu^2}{\Lambda^2} + \order\left( \frac{m_e^2}{ \Lambda^2}, \alpha^2 \frac{m_\mu^2}{ \Lambda^2}\ln^2 \frac{m_\mu }{ \Lambda}, \frac{m_\mu^4}{ \Lambda^4} \right)\,,
\end{align}
where we have written ${\cal B}_0=B_0/A$ and ${\cal B}_1=B_1/A$. 

As an explicit example, the static-limit cross section through one-loop order is
\begin{align}
 \sigma(m_\ell) = \sigma_{\rm LO} \bigg\{ 1 + \frac{\alpha }{ \pi} \left[ \frac34 \ln \frac{\mu^2}{ 4 E_\nu^2} + \frac{17}{4} + \frac{\pi^2}{3} + \frac{m_\ell^2}{E_\nu^2}\left( \frac34 \ln\frac{4 E_\nu^2}{m_\ell^2} + \frac{19}{8} + \frac{\pi^2}{6} \right) + \order(m_\ell^4)\right] \bigg\} \,.
\end{align}
Choosing $\Lambda=2 E_\nu$ and accounting for the $m_\ell$ dependence of $\sigma_{\rm LO}$, we find 
\begin{align}
 {\cal B}_0 = -2 + \frac{\alpha}{\pi}\left( \frac{19}{2} + \frac{2\pi^2}{3} \right) \,, \quad {\cal B}_1 = -\frac{3 \alpha}{\pi} \,.
\end{align}

The general case, beyond the static limit, is treated similarly but with different numerical values of ${\cal B}_0$ and ${\cal B}_1$. For example, at an illustrative neutrino energy $E_\nu=2\,{\rm GeV}$, and setting $\Lambda = 1\,{\rm GeV}$, integrating the total charged-current elastic cross section of Eq.~(\ref{eq:xsection_CCQE}) yields
\begin{align}
{\cal B}_0(E_\nu=2\,{\rm GeV}) = -0.28 + \order(\alpha, \epsilon_{\rm nuc}) \,, \quad {\cal B}_1(E_\nu=2\,{\rm GeV}) = \order(\alpha, \epsilon_{\rm nuc}) \,, 
\end{align}
where $\epsilon_{\rm nuc}$ denotes a possible nuclear correction for applications to bound nucleons. We will compute the $\order(\alpha)$ contributions within our hadronic model as part of the phenomenological analysis in Sec.~\ref{sec:inclusive_xsection}. We also consider $\order(\epsilon_{\rm nuc})$ corrections in a standard nuclear model in Sec.~\ref{sec:nuclear_effects}.

\section{Phenomenological applications \label{sec:pheno}}

\begin{figure}[t]
 \centering
 \includegraphics[height=0.5\textwidth]{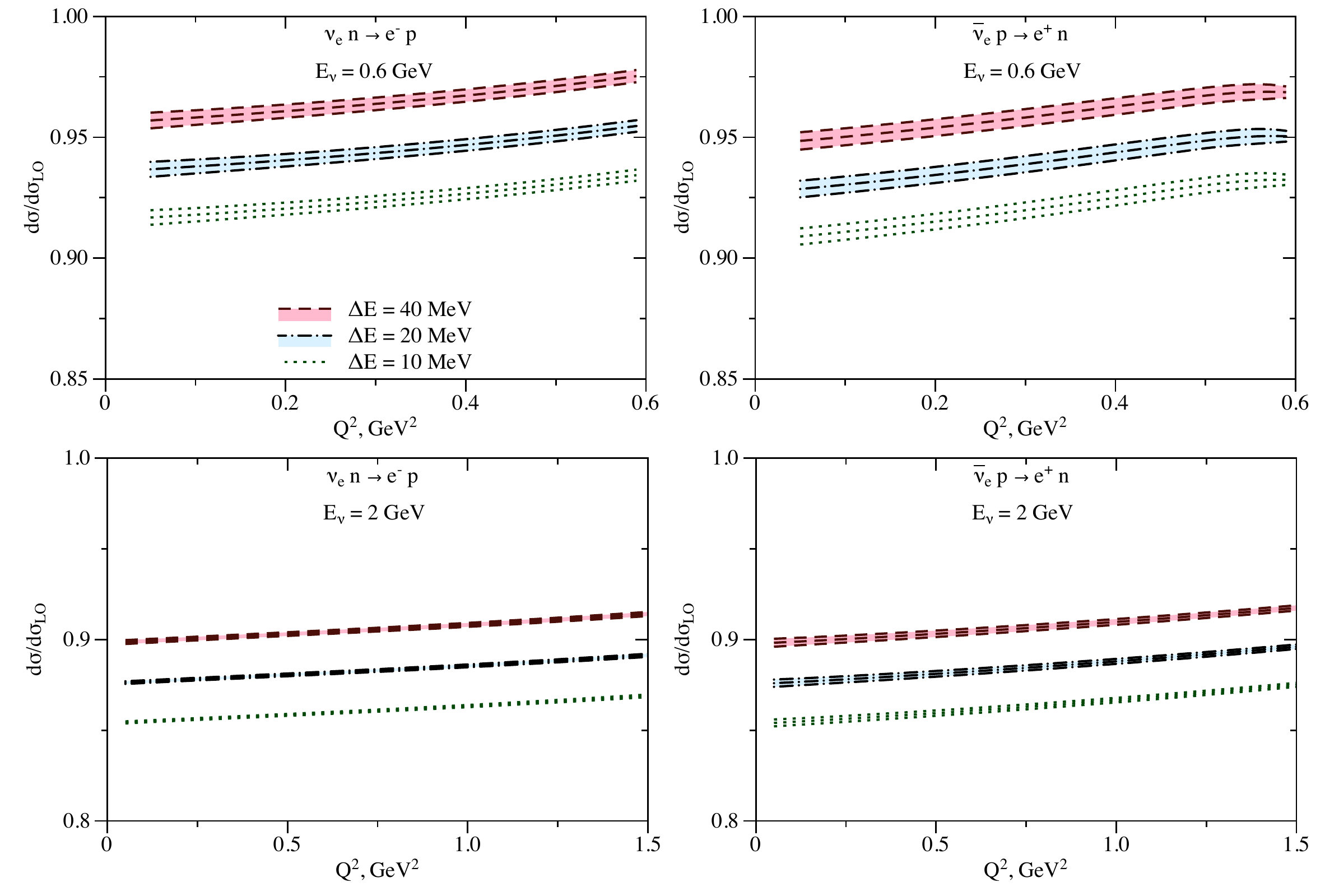}
 \caption{Differential cross-section ratio to the tree-level result for electron (anti)neutrino charged-current elastic scattering on the nucleon accompanied by soft radiation, as a function of $Q^2$ for a few different soft-photon energy cutoffs $\Delta E$. Left and right plots are for neutrino and antineutrino scattering, respectively. Upper and lower plots are for $E_\nu = 600~\mathrm{MeV}$ and $E_\nu = 2~\mathrm{GeV}$, respectively. The curves show the results for $\Delta E = 10, \, 20$, and $40\,{\rm MeV}$. \label{fig:soft_e}}
\end{figure}

\begin{figure}[htb]
 \centering
 \includegraphics[height=0.5\textwidth]{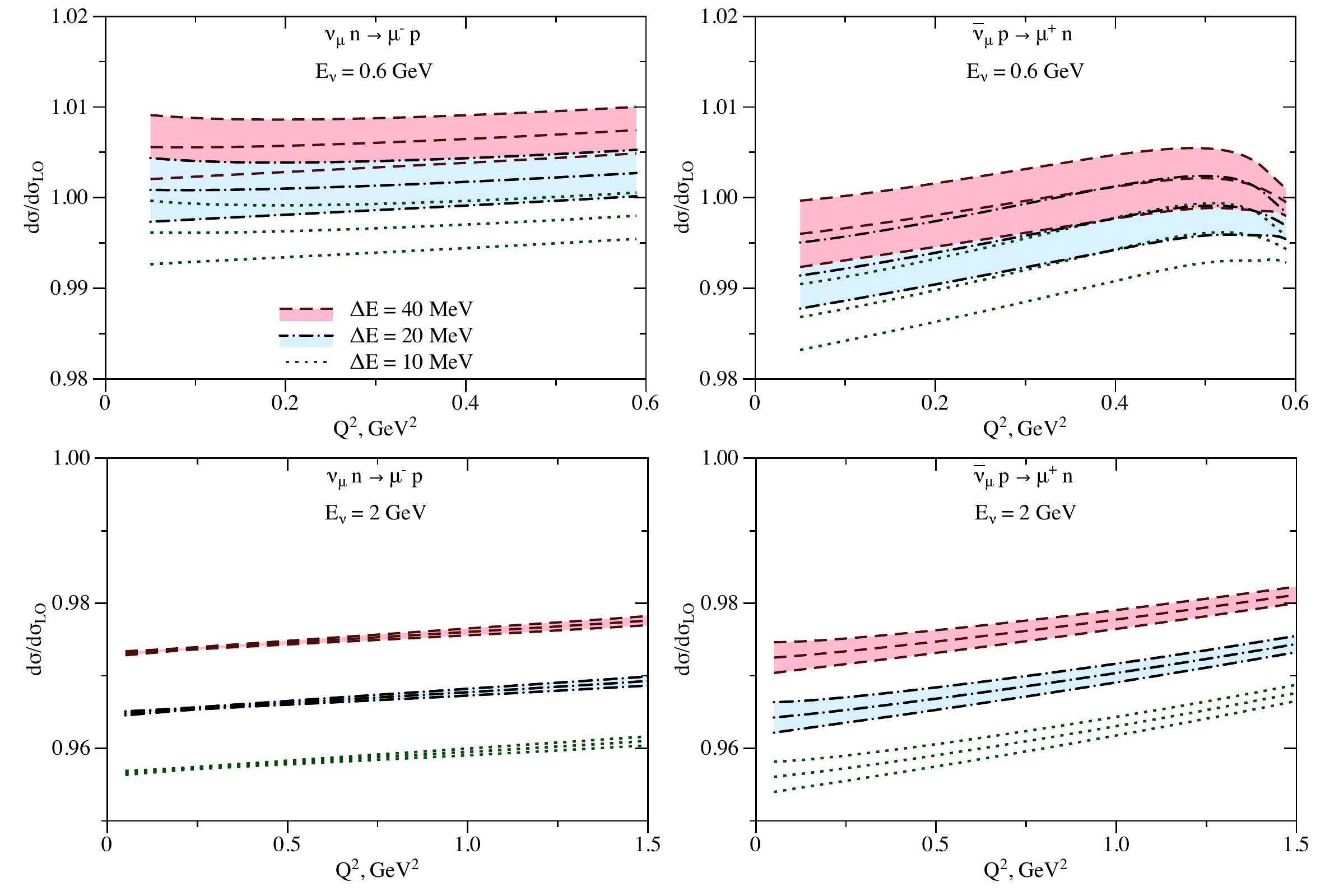}
 \caption{The same as Fig.~\ref{fig:soft_e} but for the muon flavor. \label{fig:soft_mu}}
\end{figure}

In this section, we examine in detail several variations of the observables introduced above. The observables are representative of typical event classes for charged-current elastic (anti)neutrino scattering with final-state electron or muon, with choices for real photons determined by typical detector capabilities, as discussed in Sec.~\ref{sec:overview}. In the following Sec.~\ref{sec:electron_soft_photons_xsection}, we consider a ``minimal" observable containing just the charged lepton and soft radiation. This type of observable is difficult to realize with typical neutrino detectors that cannot cleanly separate electrons from energetic photons; however, it provides a contrast with more inclusive cross sections, and also is similar to well-studied observables in elastic electron-proton scattering. Section~\ref{sec:electron_jet} considers the default observable for electronlike jet events. Section~\ref{sec:muon_jet} discusses observables for muon flavor and studies scattering with energetic photons, where muon (anti)neutrino events can be misidentified as electron flavor. Section~\ref{sec:hard_outside_jet} considers the emission of noncollinear hard photons. Section~\ref{sec:inclusive_xsection} includes such noncollinear hard-photon emission in the evaluation of inclusive observables. Section~\ref{sec:ratio} considers the ratio of electron versus muon cross sections and discusses the extent to which hadronic uncertainties cancel in the ratio. Leading nuclear effects are considered in Sec.~\ref{sec:nuclear_effects}. Finally, we compare the size of radiative corrections with existing experimental $\nu_\mu$ and $\overline{\nu}_\mu$ data in Sec.~\ref{sec:comparison_to_data}.

\subsection{Cross section with charged lepton and soft photons \label{sec:electron_soft_photons_xsection}}

In all observables that involve only soft or collinear radiation, the cross section is given by Eq.~(\ref{eq:factorization_formula}). The soft function $S$ is given by Eq.~(\ref{eq:soft_function}). The hard function at the hard renormalization scale is computed as in Sec.~\ref{sec:hadronic}, and the running between hard and soft renormalization scales is determined as in Sec.~\ref{sec:rge_evolution}. The collinear function depends on the specification of jet observable. For measurements that separate charged leptons from the accompanying collinear radiation, only soft real radiation is relevant and the collinear function is given by $J$ in Eq.~(\ref{eq:jet_function_at_fixed_x}). Precisely this observable has been used in Sec.~\ref{sec:hadronic} to illustrate the impact of hadronic uncertainties in Figs.~\ref{fig:hm_electron_error2} and \ref{fig:hm_muon_error2}, and to illustrate the convergence of resummed perturbation theory in Fig.~\ref{fig:pert_electron_error}. The ratio of this cross section to the tree-level result in the nonrelativistic limit was shown as a function of (anti)neutrino energy by the solid blue curve in Fig.~\ref{fig:static_limit}, where a default soft-photon energy cutoff $\Delta E = 20\,{\rm MeV}$ was used. The complete relativistic case, for a range of $\Delta E$ values, is shown for neutrino-neutron and antineutrino-proton scattering with electron flavor in Fig.~\ref{fig:soft_e} and with muon flavor in Fig.~\ref{fig:soft_mu}. Results are displayed for typical accelerator (anti)neutrino energies $E_\nu = 600~\mathrm{MeV}$ and $E_\nu = 2~\mathrm{GeV}$. In these figures, $Q^2$ can be identified as the momentum transfer between nucleons: $Q^2 = -(k-k')^2$.

\subsection{Electron-jet observable \label{sec:electron_jet}}

\begin{figure}[tb!]
 \centering
 \includegraphics[height=0.5\textwidth]{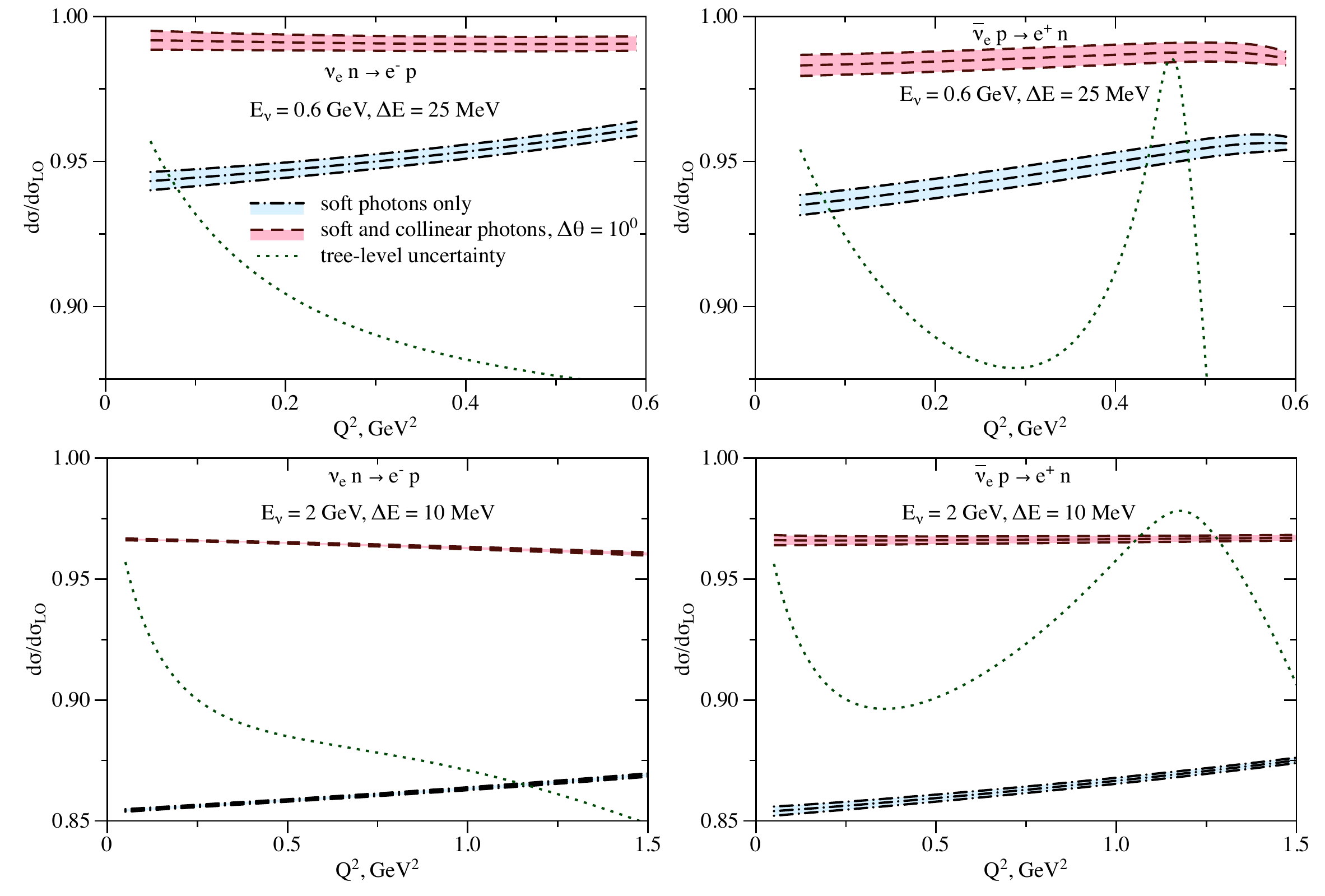}
 \caption{Ratio of the differential cross section to the tree-level result for $\nu_e$ ($\bar{\nu}_e$) including only soft or both soft and collinear photon emission. Left and right plots are for neutrino and antineutrino scattering, respectively. Upper and lower plots are for $E_\nu = 600~\mathrm{MeV}$ and $E_\nu = 2~\mathrm{GeV}$, with soft-photon energy cutoff $\Delta E = 25~\mathrm{MeV}$ and $\Delta E = 10~\mathrm{MeV}$, respectively. The lower blue dash-dotted band corresponds to curves in Fig.~\ref{fig:soft_e}. The upper pink dashed band includes also real photons within a cone of size $\Delta \theta=10^\circ$. The difference between the green dashed line and unity represents the uncertainty of the tree-level cross section itself; this uncertainty largely cancels in the displayed cross-section ratios. \label{fig:xsection_NLO_to_LO_electron}}
\end{figure}

In most practical neutrino detectors at energies of accelerator oscillation experiments, final-state electrons with and without nearly collinear photons cannot be distinguished. Neutrino detectors are typically homogeneous and are the target material in which the (anti)neutrinos themselves interact. In traveling through the detector, the final-state electrons lose energy by bremsstrahlung, which also results in nearly collinear photons. These photons, in turn, interact by producing $e^+e^-$ pairs, which then deposit energy near the primary electron.

To account for the energy deposited by real collinear photons, we must modify the observable from Sec.~\ref{sec:electron_soft_photons_xsection} that allowed only soft real photon emission. For a fixed direction of charged lepton momentum, we consider all events where either the photon is soft, i.e., has energy below $\Delta E$, or the photon is energetic and collinear, i.e., has energy above $\Delta E$ and is radiated within the angle $\Delta \theta$ to the electron direction. Such events are not distinguished from charged-current elastic events without radiation. We remark that this definition is similar to, but differs slightly from, a common scheme in the QCD literature due to Sterman and Weinberg (SW)~\cite{Sterman:1977wj}, where the angle is specified relative to the direction of total energy flow. Such fine-grained details are beyond the precision of the current generation of neutrino experiments, but could be resolved using a more elaborate detector simulation. Integrating over the phase space of energetic photons within the angle $\Delta \theta$ to the electron direction, we obtain an additional contribution to the collinear function in the factorization formula, given at one-loop order by $j(x)$ in Eq.~(\ref{eq:SCETjet}). As discussed in Sec.~\ref{sec:collinear}, in order to resum leading large logarithms from multiple collinear photon emission, we exponentiate the collinear photon correction~\cite{Mukhi:1982bk,Curci:1978kj,Smilga:1979uj}.

Figure~\ref{fig:xsection_NLO_to_LO_electron} shows the correction to unpolarized cross sections as a function of hadronic momentum transfer, $Q^2=-(k-k^\prime)^2$, over the kinematic region of the elastic $2\to2$ process. For reference, the plot includes the curve from Sec.~\ref{sec:electron_soft_photons_xsection} with the observable describing only soft-photon radiation. Also shown in the figure as a deviation from unity is the uncertainty of the tree-level process, dominated by the poorly constrained axial-vector form factor. This uncertainty dominates the error budget of the absolute cross section but cancels between the numerator and denominator in various cross-section ratios. The impact of this tree-level uncertainty and of radiative corrections, on critical $\nu_e$ versus $\nu_\mu$ cross-section ratios will be examined below.

\subsection{Electronlike muon-jet events \label{sec:muon_jet}}

\begin{figure}[t]
 \centering
 \includegraphics[height=0.5\textwidth]{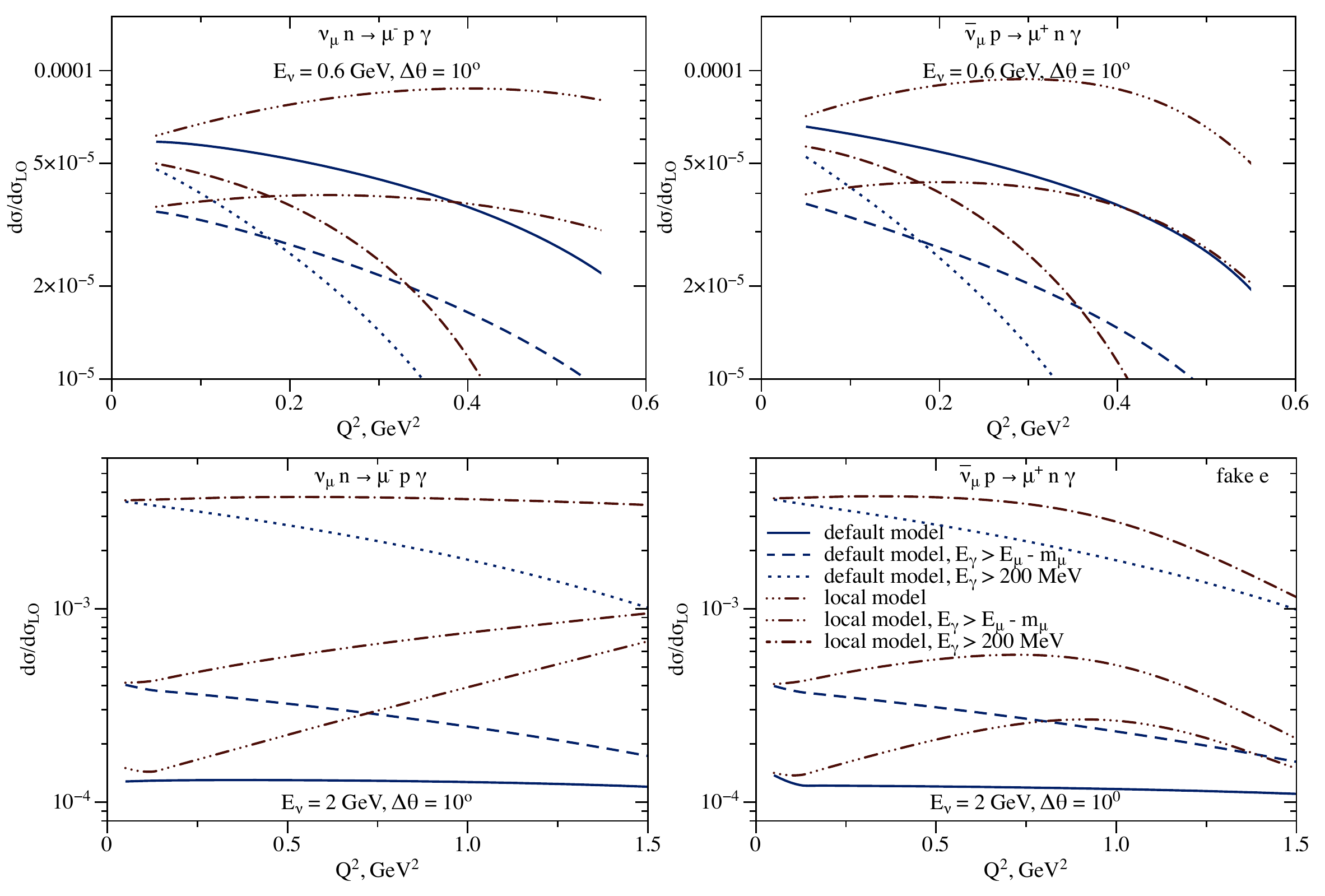}
 \caption{Ratio of unpolarized radiative charged-current-elastic-like $\nu_\mu$ and $\bar{\nu}_\mu$ differential cross sections to the tree-level result with collinear photon energy above the threshold for final muon to be misidentified as electron, as described in the text. Reference scenarios with $E_\gamma> E_\mu - m_\mu$ and $E_\gamma > 200\,{\rm MeV}$ are shown for comparison. Plots are for (anti)neutrino beam energies $E_\nu = 600~\mathrm{MeV}$ (upper plots) and $E_\nu = 2~\mathrm{GeV}$ (lower plots). The calculation based on our default electroweak vertex is labeled as default model. The calculation when the electroweak vertex is taken as in diagrams with local interactions is labeled as local model. \label{fig:real_radiation_misidentified}}
\end{figure}

Compared to the case of electrons, bremsstrahlung from muons is a rarer process because of the larger muon mass. It is precisely this difference between muons, which slowly lose energy by ionization and collision processes, and electrons, which lose energy by bremsstrahlung, that is used as the primary way to distinguish electron and muon (anti)neutrino interactions from each other in neutrino oscillation experiments. Therefore, events with muons and energetic collinear photons can be misidentified as electron events. For this confusion to happen, the photon must carry a significant fraction of the primary muon energy because low-energy collinear photons are consistent with collisional processes where muons create $\gamma$ rays in the detector. To accurately predict this confusion, a detailed detector simulation is required. However, we may estimate the effect in a simplistic model in which a collinear muon and photon will only be confused with an electron if the range of the muon in the detector is less than the range of an electron with energy equal to the sum of the photon energy and muon kinetic energy. Because the electron shower range at high energies grows only logarithmically with electron energy, but muon range grows nearly linearly, the fraction of the energy that must be carried by the collinear photon will grow with muon energy. By comparing the average length of electron showers~\cite{Grindhammer:1989zg,Zyla:2020zbs} to muon range~\cite{pdg.lbl.gov/AtomicNuclearProperties}, we have developed an empirical parametrization of the collinear photon energies that would not cause this confusion: $E_\gamma \lesssim \Tilde{E} \left\{ 0.95 - 0.75/[ (\Tilde{E}/1 ~{\rm GeV}) + 0.85 ] \right\}$, where $\Tilde{E} = E_\nu-{Q^2}/{2M}-m_\mu$. This parametrization is independent of the choice of detector materials considered in this work -- scintillator, water, and argon -- and is valid for $E_c\ll \Tilde{E}\lesssim 5$~GeV, where $E_c$ is the electron critical energy in the target medium, i.e., the energy above which electron energy loss is dominated by bremsstrahlung~\cite{Rossi:1952kt}. $E_c$ is approximately $90$~MeV in polystyrene scintillator, $75$~MeV in water, and $32$~MeV in liquid argon.

A quantity of interest is the cross section for muon neutrino and antineutrino events with collinear photon energies large enough so that the final-state muon could be mistaken for an electron and therefore lead to flavor misidentification. This is a potential concern, especially since electron (anti)neutrino events are much less common than muon (anti)neutrino events in the relevant (anti)neutrino beams. The ratio of this cross section to the tree-level charged-current elastic cross section is shown as the solid curve in Fig.~\ref{fig:real_radiation_misidentified}, using our default hadronic model (``default model" in the figure). As the figure illustrates, the probability for such misidentification is small, of order one part in $10^4$, and this is a subleading effect in the current generation of experiments. This effect is potentially much larger for experiments whose particle identification algorithms are sensitive to lower-energy collinear photons than were presumed to cause misidentification in this study. To illustrate the dependence on the assumed photon energy threshold for misidentification as electron (anti)neutrinos, the figure also shows the cross section for the cases where $E_\gamma>E_\mu-m_\mu$ and $E_\gamma>200~\mathrm{MeV}$. As remarked in Sec.~\ref{sec:hadronic}, a naive form-factor insertion model for the hard matching condition violates electromagnetic current conservation, leading, e.g., to spurious $\sim \ln(m_\ell)$ collinear singularities. We illustrate the importance of ensuring this conservation condition by comparing our default hadronic model to the naive form-factor insertion model (``local model" in the figure).

\subsection{Energetic photons isolated from charged lepton jets \label{sec:hard_outside_jet}}

\begin{figure}[htb]
 \centering
 \includegraphics[height=0.5\textwidth]{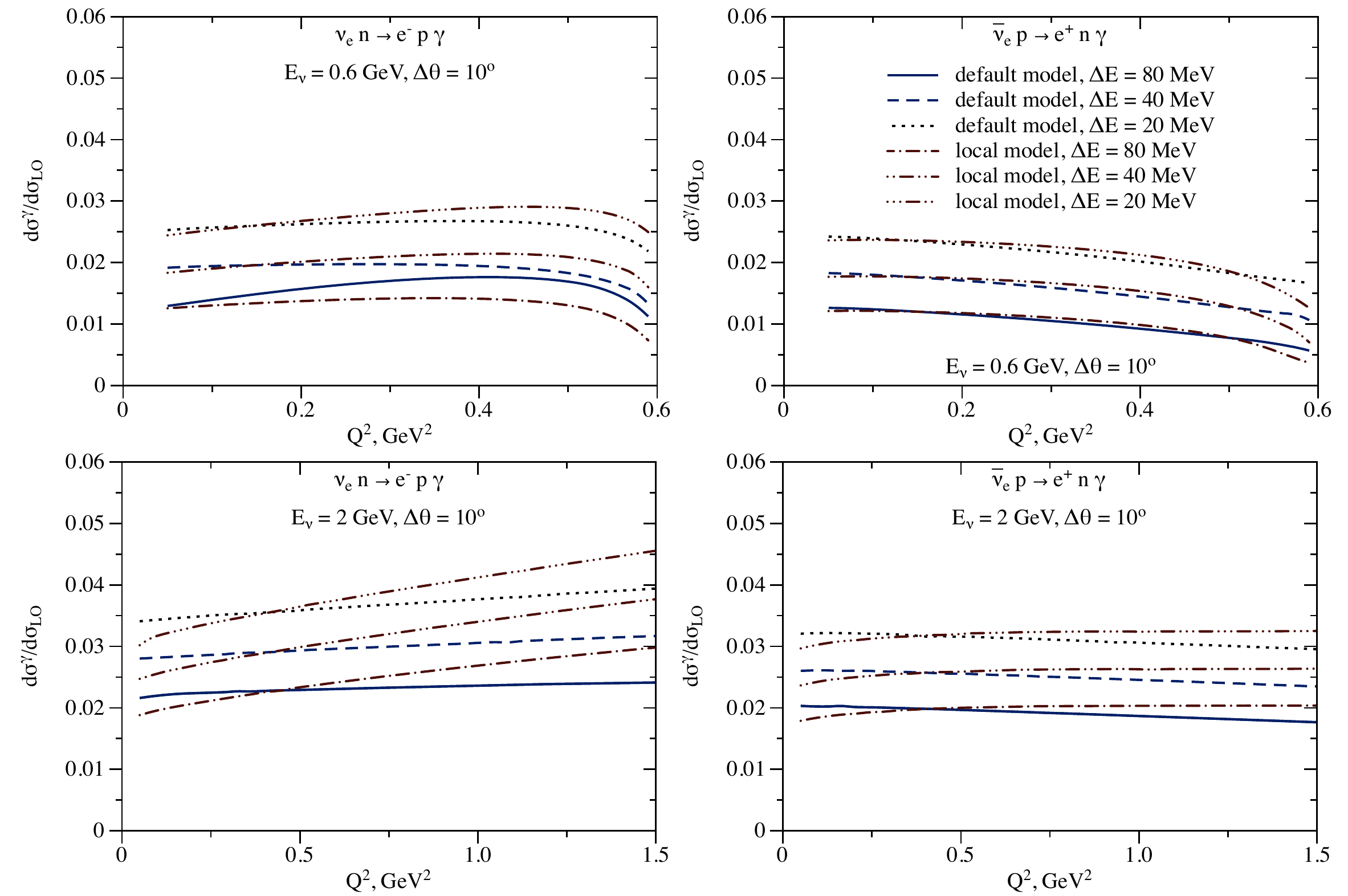}
 \caption{Ratio of unpolarized radiative charged-current-elastic-like $\nu_e$ and $\bar{\nu}_e$ differential cross sections to the tree-level result with hard photons outside the jet cone. Curves show $E_\gamma > 20~\mathrm{MeV}$, $40$, and $80$~MeV, for the angle between the electron and photon $\theta > 10^\circ$. Plots are for (anti)neutrino beam energies $E_\nu = 600~\mathrm{MeV}$ (upper plots) and $E_\nu = 2~\mathrm{GeV}$ (lower plots). The calculation based on our default electroweak vertex is labeled as default model. The calculation when the electroweak vertex is taken as in diagrams with local interactions is labeled as local model. \label{fig:real_radiation_hard_noncollinear_photons_electron}}
\end{figure}
\begin{figure}[tb!]
 \centering
 \includegraphics[height=0.5\textwidth]{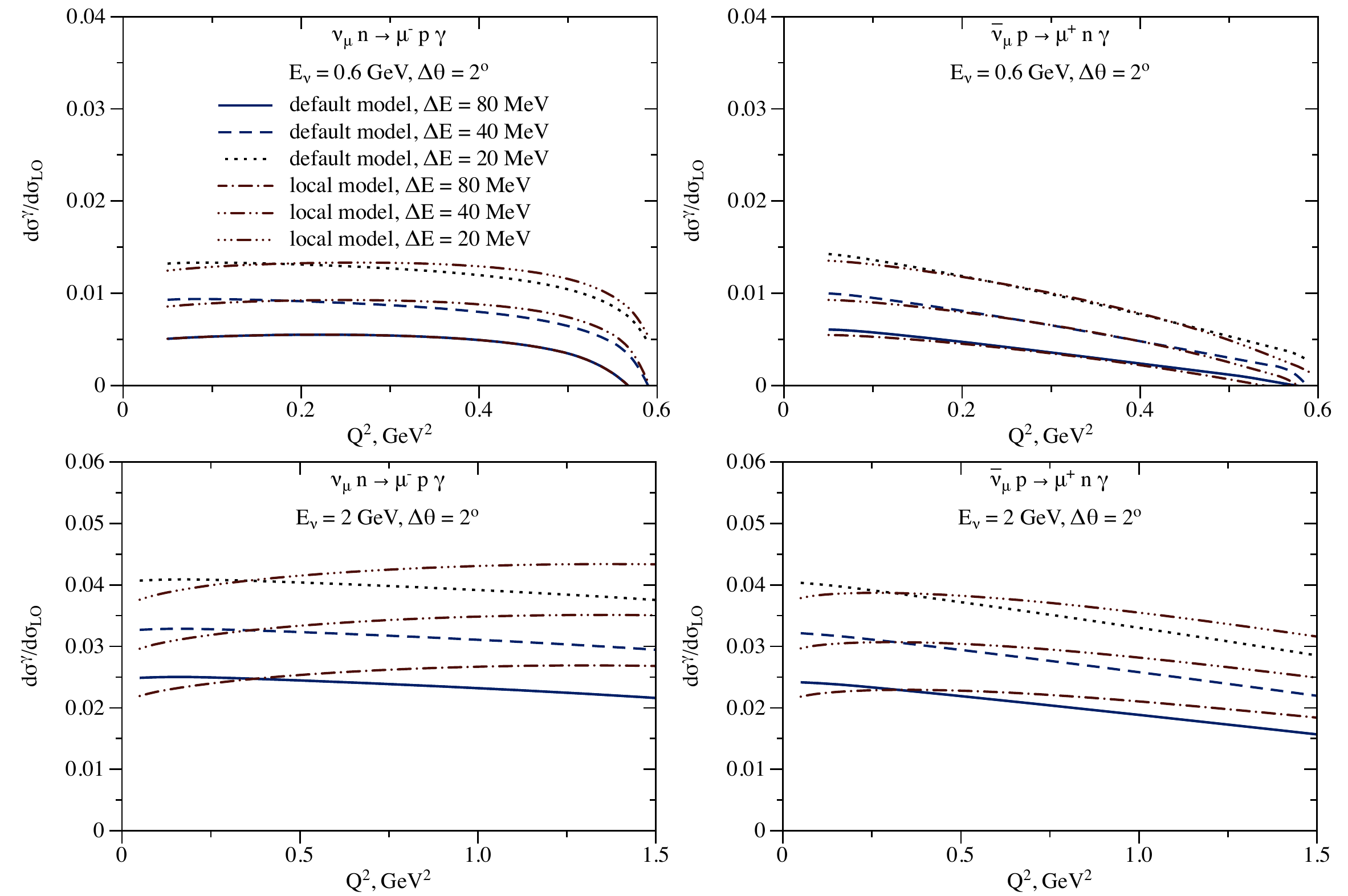}
 \caption{The same as Fig.~\ref{fig:real_radiation_hard_noncollinear_photons_electron} but for the muon flavor charged-current-elastic-like cross section, with $\theta > 2^\circ$. \label{fig:real_radiation_hard_noncollinear_photons_muon}}
\end{figure}

Hard photons outside of electron jets can be misidentified as coming from $\pi^0\to\gamma\gamma$ and may be removed from many oscillation analyses in an attempt to reduce such backgrounds. In the case of low-energy oscillation experiments, attempts to select quasielastic muon neutrino or antineutrino interactions may also remove events with a visible photon that is not collinear with the muon. We estimate the relative contributions of such events in the following.

To evaluate one-photon events with energy above $\Delta E$ outside the cone of angle $\Delta \theta$, we use the hadronic model of Secs.~\ref{sec:model} and~\ref{sec:inclusive_model}, labeled default model in our figures. Recall that this model differs from a naive ``form-factor insertion" ansatz by the choice of form-factor arguments: $Q^2$ at the weak vertex is determined by the external nucleon leg kinematics. In the phase space of photons {\it outside} the collinear region, the naive model does not give rise to spurious $\sim \ln{m_\ell}$ singularities, and we may use the difference between this local model and our default hadronic model as a crude measure of model dependence. Note that the cross section including these one-photon events is exclusive of the jet cross section of Sec.~\ref{sec:electron_jet} with only one photon, for the same $\Delta E$ and $\Delta \theta$.

The results in Figs.~\ref{fig:real_radiation_hard_noncollinear_photons_electron} and \ref{fig:real_radiation_hard_noncollinear_photons_muon} for photon energy thresholds $\Delta E$ of $20$~MeV, $40$, and $80$~MeV show that the cross section for hard photons outside of electron jets is of order $1\%$ to several~$\%$ of the tree-level cross section and potentially significant for current and future neutrino oscillation experiments.

\subsection{Inclusive observables \label{sec:inclusive_xsection}}

\begin{figure}[thb!]
 \centering
 \includegraphics[height=0.5\textwidth]{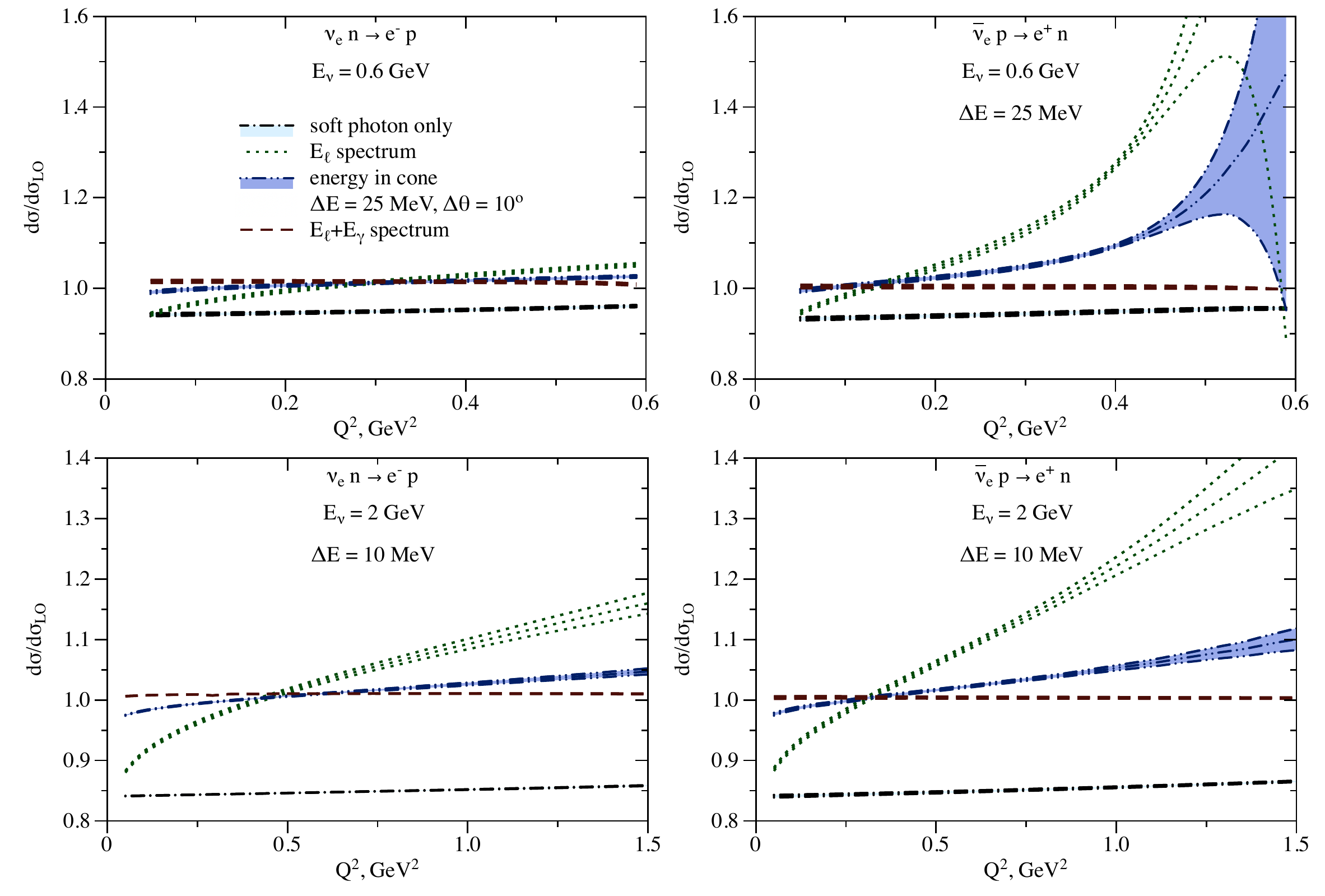}
 \caption{Ratio of the inclusive differential cross section including noncollinear emission of one hard photon to the tree-level result is presented as a function of $Q^2$. Electron neutrino-neutron scattering is shown on left plots and antineutrino-proton scattering on right plots. For comparison, the result including only soft photon radiation with energy below $\Delta E$ is shown by the black dash-dotted lines and light blue band. The $E_\ell+E_\gamma$ energy spectrum is shown by the red dashed lines; the spectrum of events reconstructed from the energy within the $\Delta \theta = 10^\circ$ cone is shown by the dark-blue dashed double-dotted lines and blue band; and the energy spectrum reconstructed from the electron energy is shown by the green dotted lines. (Anti)neutrino beam energies are taken as $E_\nu = 600~\mathrm{MeV}$ (upper plots) and $E_\nu = 2~\mathrm{GeV}$ (lower plots), and the soft-photon energy cutoff is $\Delta E = 25~\mathrm{MeV}$ and $\Delta E = 10~\mathrm{MeV}$, respectively. $Q^2$ is defined according to the spectrum in figures. The displayed $Q^2$ range for $600~\mathrm{MeV}$ (anti)neutrino beam corresponds to the maximum for tree-level elastic kinematics. For $2~\mathrm{GeV}$ (anti)neutrinos, the displayed $Q^2$ range is truncated to focus on the phenomenologically relevant regime. Central value cross sections for the entire $Q^2$ region are displayed in Fig.~\ref{fig:Q2_issues}. \label{fig:xsection_inclusive_plots_electron}}
\end{figure}
\begin{figure}[htb!]
 \centering
 \includegraphics[height=0.5\textwidth]{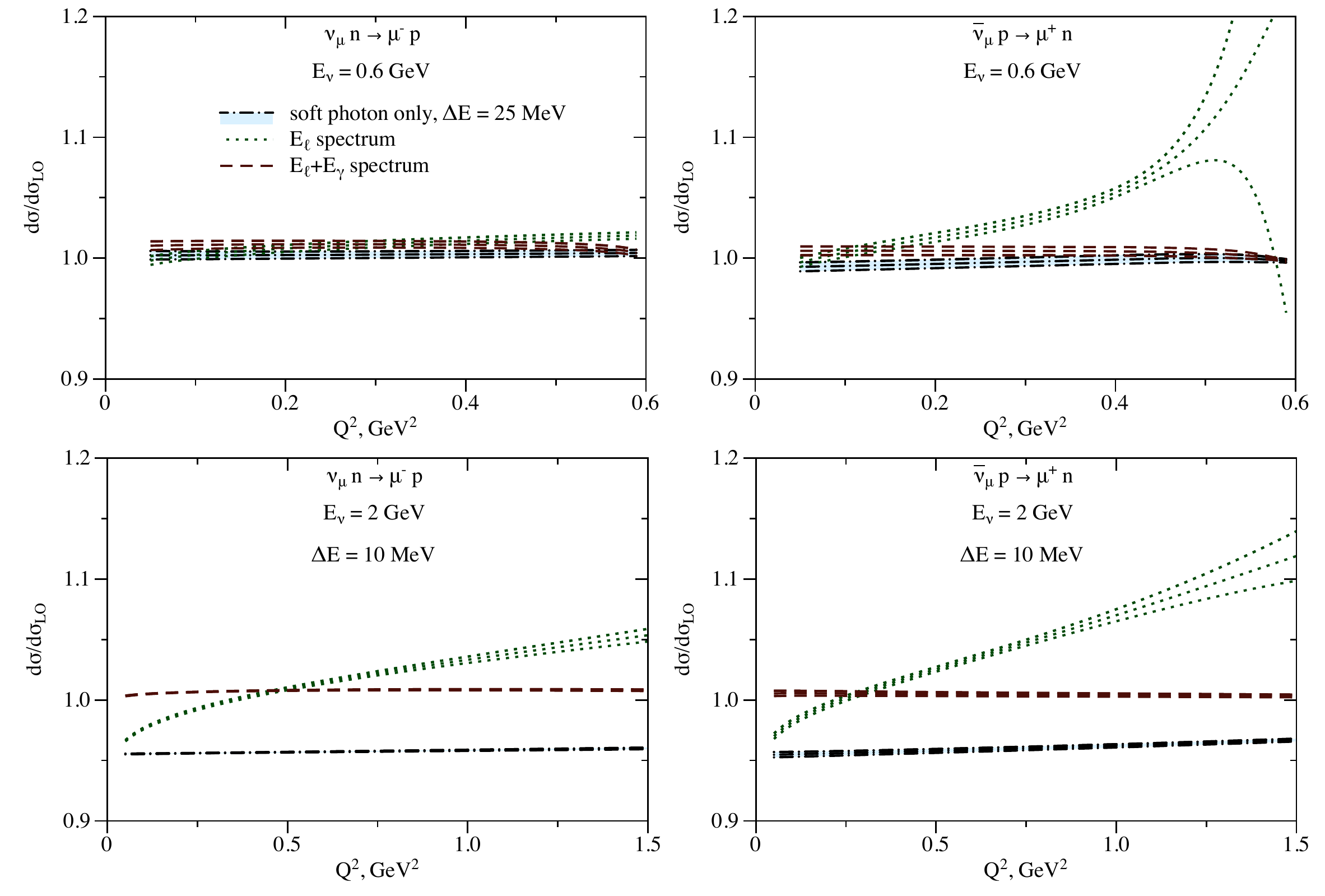}
 \caption{The same as Fig.~\ref{fig:xsection_inclusive_plots_electron} but for the muon flavor. The cone observable is not presented since the radiation inside the cone is negligible for the muon flavor. \label{fig:xsection_inclusive_plots_muon}}
\end{figure}
\begin{figure}[htb!]
 \centering
 \includegraphics[height=0.5\textwidth]{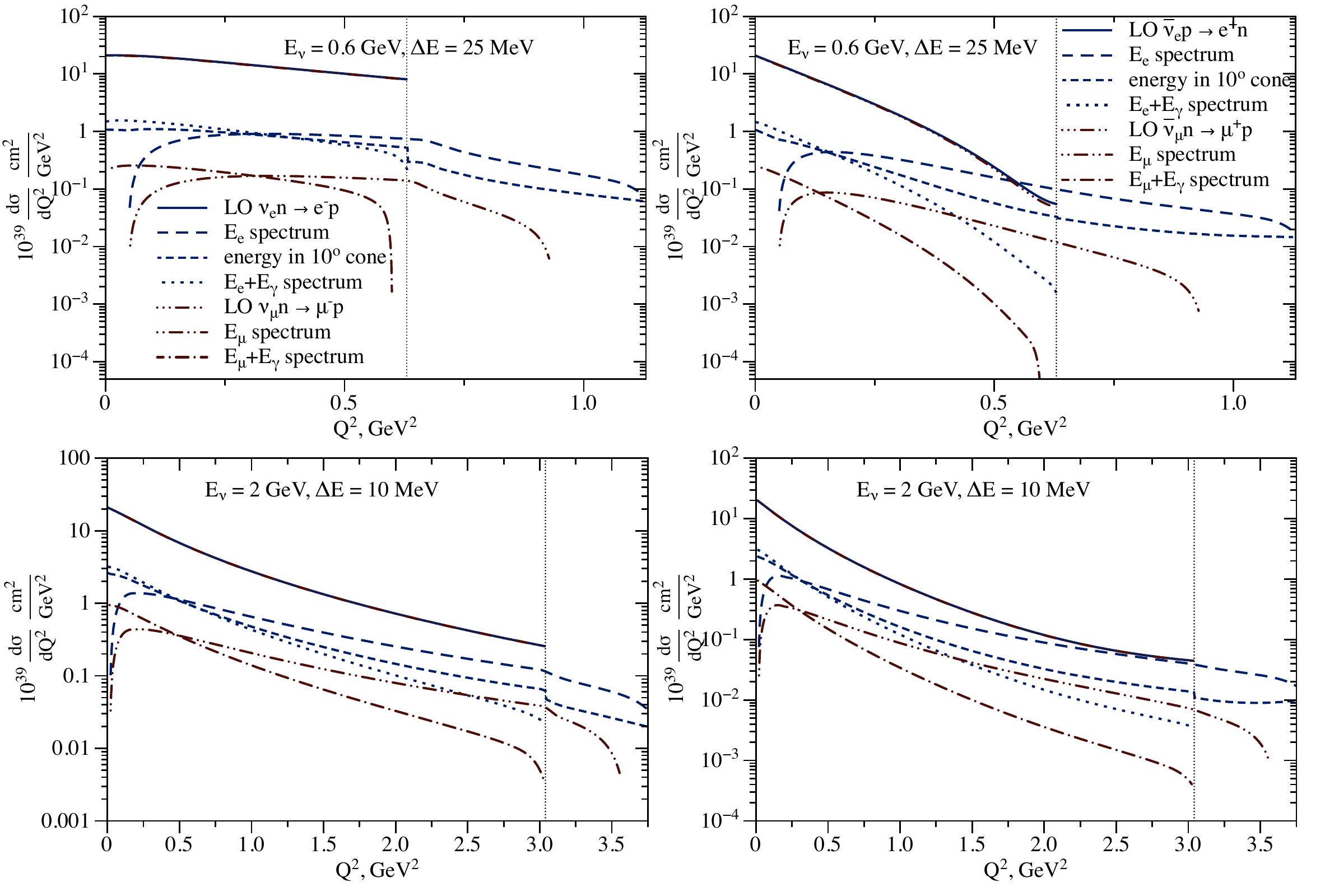}
 \caption{Tree-level differential cross section for the (anti)neutrino beam energies $E_\nu = 600~\mathrm{MeV}$ and $E_\nu = 2~\mathrm{GeV}$ is compared to the spectra of radiated events, with photons above the energy $\Delta E = 25~\mathrm{MeV}$ and $\Delta E = 10~\mathrm{MeV}$, respectively, when $Q^2$ is defined from the lepton energy $E_\ell$ or from the sum of lepton and photon energies $E_\ell + E_\gamma$. Results are shown for electron and muon flavors, for neutrino-neutron scattering on the left plot and for antineutrino-proton scattering on the right plot. The flavor difference of tree-level cross sections is unresolved on this figure over the wide range of allowed kinematics; see upper curves on all plots. We also present the spectrum of events reconstructed from the energy within the cone of angle $\Delta \theta = 10^\circ$ for the electron flavor. $Q^2$ and its range are defined according to the spectrum in figures. Integrals over all $Q^2$ values for $E_\ell$, energy in 10$^{\circ}$ cone, and $E_\ell + E_\gamma$ spectra are the same. \label{fig:Q2_issues}}
\end{figure}

Sections~\ref{sec:electron_soft_photons_xsection}, \ref{sec:electron_jet}, and \ref{sec:muon_jet} restricted attention to event classes that excluded noncollinear hard photons. In Sec.~\ref{sec:hard_outside_jet}, we considered the contribution of noncollinear hard photons to the cross section for charged-current (anti)neutrino-nucleon scattering, finding that such events can contribute significantly in current and future neutrino experiments. These photons may not be explicitly reconstructed but, if not vetoed, would contribute to the observed event rate and would affect (anti)neutrino energy reconstruction based solely on the visible energy in the leptonic shower.

For measurements that include (i.e., do not veto) such hard photons, the relevant cross section involves integration over the full phase space of radiated photons. For such observables, we perform a calculation within our default hadronic model described in Secs.~\ref{sec:model}~and~\ref{sec:inclusive_model}. In contrast to Secs.~\ref{sec:electron_soft_photons_xsection}~and~\ref{sec:electron_jet}, we compute at fixed order in QED perturbation theory. We assign the same tree-level form-factor uncertainty and the same uncertainty assignment for virtual contributions (cf. Sec.~\ref{sec:hard_outside_jet} regarding uncertainty of the hard real photon contribution). Figures~\ref{fig:xsection_inclusive_plots_electron} and~\ref{fig:xsection_inclusive_plots_muon} provide results for distributions when the independent variable $Q^2$ is defined using different measures of leptonic energy. For example, in the case of electrons, $Q^2$ may be defined using the electron energy itself, $Q^2 = 2 M (E_\nu - E_\ell)$, or using the sum of electron and photon energies, $Q^2 = 2 M (E_\nu - E_\ell - E_\gamma)$, or perhaps as in a realistic experiment, using the energy reconstructed within a $\Delta \theta = 10^\circ$ cone whose axis is around the electron direction. Note that for the $E_\ell$ spectrum, parameter $\Delta E$ does not enter: the observable sums over {\it all} ``unobserved" photons, not just those below a threshold $\Delta E$. Similarly for the $E_\ell+E_\gamma$ spectrum, $\Delta E$ does not enter, since the spectrum is equivalent to the spectrum in hadronic momentum transfer $Q^2$, which sums over all possible photon energies.\footnote{Parameter $\Delta E$ enters as a regulator parameter during the calculation, separating an IR-divergent soft region from the remainder, but is taken negligibly small for the final results.} For the ``energy in cone" observable, we use $\Delta E = 20~\mathrm{MeV}$ to specify the selection, assuming that photons below this value are not reconstructed.

Note that the allowed region of $Q^2$ is different for different spectra and may extend beyond the tree-level kinematic limit. In particular, the $E_\ell$ spectrum extends up to $Q^2 = 2 M \left( E_\nu - m_\ell \right)$, in contrast to the $E_\ell +E_\gamma$ spectrum, which extends only to the tree-level kinematic limit. We illustrate these regions and provide corresponding cross sections for neutrino-neutron and antineutrino-proton scattering in Fig.~\ref{fig:Q2_issues}.

\subsection{Cross-section ratios \label{sec:ratio}}

\begin{figure}[htb]
 \centering
 \includegraphics[height=0.5\textwidth]{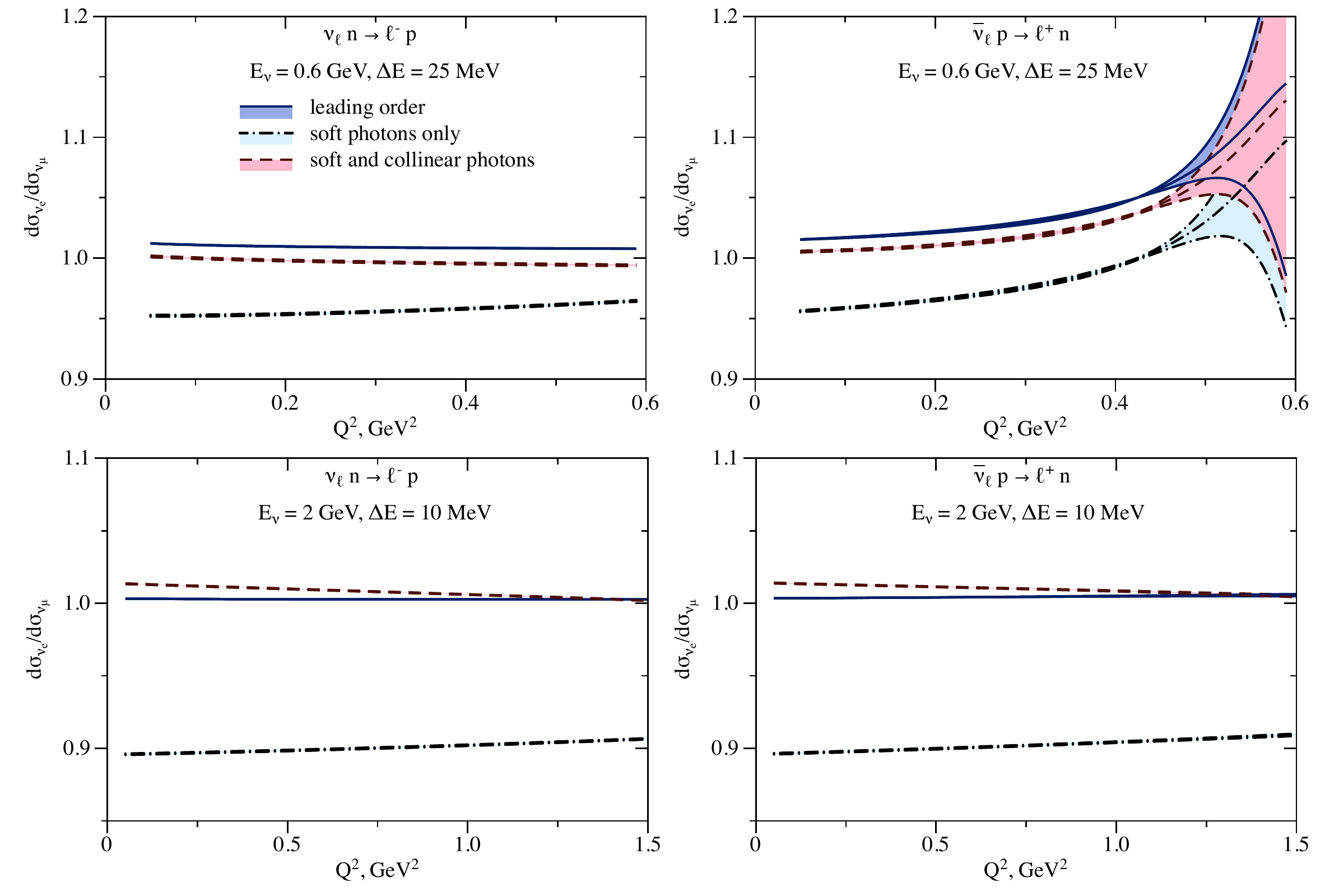}
 \caption{Electron over muon flavor ratios of unpolarized differential cross section in neutrino-neutron (left plots) and antineutrino-proton (right plots). Blue solid lines and blue band, tree level; black dash-dotted lines and light blue band, including radiation of soft photons with energy below $\Delta E$; red dashed lines and pink band, including radiation of soft photons with energy below $\Delta E$ and including radiation of collinear photons for the electron flavor as described in Sec.~\ref{sec:electron_jet} but not for the muon flavor. For (anti)neutrino beam energies $E_\nu = 600~\mathrm{MeV}$ (upper plots) and $E_\nu = 2~\mathrm{GeV}$ (lower plots), the soft-photon energy cutoff is $\Delta E = 25~\mathrm{MeV}$ and $\Delta E = 10~\mathrm{MeV}$, respectively.\label{fig:xsection_ratio_muon_over_electron}}
\end{figure}
\begin{figure}[htb]
 \centering
 \includegraphics[height=0.5\textwidth]{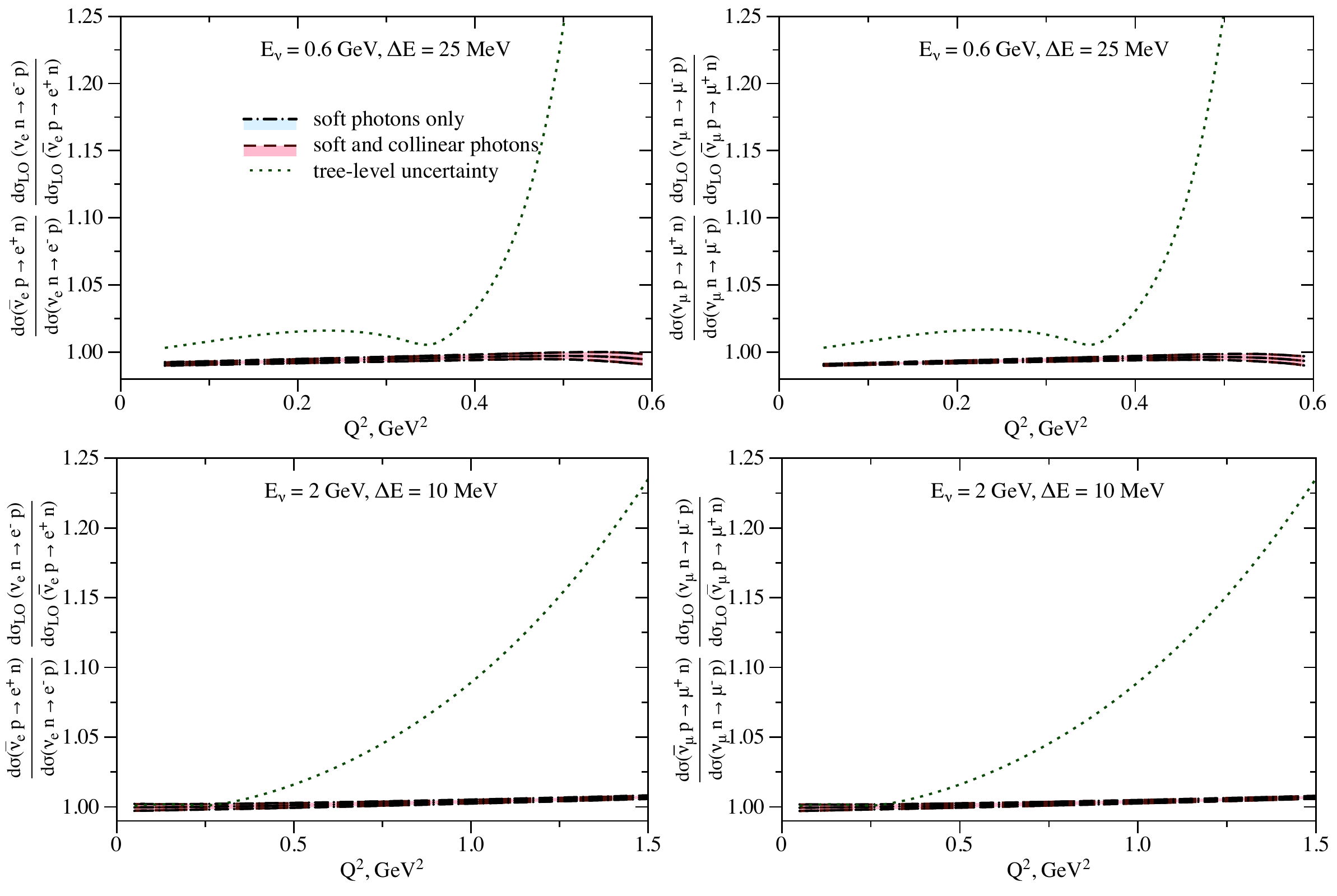}
 \caption{Antineutrino-proton over neutrino-neutron differential cross-section double ratios to the tree-level results with electron (left plots) and muon (right plots) flavors including the radiation of soft photons with energy below $\Delta E$, shown by the black dash-dotted lines and light blue band, and accounting also for the radiation of collinear photons in case of electron flavor as described in Sec.~\ref{sec:electron_jet}, shown by the red dashed lines and pink band. For (anti)neutrino beam energies $E_\nu = 600~\mathrm{MeV}$ (upper plots) and $E_\nu = 2~\mathrm{GeV}$ (lower plots), the soft-photon energy cutoff is $\Delta E = 25~\mathrm{MeV}$ and $\Delta E = 10~\mathrm{MeV}$, respectively. Differences between the double ratios with and without collinear photons arise from subleading effects and are indistinguishable in the figure (see text). The relative uncertainty of the tree-level antineutrino over neutrino cross-section ratio is represented by the green dotted line as a deviation from unity.\label{fig:xsection_ratio_anti}}
\end{figure}
\begin{figure}[htb]
 \centering
 \includegraphics[height=0.5\textwidth]{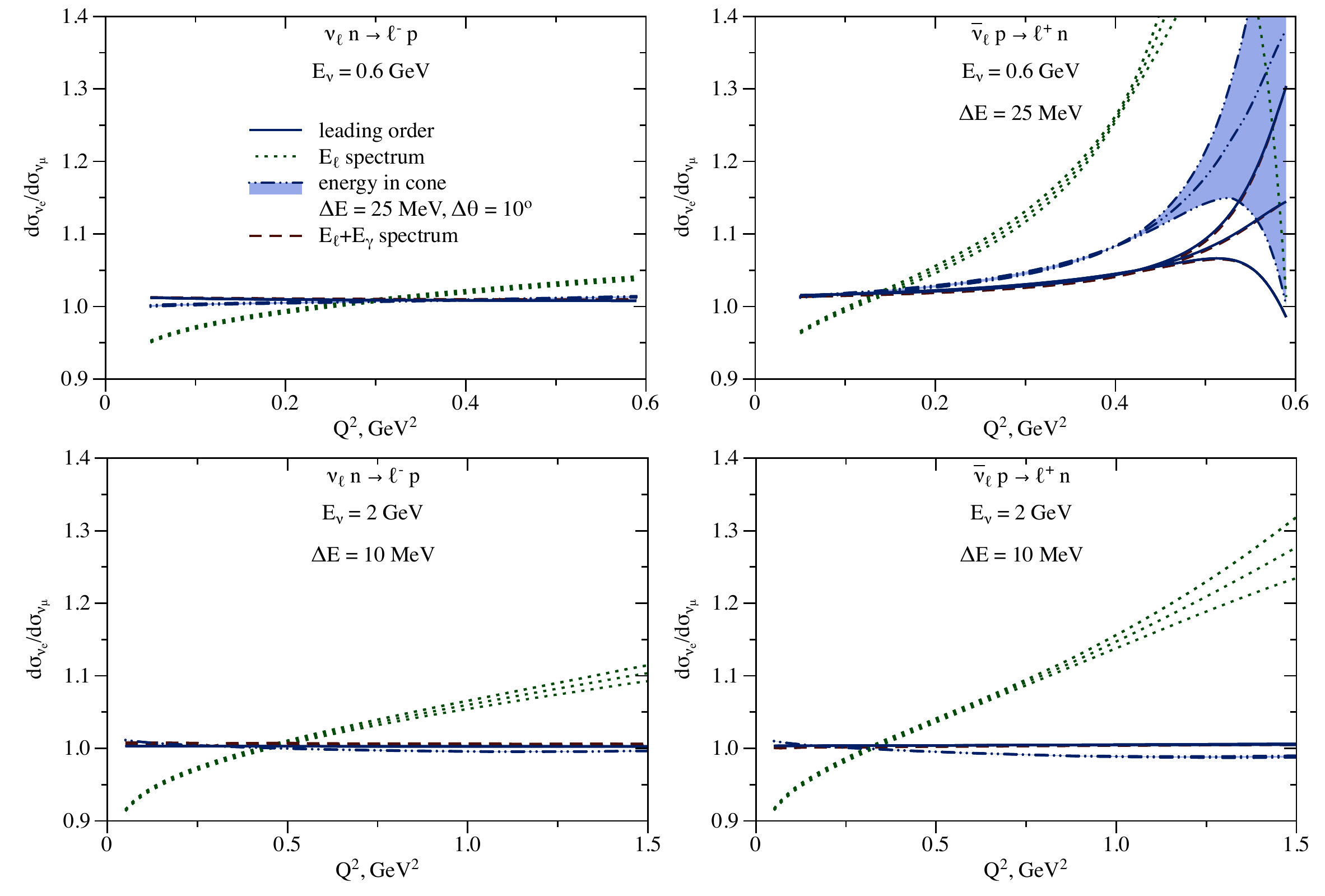}
 \caption{(Inclusive) Electron over muon flavor ratios of unpolarized differential cross sections in neutrino-neutron (left plots) and antineutrino-proton (right plots) scattering at tree level, shown by the blue solid lines; the ratio of $E_\ell$ spectra is shown by the green dotted lines; the ratio of spectrum of events reconstructed from the energy within the $10^\circ$ cone for the electron flavor to the $E_\ell$ spectrum for the muon flavor is shown by the dark-blue dashed double-dotted lines and blue band; and the ratio of $E_\ell+E_\gamma$ energy spectra is shown by the red dashed lines. (Anti)neutrino beam energies are taken as $E_\nu = 600~\mathrm{MeV}$ (upper plots) and $E_\nu = 2~\mathrm{GeV}$ (lower plots); the soft-photon energy cutoff for jet events with electron (anti)neutrinos is $\Delta E = 25~\mathrm{MeV}$ and $\Delta E = 10~\mathrm{MeV}$, respectively. $Q^2$ is defined according to the spectrum in figures. \label{fig:xsection_ratio_muon_over_electron_inclusive}}
\end{figure}
\begin{figure}[htb]
 \centering
 \includegraphics[height=0.5\textwidth]{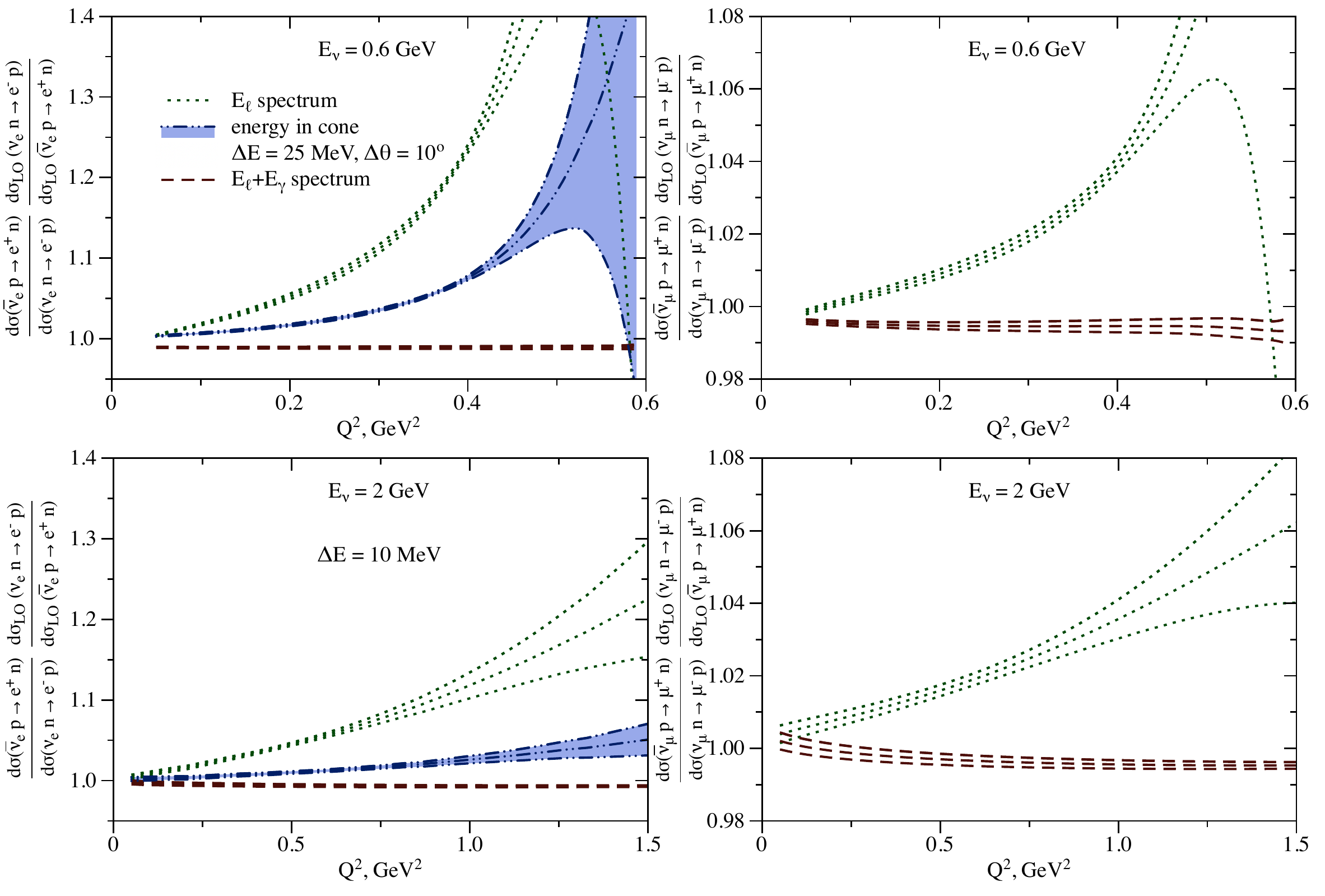}
 \caption{(Inclusive) The same as Fig.~\ref{fig:xsection_ratio_anti} but for inclusive observables as specified in the caption to Fig.~\ref{fig:xsection_ratio_muon_over_electron_inclusive}. \label{fig:xsection_ratio_anti_inclusive}}
\end{figure}

We have noted that, after the inclusion of radiative corrections, the uncertainty of absolute cross sections for either electron or muon flavor, and for either neutrino or antineutrino, is dominated by hadronic uncertainties of the tree-level process. Cross-section ratios, for electron versus muon and for neutrino versus antineutrino, are critical to oscillation analyses. Within our formalism, we can systematically investigate the impact of radiative corrections on these ratios, determining the absolute size of corrections, and the extent to which hadronic uncertainties cancel.

Figure~\ref{fig:xsection_ratio_muon_over_electron} displays the cross-section ratio of electron to muon flavor, comparing the tree-level case to the default observable for exclusive electron and muon charged-current events. For comparison, we show also the ratio of the soft-photon-only cross sections. For observables taken at the same value of hadronic $Q^2$, the hard function $H$ cancels identically in the ratio in our leading-power analysis. The remaining soft and jet functions depend on the charged lepton mass and the kinematic parameter $\Delta \theta$.

Figure~\ref{fig:xsection_ratio_anti} displays the double ratio of $\sigma/\sigma^{\rm LO}$ for neutrino versus antineutrino events. The deviation from unity is small compared to the magnitude of the total radiative correction to either neutrino or antineutrino separately, reflecting the structure of the factorization formula, as follows. First, for a given charged lepton mass, the jet function is identical when the same event selection is used. Second, the soft function differs only in the kinematics of the proton, and identical leading logarithmically enhanced terms appear for neutrino and antineutrino cases at the same charged lepton kinematics (the soft functions become identical in the static limit; cf. Sec.~\ref{sec:static}). Finally, deviations of the ratio from unity from the hard function $H$ also do not contain large logarithm enhancements. Note that unlike the case of $e$ versus $\mu$ cross sections for either neutrino or antineutrino, hadronic uncertainties in the hard function do not cancel. Since only the ratio of antineutrino over neutrino cross sections but not the double ratio can be accessed experimentally, we present the relative uncertainty of the tree-level ratio as a deviation from unity in Fig.~\ref{fig:xsection_ratio_anti}. This uncertainty from the hard function presently limits our knowledge of antineutrino over neutrino cross-section ratios and can be reduced only with improved knowledge of elementary (anti)neutrino-nucleon scattering amplitudes.

Figures~\ref{fig:xsection_ratio_muon_over_electron_inclusive} and \ref{fig:xsection_ratio_anti_inclusive} show the analog of Figs.~\ref{fig:xsection_ratio_muon_over_electron} and \ref{fig:xsection_ratio_anti}, but calculated for the inclusive event sample, as discussed in Sec.~\ref{sec:inclusive_xsection}. For each spectrum here, $Q^2$ is reconstructed from the corresponding energy as described in Sec.~\ref{sec:inclusive_xsection}. For this case, there is no exact cancellation between hard functions in electron versus muon cross sections, since the events for a given value of (reconstructed) $Q^2$ do not necessarily correspond to the same nucleon kinematics. Moreover, the double ratio deviates significantly from unity for the $E_\ell$ spectrum, for which there is no direct correspondence to the events of the $E_\ell+E_\gamma$ spectrum.

\subsection{Leading nuclear effects \label{sec:nuclear_effects}}

In current neutrino detectors, most scattering events take place on nucleons bound inside nuclei. In the following, we estimate the relative importance of leading nuclear corrections. We represent these corrections by an ``impulse" model, where the nucleon-level process is described by assigning modified kinematics to the initial-state nucleon, and by enforcing constraints on the final-state nucleon kinematics. This model does not represent a complete description of complicated nuclear dynamics.\footnote{For work on nuclear structure effects for interaction cross sections, and further references, see Refs.~\cite{Day:1990mf,Capuzzi:1991qd,Donnelly:1998xg,DePace:2003spn,DePace:2004cr,Amaro:2010sd,Meucci:2011vd,Bodek:2011ps,Martini:2011wp,Martini:2013sha,Nieves:2013fr,Carlson:2014vla,Benhar:2015wva,Benhar:2015ula,Rocco:2015cil,VanCuyck:2016fab,Pastore:2017uwc,NuSTEC:2017hzk,Lynn:2019rdt}.} However for GeV (anti)neutrino energies and for the nuclei of interest (e.g., carbon, oxygen, and argon), it is expected to capture the leading effects and is sufficient for our purposes. In particular, we demonstrate that the dominant corrections cancel in critical cross-section ratios, in accordance with the general arguments in Sec.~\ref{sec:mass_expansion}. We evaluate separately the effects due to the binding/removal energy, Fermi motion, and Pauli blocking. A detailed comparison to recent developments in nuclear physics would require more elaborate calculation including final-state interactions, long-range correlations, and meson exchange currents. Some of these effects are taken into account in modern neutrino event generators~\cite{Andreopoulos:2009rq,GENIE:2021npt,Buss:2011mx,Hayato:2009zz,Hayato:2002sd,Golan:2012wx,Gran:2013kda,Megias:2016fjk}. 

According to the prescription of Smith and Moniz~\cite{Smith:1972xh}, a bound nucleon is described by changing the free-nucleon calculation by the replacement of the initial nucleon mass $M \to M-E_b$.\footnote{The precise replacement ansatz depends on a conventional choice of hadronic amplitude basis and of independent kinematic variables~\cite{Bhattacharya:2011ah}. This choice does not affect the present discussion.} Since the limit $E_b \to 0$ is smooth, the leading term in the expansion is of order $E_b/M$ and cancels in the flavor ratio when lepton-mass corrections are ignored. For the relevant range of $E_b$, of order ${\rm few} \times 10\,{\rm MeV}$, binding effects on the absolute cross sections are $\sim {\rm few} \times 1\, \%$ or below.

To estimate the corrections due to Fermi motion, we boost to the rest frame of the initial-state struck nucleon. Instead of the lab frame (anti)neutrino energy $E_\nu$, the nucleon in its rest frame sees a neutrino with energy 
\begin{equation}
\sqrt{E_\nu^2 \sin^2 \theta + \left(1+\frac{\vec{k}^2}{M^2}\right) \left(E_\nu \cos \theta - |\vec{k}|\right)^2},
\end{equation}
and the nuclear cross section is given by an average over the initial nucleon angle $\theta$ and summation over all nucleon states up to the Fermi momentum, $|\vec{k}| \le k_F$ (for numerical evaluations, we take a value for the $^{40}\mathrm{Ar}$ nucleus $k_F \approx 240~\mathrm{MeV}$). Consequently, Fermi motion results in relative corrections of order $k_F^2/M^2$ and $k_F^2/E_\nu^2$. Such corrections could become large at very small (anti)neutrino energies close to the muon production threshold, which we illustrate in Table~\ref{tab:Fermi_motion}. For flavor ratios, relative changes are much smaller than for cross sections themselves. A similar effect applies to all distributions presented in this paper for the (anti)neutrino energies of interest.
\begin{table}[tb]
 \centering 
 \begin{tabular}{c|c|c|c|c|c|c}
 \begin{tabular}{c}$E_\nu$, GeV\end{tabular} & $ \frac{\delta \sigma}{\sigma}$, \% & $\left(\frac{\delta \left( {\sigma_e}/{\sigma_\mu} \right)}{{\sigma_e}/{\sigma_\mu}}\right)_\mathrm{LO}$, \% & $\left(\frac{\delta \left( {\sigma_e}/{\sigma_\mu} \right)}{{\sigma_e}/{\sigma_\mu}}\right)_\mathrm{NLO}$, \% \\ \hline
 $0.6$ & $\lesssim2.5$ & $\lesssim 0.06$ & $\lesssim 0.05$ \\
 $2.0$ & $\lesssim0.3$ & \hspace{1.9mm}$\lesssim 0.003$ & $\lesssim 0.01$
 \end{tabular}
 \caption{Relative effects of the Fermi motion on the total unpolarized cross sections and flavor ratios at leading (LO) and next-to-leading (NLO) orders for (anti)neutrino beam energies $600$ MeV and $2$ GeV. Upper limits for both neutrino and antineutrino (and for both muon and electron flavor in the second column) are shown in this table. \label{tab:Fermi_motion}}
\end{table}

Pauli blocking effects similarly scale as $k_F^2$, but because they are concentrated at low $Q^2$ where the cross section is sizable, the impact on total cross sections can be significant. In Table~\ref{tab:Pauli_blocking}, we present the effect on the cross section as well as the change of flavor ratios. We represent the Pauli blocking effect by retaining only that region of phase space where the final-state nucleon momentum has $|\vec{k}| > k_F$~\cite{Smith:1972xh}. This effect only slightly depends on the (anti)neutrino flavor at energies much larger than the muon mass. Going to lower (anti)neutrino energies, the effect increases~\cite{Martini:2016eec,Ankowski:2017yvm,Nikolakopoulos:2019qcr}. Pauli blocking corrections to flavor ratios are below or comparable to hadronic uncertainties propagated from total inclusive cross sections.
\begin{table}[htb]
 \centering 
 \begin{tabular}{c|c|c|c|c|c|c}
 \begin{tabular}{c}$E_\nu$, GeV\end{tabular} & & $ \frac{\delta \sigma}{\sigma}$, \% & $\left(\frac{\delta \left( {\sigma_e}/{\sigma_\mu} \right)}{{\sigma_e}/{\sigma_\mu}}\right)_\mathrm{LO}$, \% & $\Bigg | \left(\frac{\delta \left( {\sigma_e}/{\sigma_\mu} \right)}{{\sigma_e}/{\sigma_\mu}}\right)_\mathrm{NLO} \Bigg |$, \% \\ \hline
 $0.6$ & \begin{tabular}{c} $\nu$ \\ $\bar{\nu}$\end{tabular} & \begin{tabular}{c} $ -13$ \\ $-37$\end{tabular} & \begin{tabular}{c}\hspace{-2.9mm} $0.15$\\\hspace{-2.9mm} $0.21$ \end{tabular} & \begin{tabular}{c}$\lesssim0.2$\\ $\lesssim 0.2$ \end{tabular} \\ $2.0$ & \begin{tabular}{c} $\nu$ \\ $\bar{\nu}$\end{tabular} & \begin{tabular}{c} $-11$ \\ $-18$ \end{tabular} & \begin{tabular}{c} \hspace{-3.9mm} $-0.002$ \\ $0.009$ \end{tabular} & \begin{tabular}{c} $\lesssim 0.01$ \\ $\lesssim 0.01$ \end{tabular} 
 \end{tabular}
 \caption{Relative effects of Pauli blocking on the total unpolarized cross sections and flavor ratios for (anti)neutrino beam energies $600$ MeV and $2$ GeV. \label{tab:Pauli_blocking}}
\end{table}

In summary, while nuclear effects can have a sizable impact on absolute cross sections, they do not significantly change flavor ratios of total inclusive cross sections and of spectra with respect to the hadronic momentum transfer. This statement remains true in the presence of radiative corrections, following from the results in Sec.~\ref{sec:inclusive_xsection}. Returning to the discussion of Sec.~\ref{sec:mass_expansion}, we may summarize the impact of nuclear corrections on the flavor ratio as a shift in ${\cal B}_0(E_\nu = 2\,{\rm GeV}) = -0.28 \to -0.32$~\cite{Bhattacharya:2011ah}, for the combined effects of binding energy, Fermi motion and Pauli blocking.

\subsection{Comparison to (anti)neutrino-nucleus quasielastic data \label{sec:comparison_to_data}}

Our results predict significant radiative corrections to the charged-current elastic scattering process. These corrections are especially important in $\nu_e$ versus $\nu_\mu$ flavor ratios owing to the sparsity of $\nu_e$ cross-section measurements, the potentially larger contributions for $\nu_e$ versus $\nu_\mu$, and the direct impact of corrections on $\nu_\mu \to \nu_e$ oscillation analyses~\cite{Tomalak:2021qrg}. However, radiative corrections are also present in the more-abundant $\nu_\mu$ scattering data. The elastic scattering process has been studied with muon flavor neutrinos and antineutrinos scattering off free protons~\cite{TejinNature}, weakly bound light nuclei~\cite{Mann:1973pr,Barish:1977qk,Baker:1981su,Miller:1982qi,Kitagaki:1983px,Allasia:1990uy,Kitagaki:1990vs,Meyer:2016oeg}, and tightly bound medium-sized nuclei~\cite{Gran:2006jn,MiniBooNE:2010bsu,MINERvA:2013kdn,MINERvA:2014ypj,MINOS:2014axb,T2K:2014hih,T2K:2015ujp,T2K:2016jor,T2K:2017qxv,MINERvA:2017dzh,MINERvA:2018hba,MINERvA:2018hqn,T2K:2018rnz,MINERvA:2019ope,MINERvA:2019gsf,T2K:2020jav,T2K:2020sbd,MicroBooNE:2020akw,MicroBooNE:2020fxd}. Here, we examine how radiative corrections affect the interpretation of these measurements, which were analyzed using tree-level cross-section models.

We consider radiative corrections in several large datasets, all of which have been collected using hydrocarbon targets. A complication in implementing the radiative corrections is that, for measurements on nuclei, the same observable hadronic final state can originate from different nucleon-level processes: events originating from nucleon-level elastic scattering are observed together with other processes such as multinucleon knockout, or pion production followed by pion absorption in the target nucleus. The collection of these events is referred to as ``CCQE-like" or ``CC0$\pi$" in the literature. In this study of CC0$\pi$ $\nu_\mu$ and $\overline{\nu}_\mu$ interactions, we define CC$0\pi$ to be an event with a single appropriately charged muon and no other charged leptons, no mesons, and any number of nucleons in the final state, after final-state interactions in the nucleus have occurred. To compare the effects from radiative corrections with experimental data, we embed our predictions into neutrino event generators that include a default model for nuclear effects and predictions for inelastic feed-down into the signal.

\begin{figure}[b!]
\centering
\includegraphics[width=0.4\textwidth,page=1,trim=4mm 4mm 0mm 12mm,clip]{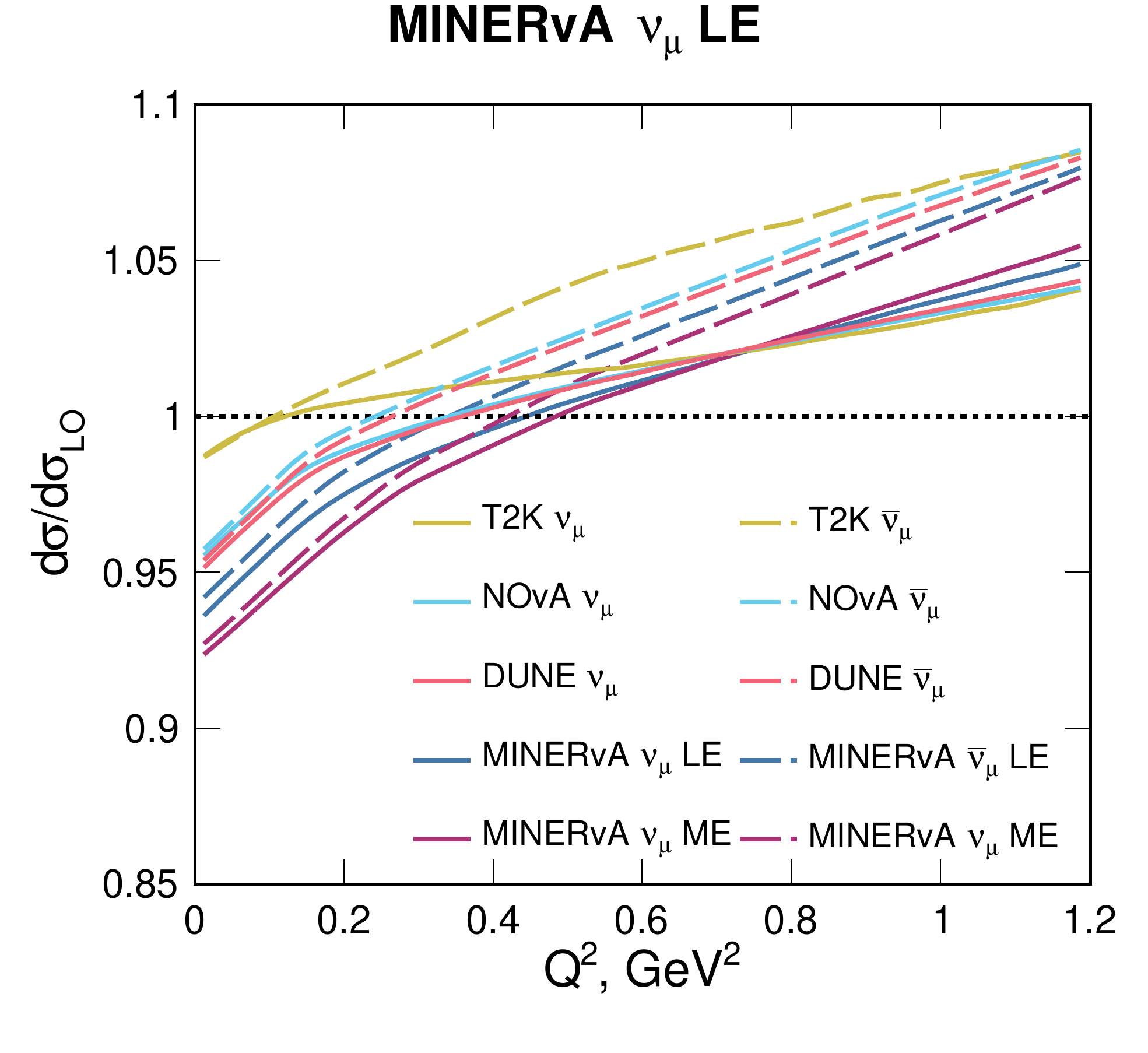}
\caption{Estimated cross-section ratio to the tree-level result for the CC$0\pi$ process in T2K, NO$\nu$A, DUNE, and MINERvA (LE and ME flux configurations) experiments after averaging over the corresponding neutrino flux, as a function of $Q^2 = -\left(p_\nu - p_\mu\right)^2$. See text for details.}
\label{fig:CCQElikeRatios}
\end{figure}

We implement radiative corrections as follows. For each neutrino energy, we compute the unpolarized differential neutrino-nucleon cross section for $\nu_\mu$ (and $\overline{\nu}_\mu$) scattering, with $Q^2$ defined using lepton kinematics [i.e., $Q^2 \equiv 2M (E_\nu - E_\ell)$]. We form the ratio of this result to the tree-level cross section. For $E_\nu= 600\,{\rm MeV}$ and $2\,{\rm GeV}$, this ratio is displayed in Fig.~\ref{fig:xsection_inclusive_plots_muon}, labeled ``$E_\ell$ spectrum''. In the sum over initial-state nucleons of the nuclear model, radiative corrections are applied by multiplying the tree-level result by this ratio, in the rest frame of each struck nucleon. A flux average is then performed over neutrino energy for the relevant neutrino beam. The logarithmically enhanced contributions to radiative corrections, $\sim \ln(E_\nu^2/m_\mu^2)$, depend only on lepton kinematics and are independent of hadronic structure. We therefore apply the same $Q^2$-dependent correction determined from ``true CCQE'' events to all events that pass a CC0$\pi$ selection. For multinucleon events in the nuclear model, the struck nucleon (neutron for $\nu$, proton for $\overline{\nu}$) is the one considered participating in the interaction, and radiative corrections are again applied in the rest frame of the struck nucleon. In this evaluation for muon flavor, we neglect events in the phase-space region outside elastic kinematics (cf. Fig.~\ref{fig:Q2_issues}); explicit evaluation shows that such events change muon (electron) flavor cross sections in the relevant energy range at permille (percent) level or below.

Neutrino interactions are generated with the NEUT~\cite{Hayato:2021heg} and GENIE~\cite{Andreopoulos:2015wxa} neutrino-event generators\footnote{NEUT 5.5.0 with Valencia model 1p1h and 2p2h and GENIEv3 G18\_10a\_02\_11a.} for the MINERvA, NO$\nu$A, and T2K muon (anti)neutrino fluxes on a plastic scintillator (CH) target, and for DUNE~\cite{DUNE:2021tad} on a $^{40}\text{Ar}$ target, using NUISANCE~\cite{Stowell:2016jfr}. For MINERvA, both the low-energy (LE) and medium-energy (ME) neutrino flux configurations are included. NEUT and GENIE show very similar features, and only predictions using NEUT are presented in the following figures. The resulting ratios to tree-level cross sections for CC0$\pi$ events from the T2K, NO$\nu$A, MINERvA, and DUNE experiments are shown in Fig.~\ref{fig:CCQElikeRatios}. A $4-7~\%$ suppression is observed at low $Q^2$ for the NO$\nu$A, MINERvA, and DUNE experiments, all of which have significant neutrino flux in the $E_\nu > 1~\text{GeV}$ region.

\begin{figure}[hbtp]
\centering
\begin{subfigure}[b]{0.45\textwidth}
\includegraphics[width=\textwidth,page=3,trim=0mm 0mm 10mm 5mm,clip]{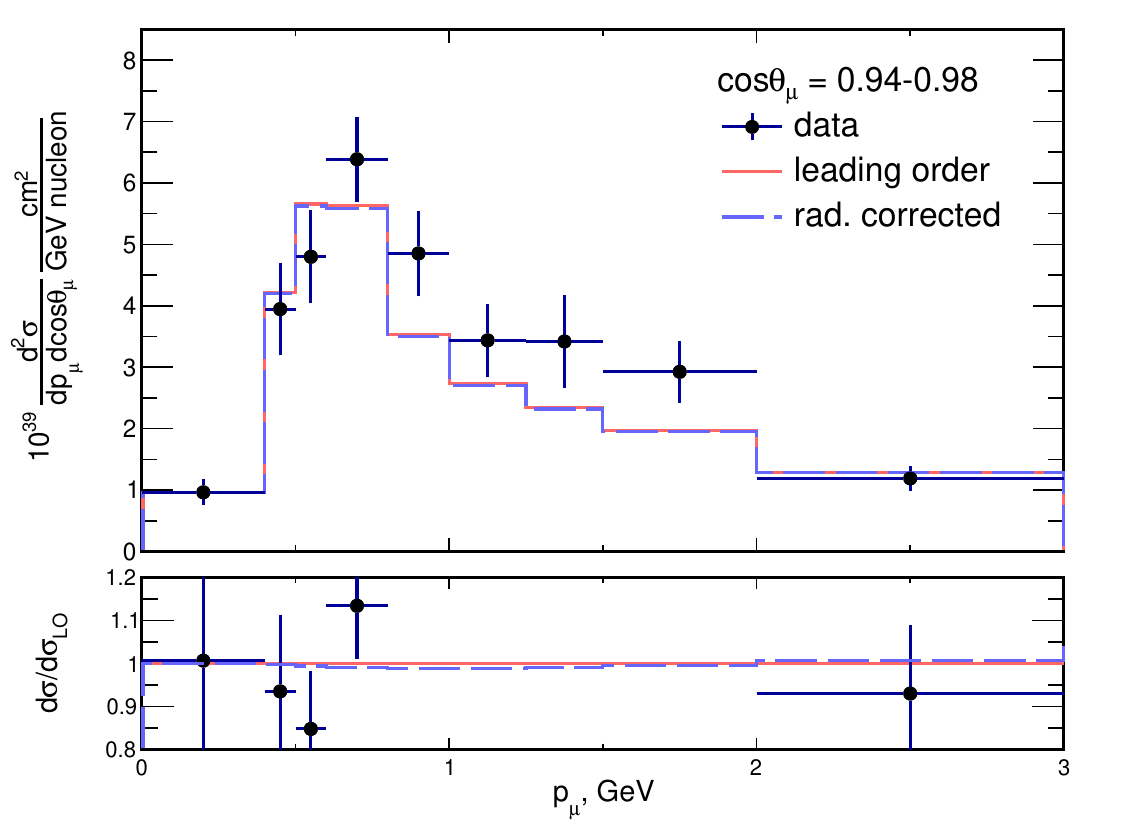}
\end{subfigure}
\begin{subfigure}[b]{0.45\textwidth}
\includegraphics[width=\textwidth,page=5,trim=0mm 0mm 10mm 5mm,clip]{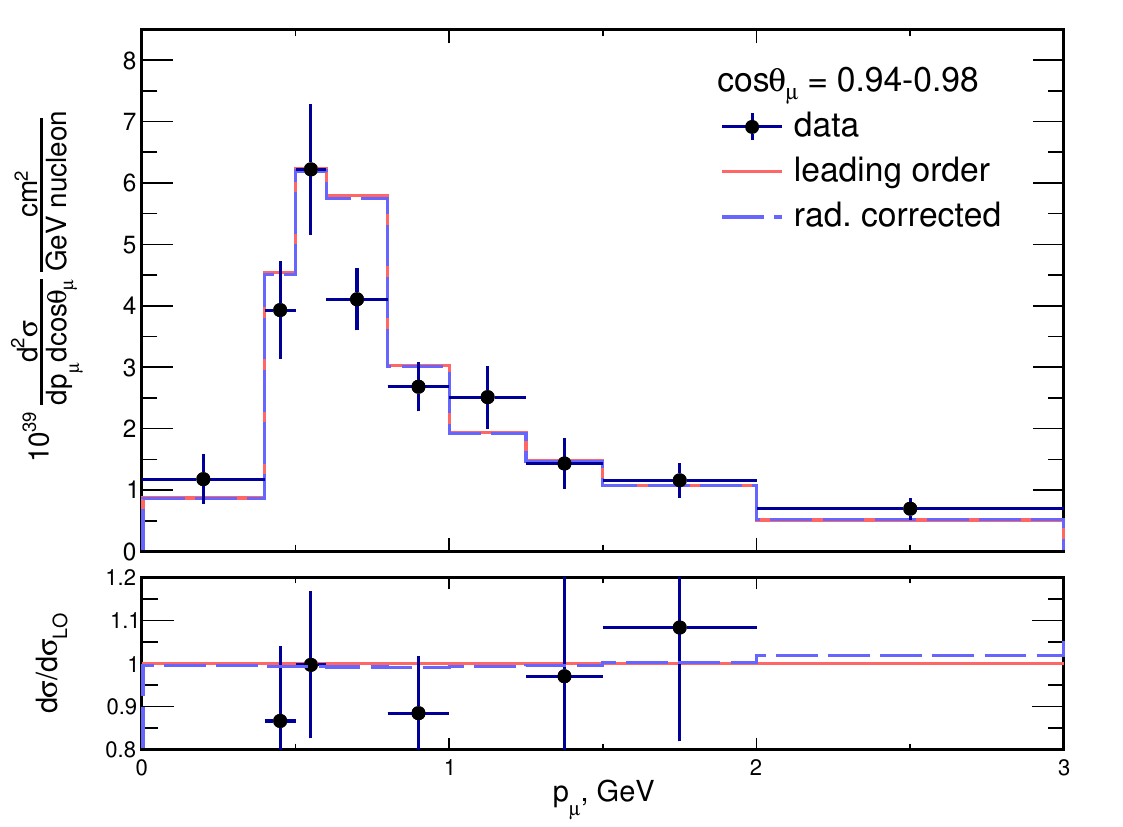}
\end{subfigure}
\caption{The effect of radiative corrections to the tree-level prediction for the $\cos\theta_\mu=0.94-0.98$ bin of the T2K ND280 $\nu_\mu$ (left) and $\overline{\nu}_\mu$ (right) CCQE-like data~\cite{T2K:2020sbd} is shown for the cross section per nucleon of a CH target.}
\label{fig:CCQElikeT2Kfwd}
\end{figure}

\begin{figure}[hbtp]
\centering
\begin{subfigure}[b]{0.45\textwidth}
\includegraphics[width=\textwidth,page=4,trim=0mm 0mm 10mm 5mm,clip]{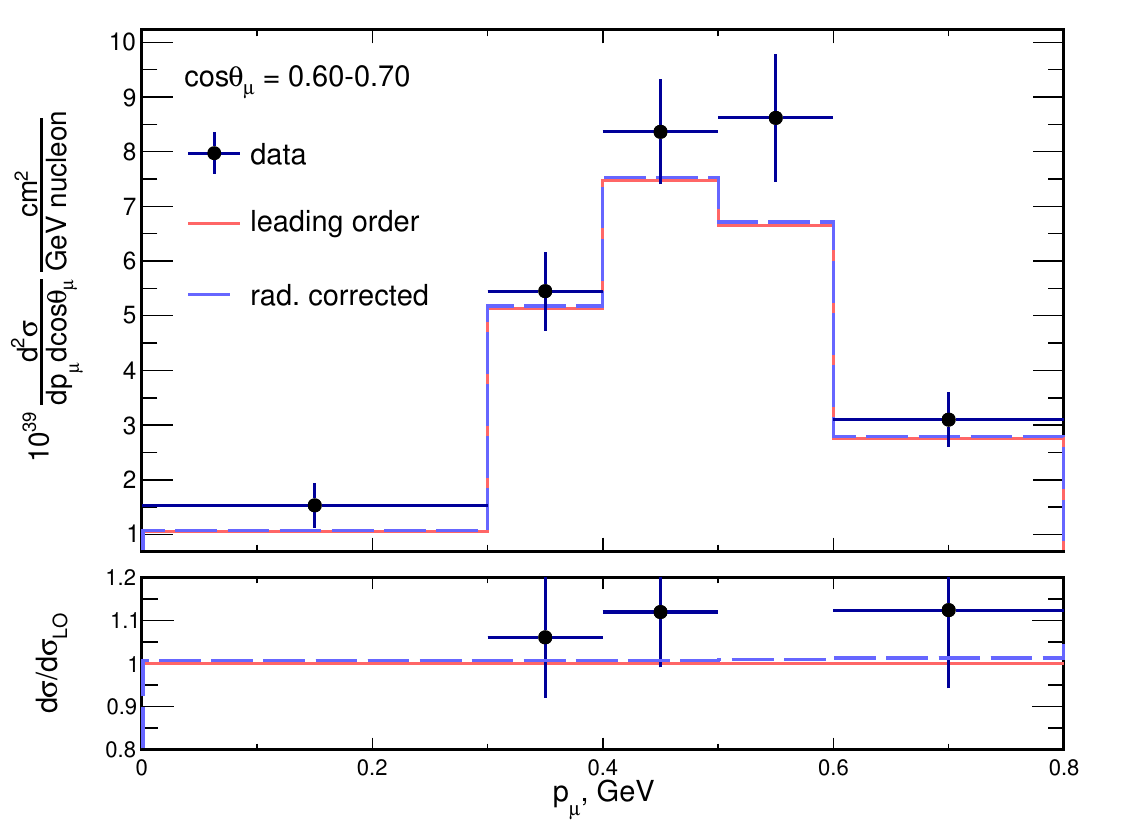}
\end{subfigure}
\begin{subfigure}[b]{0.45\textwidth}
\includegraphics[width=\textwidth,page=6,trim=0mm 0mm 10mm 5mm,clip]{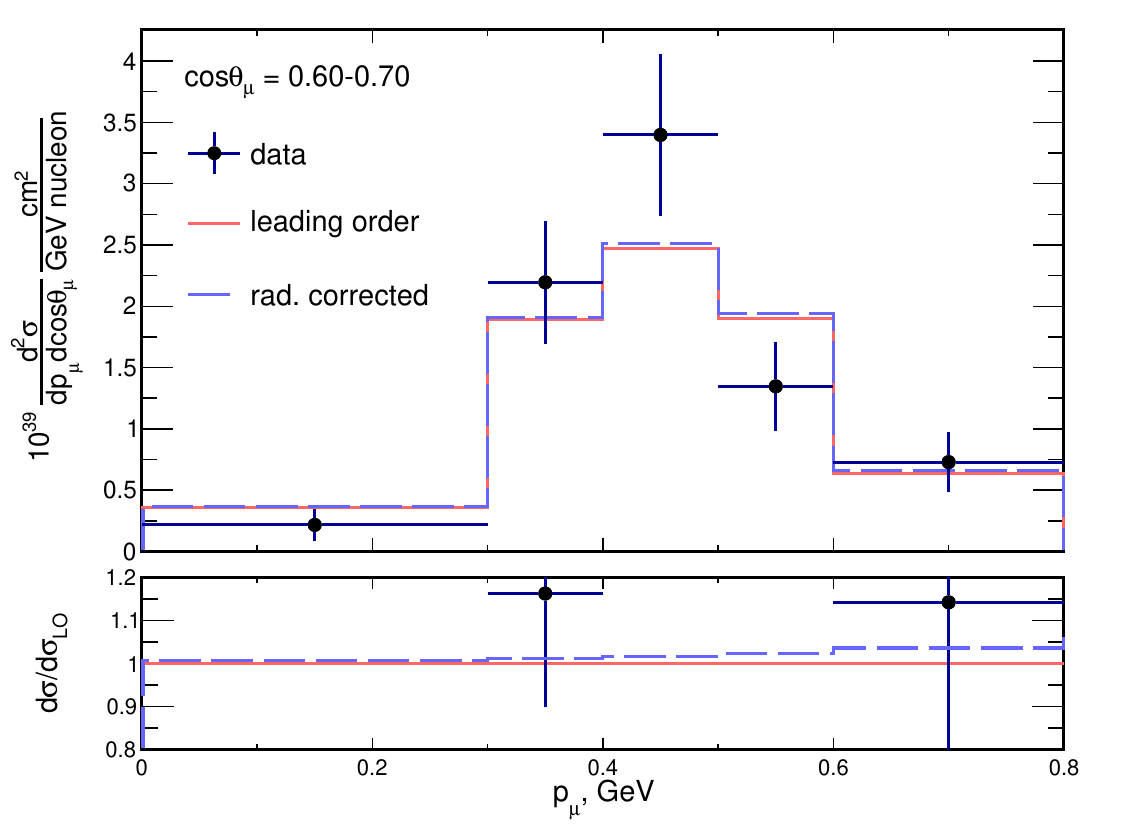}
\end{subfigure}
\caption{The effect of radiative corrections to the tree-level prediction for the $\cos\theta_\mu=0.60-0.70$ bin of the T2K ND280 $\nu_\mu$ (left) and $\overline{\nu}_\mu$ (right) CCQE-like data~\cite{T2K:2020sbd} is shown for the cross section per nucleon of the CH target.}
\label{fig:CCQElikeT2Kbwd}
\end{figure}

\begin{figure}[ptb]
\centering
\begin{subfigure}[b]{0.45\textwidth}
\includegraphics[width=\textwidth,page=1,trim=0mm 0mm 10mm 6mm,clip]{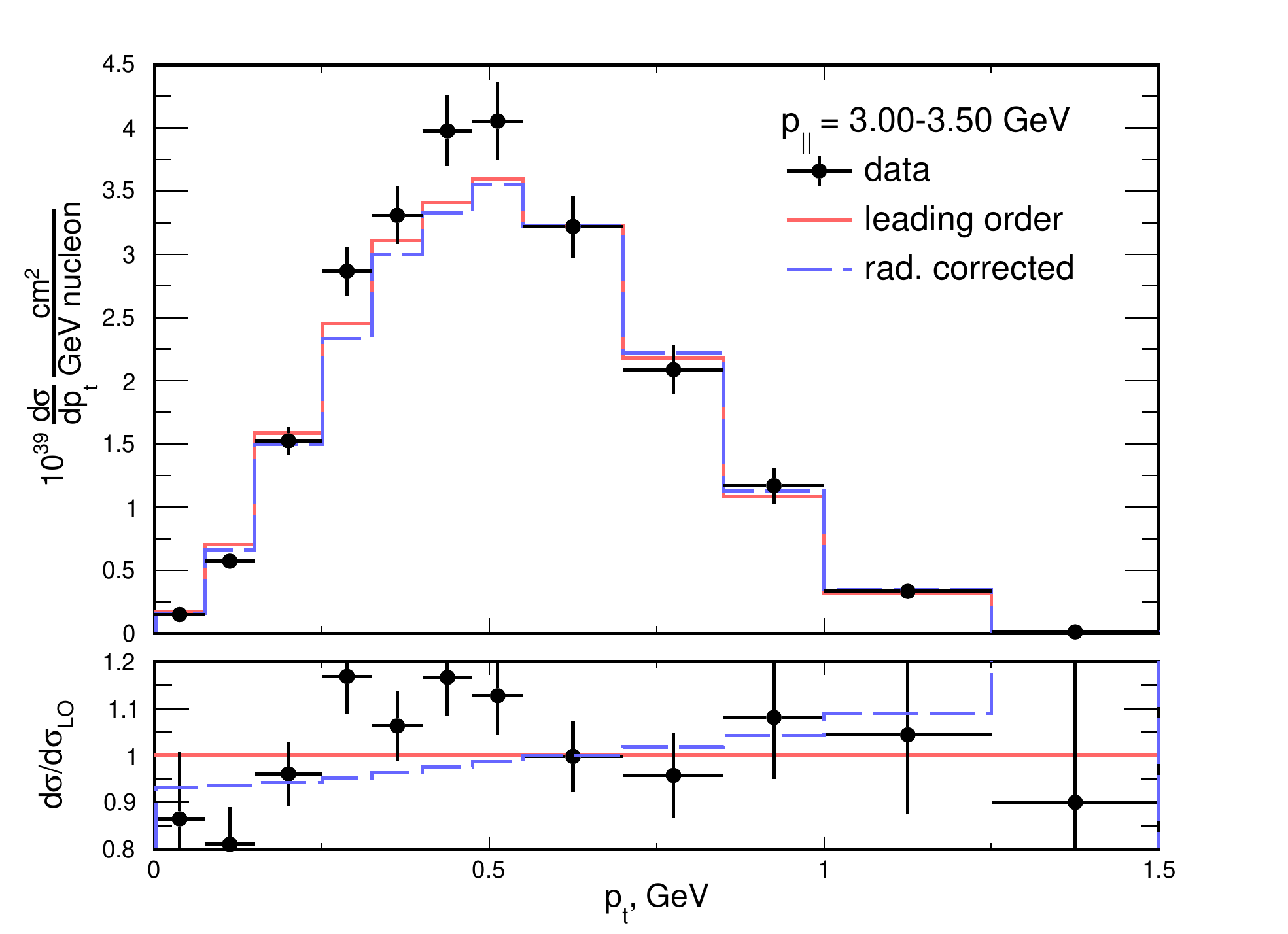}
\end{subfigure}
\begin{subfigure}[b]{0.45\textwidth}
\includegraphics[width=\textwidth,page=2,trim=0mm 0mm 10mm 6mm,clip]{minerva_t2k_allw.pdf}
\end{subfigure}
\caption{The effect of radiative corrections to the tree-level prediction for one slice of the MINERvA LE $\nu_\mu$ CCQE-like data~\cite{MINERvA:2018hqn} (left) and $\overline{\nu}_\mu$ CCQE-like data~\cite{MINERvA:2018vjb} (right) is shown for the cross section per nucleon of the CH target.}
\label{fig:CCQElikeMINERvALE_both}
\end{figure}

\begin{figure}[ptb]
\centering
\begin{subfigure}[b]{0.45\textwidth}
\includegraphics[width=\textwidth,page=2,trim=0mm 0mm 10mm 5mm,clip]{min_ch_me.pdf}
\end{subfigure}
\begin{subfigure}[b]{0.45\textwidth}
\includegraphics[width=\textwidth,page=1,trim=0mm 0mm 10mm 5mm,clip]{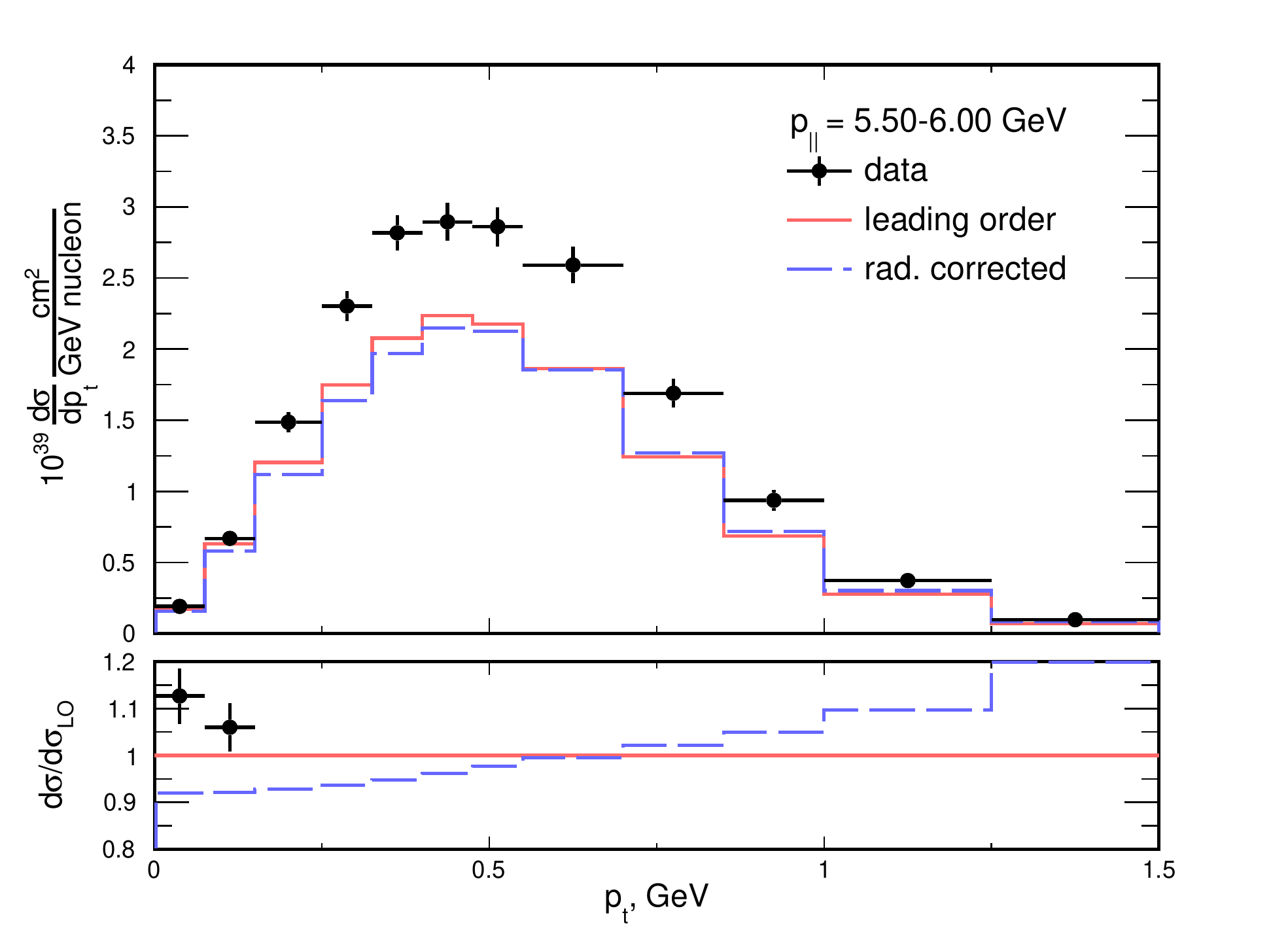}
\end{subfigure}
\caption{The effect of radiative corrections to the tree-level prediction for the same slice of the MINERvA ME $\nu_\mu$~\cite{MINERvA:2019gsf} (left) and $\bar{\nu}_\mu$~\cite{Bashyal:2021tzd,MINERvA_ME_numub_CC0pi} (right) CCQE-like data is shown for the cross section per nucleon of the CH target.}
\label{fig:CCQElikeMINERvAMEnu}
\end{figure}

The kinematic dependence of radiative corrections is compared to experimental data in Figs.~\ref{fig:CCQElikeT2Kfwd}-\ref{fig:CCQElikeMINERvAMEnu}. For the T2K experiment, the data are presented as a function of the momentum of the muon, $p_\mu$, in a fixed range of the cosine of the lepton scattering angle relative to the neutrino beam direction, $\cos\theta_\mu$. We choose illustrative $\cos\theta_\mu$ bins close to the peak in the $\cos\theta_\mu$ distribution from Ref.~\cite{T2K:2020sbd}. For the MINERvA experiment, the data are presented as a function of the transverse projection of the muon momentum along the neutrino beam direction, $p_t$, in a fixed range of the longitudinal projection $p_{||}$ (we choose illustrative $p_{||}$ bins close to the broad peak in the $p_{||}$ distributions from Refs.~\cite{MINERvA:2018hqn,MINERvA:2018vjb}). For the comparison to T2K $\nu_\mu$ and $\overline{\nu}_\mu$ data in Figs.~\ref{fig:CCQElikeT2Kfwd} and ~\ref{fig:CCQElikeT2Kbwd}, experimental error bars are large compared to the small radiative corrections displayed in Fig.~\ref{fig:CCQElikeRatios}. For the LE MINERvA data, opposite effects are observed in the low-$p_t$ and high-$p_t$ regions, corresponding to the low-$Q^2$ and high-$Q^2$ regions of Fig.~\ref{fig:CCQElikeRatios}. For the ME MINERvA data, the effects are enhanced further at low $p_t$ due to the larger neutrino energy (cf. Fig.~\ref{fig:CCQElikeRatios}); as for the LE MINERvA data, the effect for $\overline{\nu}_\mu$ is stronger than for $\nu_\mu$.

Previous studies have used the kinematic dependence of data-theory comparisons to constrain uncertain nucleon and nuclear parameters, e.g., providing measurements of axial nucleon form factors~\cite{Mann:1973pr,Barish:1977qk,Miller:1981fa,Baker:1981su,Kitagaki:1983px,Belikov:1983kg,Bodek:2007ym,Kuzmin:2007kr,Bhattacharya:2011ah,Meyer:2016oeg,Alvarez-Ruso:2018rdx,Megias:2019qdv,TejinNature}, multinucleon processes, or other nuclear effects~\cite{Rodrigues:2015hik,MINERvA:2021wjs}. The corrections to tree-level predictions in this work can be used to determine the impact of radiative corrections on these measurements and will be important for the interpretation of future precise measurements with neutrinos.

\section{Summary and outlook \label{sec:summ}}

In this paper, we have developed the framework for radiative corrections in charged-current (anti)neutrino-nucleon elastic scattering at GeV (anti)neutrino energies. Exploiting effective field theories, we have shown that scattering cross sections factorize into soft, collinear, and hard functions. The soft and collinear functions contain flavor-dependent large logarithm enhancements and depend on detailed experimental conditions but can be computed perturbatively. The hard function is subject to hadronic uncertainty but is independent of the charged lepton mass and cancels in ratios of cross sections for different lepton flavors involving the same hadronic kinematics.

We have provided analytic expressions for the soft function and its small lepton-mass limit. We have performed the first calculations of the collinear function separately for virtual corrections and for the radiation of one collinear photon, and have provided small-angle and small lepton-mass limits. We have illustrated the factorization theorem in an exactly calculable model corresponding to a nonrelativistic nucleon and ultrarelativistic lepton.

We have expressed the hard function for arbitrary kinematics in terms of a model-independent amplitude decomposition for charged-current (anti)neutrino-nucleon elastic scattering. We have determined the invariant amplitudes in a gauge-invariant hadronic model that reproduces soft and collinear regions of the one-loop charged-current elastic process. We have performed detailed error analysis including hadronic and perturbative uncertainties within this model. The tree-level hard function is the main source of uncertainty in absolute cross sections. However, this uncertainty largely cancels in important cross-section ratios, in particular the ratio of electron to muon cross sections. Remaining uncertainties are at permille level.

Our results confirm a naive estimate of radiative corrections by powers of leading logarithms. Exclusive cross sections at GeV neutrino energies with electron flavor can change by $\sim10-20$~\% when only soft photons are included in the observable and up to $\sim$5~\% when both soft and collinear radiation are included. The corresponding changes in muon (anti)neutrino cross sections are typically smaller and can reach up to $\sim$5~\%. Inclusive cross sections at fixed hadronic momentum transfer are subject to smaller radiative corrections, due to the cancellation between virtual and real contributions after the inclusion of hard photons. However, inclusive results vary by $\sim10-20$~\% ($5-10$~\%) for electron (muon) flavor depending on the way that kinematics is reconstructed from leptons and photons.

An important result from our studies for precision accelerator neutrino oscillation program is that the total cross section as a function of (anti)neutrino energy, inclusive of real photon emission, is very similar for electron and muon (anti)neutrino events, as Figs.~\ref{fig:xsection_inclusive_plots_electron}, \ref{fig:xsection_inclusive_plots_muon}, and \ref{fig:xsection_ratio_muon_over_electron_inclusive} illustrate. However this simple result is achieved only after summing inclusively over distinct kinematic configurations. Electron-flavor and muon-flavor cross sections receive significant, and different, corrections as a function of kinematics that must be carefully accounted for when experimental cuts and efficiency corrections are applied in a practical experiment. It is also important to carefully match the theoretical calculation of radiative corrections to experimental conditions since radiative corrections depend strongly on the treatment of real photon radiation.

The double ratio of the neutrino over antineutrino cross sections to the tree-level result is very close to unity for exclusive observables and for inclusive observables corresponding to the same value of hadronic momentum transfer. This situation can be traced to the fact that the collinear function is the same in neutrino and antineutrino scattering and that enhanced perturbative contributions to the soft function are similarly the same. However, differences in the kinematic reconstruction of electron-flavor and muon-flavor events in practical detectors lead to significant deviations of this double ratio from the unity.

Our studies have shown that the probability for a muon (anti)neutrino event to be misidentified as an electron (anti)neutrino event due to the presence of energetic collinear photons is of order $10^{-4}$ or below. Radiative cross sections with noncollinear hard photons have typically percent level and should be accounted for in precise measurements with accelerator (anti)neutrino beams.

Our work can be directly applied to analyze (anti)neutrino-nucleon scattering processes, accounting for QED radiative corrections for the first time. An important application is the extraction of the nucleon axial form factor and corresponding axial radius from neutrino scattering data at GeV energies.

A primary motivation for our work is the analysis of neutrino oscillation signals using nuclear targets. Although the study was performed with (anti)neutrino-nucleon scattering, important cross-section ratios are insensitive to the explicit form of the nonperturbative hard function and similar conclusions are valid for scattering on nuclei. First, the radiative corrections to exclusive cross sections at large momentum transfer, and the corresponding flavor ratios, are dominated by large perturbative logarithms that are independent of nuclear or hadronic parameters. Second, the radiative corrections to inclusive cross sections are governed by an expansion in small lepton mass [cf. Eq.~(\ref{eq:sigmaKLN})], which implies small modifications to radiative corrections from nuclear effects and small theoretical uncertainties on flavor ratios of inclusive cross sections.

Our work paves the way for the rigorous incorporation of radiative corrections at accelerator neutrino experiments, an important missing ingredient for the achievement of the desired precision in extractions of the Pontecorvo-Maki-Nakagawa-Sakata matrix parameters, squared mass differences, and discovery of the $CP$ violation in the lepton sector.

\vskip 0.2in
\noindent

{\bf Acknowledgements}
\vskip 0.1in
\noindent We thank Kaushik Borah for an independent validation of antineutrino-proton cross-section expressions and Emanuele Mereghetti and Ryan Plestid for discussions. This work was supported by the U.S. Department of Energy, Office of Science, Office of High Energy Physics, under Grants No. DE-SC0019095 and No. DE-SC0008475. Fermilab is operated by Fermi Research Alliance, LLC under Contract No. DE-AC02-07CH11359 with the United States Department of Energy. O.T. acknowledges support by the Visiting Scholars Award Program of the Universities Research Association, and thanks theory groups at Fermilab and Institute for Nuclear Physics at JGU Mainz for warm hospitality. O.T. is supported by the U.S. Department of Energy through the Los Alamos National Laboratory (LANL). Los Alamos National Laboratory is operated by Triad National Security, LLC, for the National Nuclear Security Administration of U.S. Department of Energy (Contract No. 89233218CNA000001). This research is funded by LANL’s Laboratory Directed Research and Development (LDRD/PRD) program under Project No. 20210968PRD4. Q.C. and O.T. acknowledge KITP Graduate Fellow program supported by the Heising-Simons Foundation, the Simons Foundation, and National Science Foundation Grant No. NSF PHY-1748958. R.J.H. acknowledges support from the Neutrino Theory Network at Fermilab during the early stages of this work. K.S.M. acknowledges support from a Fermilab Intensity Frontier Fellowship during the early stages of this work and from the University of Rochester's Steven Chu Professorship in Physics. FeynCalc~\cite{Mertig:1990an,Shtabovenko:2016sxi}, LoopTools~\cite{Hahn:1998yk}, Mathematica~\cite{Mathematica}, and DataGraph were useful in this work.

\bibliography{neutrino_radcorr_theory}

\end{document}